\documentclass[preprint,nofootinbib]{revtex4}
\usepackage[dvips]{graphicx}
\usepackage{amsmath,amssymb,axodraw,graphicx,color}

\newcommand{\be}{\begin{equation}}
\newcommand{\ee}{\end{equation}}
\newcommand{\bea}{\begin{eqnarray}}
\newcommand{\eea}{\end{eqnarray}}
\newcommand{\bref}[1]{(\ref{#1})}
\newcommand{\ra}{\rangle}
\newcommand{\bsigma}{\boldsymbol{\sigma}}
\newcommand{\la}{\langle}

\newcommand{\pa}{\partial}
\newcommand{\bz}{\boldsymbol{\zeta}}
\newcommand{\bomega}{\boldsymbol{\Omega}}
\begin{document}
\begin{flushright}
\today
\end{flushright}
\title{Searching for New Physics beyond the Standard Model in Electric Dipole Moment}
%%%%%%%%%%%%%%%%%%%%%%%%%%%%%%%%%%%%%%%%%%%%%%%%%%%%%%%%%%%%%%%%%%%%%%
\author{Takeshi Fukuyama}
\email{fukuyama@se.ritsumei.ac.jp}
\affiliation{Department of Physics and R-GIRO,
Ritsumeikan University, Kusatsu, Shiga,
525-8577, Japan}
%%%%%%%%%%%%%%%%%%%%%%%%%%%%%%%%%%%%%%%%%%%%%%%%%%%%%%%%%%%%%%%%%%%%%%
%\author{Hiroaki Sugiyama}
%\email{hiroaki@fc.ritsumei.ac.jp}
%\affiliation{Department of Physics and R-GIRO,
%Ritsumeikan University, Kusatsu, Shiga,
%525-8577, Japan}
\maketitle
\textbf{Abstract}\\
This is the theoretical review of exploration of new physics beyond the Standard Model (SM) in electric dipole moment (EDM) in elementary particles, atoms, and molecule.
EDM is very important CP violating phenomenon and sensitive to new physics.

Starting with the estimations of EDM of quarks-leptons in the SM, we explore new signals beyond the SM. However, these works drive us to wider fronteer where we search fundamental physics using atoms and molecules and vice versa.

Paramagnetic atoms and molecules have great enhancement factors on electron EDM.
Diamagnetic atoms and molecules are very sensitive to nuclear P and T odd processes.

Thus EDM becomes the key word not only of New Physics but also of unprecedented fruitful collaboration among particle, atomic and molecular physics.

This review intends to help such collaboration over the wide range of physicists.
\section{Introduction}	
This article is a review of the search of new physics beyond the Standard Model (SM) concentrating on electric dipole moments (EDM) of elementary particles like neutron, proton, leptons, quarks as well as atoms and molecules. The presence of EDM implies T-odd and P-odd interactions. 
So if it exists, it indicates the direct T noninvariance as well as CP violation if CPT invariance is assumed.

It is very important that these fundamental EDMs are enhanced in paramagnetic atoms ($d_{atom}$) and molecules ($d_{molecule}$) which have an unpaired electron. Also in diamagnetic atoms and molecules, proton and neutron EDMs appear via Schiff moment due to CP-violating hadron interactions.

The discovery of CP violation in $K_L^0\rightarrow \pi^+\pi^-$ decay \cite{Christenson} in 1964 was an amazing event for the majority of theorists since the model at that time could not produce CP violation.
The introduction of CP phase in the mixing matrix by Kobayashi-Maskawa \cite{K-M} was 7 years after that, which becomes the unique origin of CP violation in the SM \cite{SM}.  This CP phase opened the door to new frontiers in a vast range of physics, especially in B factories: Belle at KEKB (KEK) and BaBar at PEP-II (SLAC).

CP violation in B mesons is measured by observing the asymmetry
\begin{eqnarray}
a_f(\tau)&\equiv & \frac{\Gamma (B^0(\tau)\rightarrow f)-\Gamma (\overline{B}^0(\tau)\rightarrow f)}{\Gamma (B^0(\tau)\rightarrow f)+\Gamma (\overline{B}^0(\tau)\rightarrow f)}\nonumber\\
&=&C_f\mbox{cos}(\Delta m\tau)-S_f\mbox{sin}(\Delta m\tau), 
\end{eqnarray}
where
\begin{equation}
C_f\equiv \frac{1-|\lambda|^2}{1+|\lambda_f|^2},~~S_f\equiv \frac{2\Im \lambda_f}{1+|\lambda_f|^2}.
\end{equation}
Here $B=B_d^0=|d\overline{b}>$ or $B=B_s^0=|s\overline{b}>$, and $f$ is a CP eigenstate such as $J/\psi K_S,~\pi^+\pi^-,~\rho K_S$. $\Im$ means an imaginary part.
For
$B_d^0\rightarrow J/\psi K_s$ \cite{BELLE} \cite{BABAR}, 
\begin{equation}
\Im \lambda_{\psi K}=0.734\pm0.054 ,
\end{equation}
with world average.
It may be more advantageous in searching for new physics to consider the SM loop suppressed process like
$B^0\rightarrow \phi K^0$ etc. However, these results seem to  be consistent with the CKM mechanism \cite{Nir}. 

As we will show, the EDM values predicted by the SM are very tiny because they appear first in three loops (quarks) and four loops (leptons) and are far smaller than the upper limit of the present and near future experiments. \footnote{ In the broad sense, there is another CP phase called the $\theta$ term \cite{t'Hoft} in the SM, playing an essential role in especially the EDM of diamagnetic atoms (see section {\bf \bref{condensate}}).}
On the other hand, there are some physical phenomena which suggest new physics beyond the SM other than neutrino oscillation experiments.
The anomalous muon magnetic moment, $a_\mu\equiv (g_\mu-2)/2$ ($g$ is defined by \bref{g}) , is one such example \cite{Hagiwara}
\begin{equation}
a_\mu^{EXP}-a_\mu^{SM}=(26.1\pm 8.0)\times 10^{-10}, 
\end{equation}
corresponding to a $3.3\sigma$ discrepancy from the SM.
There are also other indirect problems of the SM like the observed baryon asymmetry, $n_B/n_\gamma=1\times 10^{-10}$. Indeed,  in the SM, CP violation is parametrized by the Jarlskog invariant, which is too tiny to produce this amount of asymmetry; we need other CP violating terms.
The other implicit deficiencies of the SM are Dark Matter candidates and the hierarchy problem etc.

Under these situations, the EDM is very important since some models beyond the SM give rather marginal predictions on the  electron and neutron EDMs on the upper bounds of ongoing experiments.

So new models are required to recover all such discrepancies. Furthermore, it must reproduce much larger phenomena which the SM predicts beautifully like, for instance, flavour changing neutral currents (FCNCs) and other vast low energy physics phenomena.

Here we point out a peculiar property of the EDM:

As is well known, EDMs of elementary particles are enhanced in atoms and molecules. In this sense, the EDM provides an unprecedented strategy of using atoms and molecules for the search of fundamental properties of elementary particles.

Several review works on this subjects have been already published \cite{Khriplovich}\cite{G-F}\cite{P-R}\cite{K-L}.
New features of this review is that it is written by the author who is studying new physics beyond the SM, and, therefore, emphasis is on this point. However, EDM studies drive us necessarily to a wide range of physics (and chemistry), particle physics, atomic and molecular physics.
The great achievements are possible only by the collaboration of theoretical and experimental scientists over this wide range of fields.
Under these situations, we try in this review to make a small but significant bridge between these wide communities of scientists.

Accordingly, we endeavor to give a self-complete concept of EDMs as far as possible, sometimes sacrificing the exhaustive citation of important references.

\section{Basics of EDM}
In this section we give the definitions and conventions used in this review, and basic formulae useful for the EDM.
\subsection{Definitions and Conventions}
\noindent
Metric:
\begin{eqnarray}
g^{\mu\nu}
\equiv
 \left(
  \begin{array}{cc}
   1 & 0\\
    0 & -{\bf 1}_{3\times 3}
  \end{array}
 \right).
\end{eqnarray}

\noindent
Pauli matrices and spin matrices:
\begin{eqnarray}
&&
\sigma^1
\equiv
 \left(
  \begin{array}{cc}
   0 & 1\\
   1 & 0
  \end{array}
 \right), \ \ \
\sigma^2
\equiv
 \left(
  \begin{array}{cc}
   0 & -i\\
   i & 0
  \end{array}
 \right), \ \ \
\sigma^3
\equiv
 \left(
  \begin{array}{cc}
   1 & 0\\
   0 & -1
  \end{array}
 \right)\\
&&
S^i \equiv \frac{1}{\,2\,} \sigma^i.
\end{eqnarray}

\noindent
Gamma matrices:
\begin{eqnarray}
\gamma^0
&\equiv&
 \left(
  \begin{array}{cc}
   {\bf 1}_{2\times 2} & {\bf 0}\\
   {\bf 0} & -{\bf 1}_{2\times 2} 
  \end{array}
 \right), \ \ \
\gamma^i
\equiv
 \left(
  \begin{array}{cc}
   {\bf 0} & \sigma^i\\
   -\sigma^i & {\bf 0}
  \end{array}
 \right), \ \ \
\gamma^5
\equiv i\gamma^0\gamma^1\gamma^2\gamma^3=
 \left(
  \begin{array}{cc}
   {\bf 0} & {\bf 1}_{2\times 2} \\
   {\bf 1}_{2\times 2} & {\bf 0}
  \end{array}
 \right).
\end{eqnarray}

\noindent
Chirality projection:
\begin{eqnarray}
P_L \equiv \frac{1}{\,2\,} ( 1 - \gamma^5 ), \ \ \
P_R \equiv \frac{1}{\,2\,} ( 1 + \gamma^5 ).
\end{eqnarray}

\noindent
Antisymmetric tensor:
\begin{eqnarray}
\sigma^{\mu\nu}
\equiv
 \frac{1}{\,2\,} [ \gamma^\mu, \gamma^\nu ]
\equiv
 \frac{1}{\,2\,}  
 \left(
  \gamma^\mu \gamma^\nu
  - \gamma^\nu \gamma^\mu
 \right).
\end{eqnarray}

\noindent
The electromagnetic field tensor is
\begin{eqnarray}
F^{\mu\nu} \equiv \partial^\mu A^\nu - \partial^\nu A^\mu, \ \ \
F^{0i} = - E^i, \ \ \
F^{ij} = -\epsilon^{ijk} B^k, ~\mbox{so}~F^{12}=-B^3~\mbox{cyclic}.
\end{eqnarray}

\noindent
The Cabbibo-Kobayashi-Maskawa matrix \cite{K-M} and Jarlskog invariant \cite{Jarlskog}:
\begin{eqnarray}
&&
V
\equiv
 \left(
  \begin{array}{ccc}
   c_{12} c_{13}
   & s_{12} c_{13}
   & s_{13} e^{-i\delta}\\
   - s_{12} c_{23} - c_{12} s_{23} s_{13} e^{i\delta}
   & c_{12} c_{23} - s_{12} s_{23} s_{13} e^{i\delta}
   & s_{23} c_{13}\\
   s_{12} s_{23} - c_{12} c_{23} s_{13} e^{i\delta}
   & -c_{12} s_{23} - s_{12} c_{23} s_{13} e^{i\delta}
   & c_{23} c_{13}
  \end{array}
 \right)\,,\\
&&
j^\mu
 = ( \bar{u}, \bar{c}, \bar{t}\, )
   \gamma^\mu P_L V
   \left(
    \begin{array}{c}
     d\\
     s\\
     b
    \end{array}
   \right)\,,\\
&&
J_{CP}
\equiv
 \left|
  \Im( V_{\alpha j} V_{\beta j}^\ast V_{\alpha k}^\ast V_{\beta k} )
 \right|
=
 s_{12} s_{23} s_{13} c_{12} c_{23} c_{13}^2 \sin\delta.
\label{Jarlskog}
\end{eqnarray}
Here $J_{CP}$ is the base independent CP phase called the Jarlskog parameter, appearing in CP violation processes via the Kobayashi-Maskawa mechanism. Hereafter, we denote the imaginary part (real part) of $O$ by $\Im (O)$ ($\Re (O)$).
Apart from the EDM process discussed later, we also mention on neutrino oscillation processes, 
\begin{eqnarray}
P(\nu_\beta\rightarrow \nu_\alpha)=\delta_{\alpha\beta}&-&4\sum_{j<k}U_{\alpha j}U_{\beta j}^*U_{\alpha k}^*U_{\beta k}\mbox{sin}^2\left(\frac{\Delta p_{jk} L}{2}\right)\nonumber\\
&+&4i\sum_{j<k}U_{\alpha j}U_{\beta j}^*U_{\alpha k}^*U_{\beta k}\mbox{sin}\left(\Delta p_{jk}L\right).
\end{eqnarray}

%\begin{align}
%P_{\nu_e \rightarrow \nu_{\mu}} &=\frac{L^2 \alpha ^2 \Delta m_{31}^4 \cos^2[\theta_{23}] %\sin^2[2 \theta_{12}]}{16 E^2}+\frac{\sin^2[\theta_{13}] \sin^2[\theta_{23}]}{E}\sin\left[%\frac{L\Delta m_{31}^2}{4 E}\right] \left(-2 A L \Delta m_{31}^2 \cos\left[\frac{L \Delta %m_{31}^2}{4 E}\right]\right.\nonumber \\
%&+\left.2 E(1+4 A+\cos[2 \theta_{13}]) \sin\left[\frac{L \Delta m_{31}^2}{4 E}\right]\righ%t) \nonumber \\
%&+\frac{L \alpha  \Delta m_{31}^2 }{E}\sin[\theta_{13}] \sin[\theta_{23}] \left(\cos\left[%\delta-\frac{L \Delta m_{31}^2}{4 E}\right] \cos[\theta_{23}] \sin\left[\frac{L \Delta m_{%31}^2}{4 E}\right] \sin[2\theta_{12}]\right. \nonumber \\
%&-\left.\sin\left[\frac{L \Delta m_{31}^2}{2 E}\right] \sin^2[\theta_{12}] \sin[\theta_{13%}]\sin[\theta_{23}]\right)
%\end{align}
Thus we can determine the CP odd term (the third term) by measuring both
$P(\nu_\beta\rightarrow \nu_\alpha)$ and $P(\overline{\nu}_\beta\rightarrow \overline{\nu}_\alpha)$.
T2K \cite{T2K} found evidence of a nonzero $\theta_{13}$ and recently the Daya Bay Collaboration \cite{DayaBay} fixed it as
\begin{equation}
\text{sin}^22\theta_{13}=0.092\pm0.016(\text{stat})\pm0.005(\text{syst}).
\end{equation}
Therefore the above mentioned CP phase experiments have become crucial.

\subsection{Effective Dipole Operator}

A permanent EDM of the electron must lie along its spin, namely ${\bf d}=d_e\boldsymbol{\sigma}$ \cite{Edmonds}.

 At tree level in the SM,
a fermion $\psi$ of mass $m_\psi$ and electric charge e (electron's charge is $e=-|e|$)
in the presence of electromagnetic field satisfies
\begin{equation}
\left(\gamma(p-eA)-m\right)\psi=0,~~ \overline{\psi}\left(\gamma(p+eA)+m\right)=0.
\end{equation}
However if we include loop corrections, the effective electromagnetic interaction Hamiltonian is given in general by
\begin{equation}
V=e\overline{u}_2(p_2)\Gamma^\mu u_1(p_1)A_\mu\equiv eJ^\mu A_\mu(k)
\end{equation}
with
\begin{equation}
P\equiv p_1+p_2,~~k\equiv p_2-p_1.
\end{equation}
Here
\begin{equation}
A^\mu=(\phi, {\bf A})
\end{equation}
is a true vector and transfoms as
\begin{equation}
A^\mu\rightarrow (\phi,~ -{\bf A}) ~~~\mbox{under P, T transformation},
\end{equation}
whereas $J^\mu$ can be either a true or a pseudo vector.

First we consider the case where the two electron lines are external and the photon line internal.  $J^\mu$ takes the general form 
\begin{equation}
J^\mu=F_1(k^2)(\overline{u}_2u_1)P^\mu+F_2(k^2)\overline{u}_2\gamma^\mu u_1+F_3(k^2)(\overline{u}_2u_1)k^\mu.
\label{vector}
\end{equation}
However, from gauge invariance, the current is conserved
\begin{equation}
k_\mu J^\mu=0
\end{equation}
and 
\begin{equation}
F_3(k^2)=0.
\end{equation}
Using Gordon's decomposition for the bilinear form of a spinor of mass m,
\begin{equation}
\left(\overline{u}_2\sigma^{\mu\nu}u_1\right)k_\nu=-2m\overline{u}_2\gamma^\mu u_1+\overline{u}_2u_1P^\mu,
\label{Gordon}
\end{equation}
the interaction term is then given by
\begin{equation}
- e_\psi \overline{\psi} \gamma^\mu \psi A_\mu
=
 - \frac{e}{2m}\,
   \overline{\psi}
    ( i\partial^\mu - i\overleftarrow{\partial}^\mu )
   \psi\, A_\mu
 - i\frac{e}{4m}\,
   \overline{\psi}
   \sigma^{\mu\nu}
   \psi\, F_{\mu\nu}.
\end{equation}

 We should note that
\be
-i\frac{e}{4m}\overline{\psi} \sigma^{\mu\nu} \psi F_{\mu\nu}\nonumber\\
= \frac{e}{2m}
 \overline{\psi}\left[\boldsymbol{\Sigma}\cdot {\bf B}-i\gamma_5\boldsymbol{\Sigma}\cdot {\bf E}\right]\psi,
\label{CPeven}
\ee
where
\begin{eqnarray}
\boldsymbol{\Sigma}\equiv \left(
   \begin{array}{cc}
    \boldsymbol{\sigma} & {\bf 0}\\
    {\bf 0} & \boldsymbol{\sigma}
   \end{array}
   \right).
\label{Sigma}
\end{eqnarray}
Here off diagonal elements are suppressed by $O(v/c)$ relative to diagonal ones.
The magnetic dipole moment (MDM) is defined as the coefficient of ${\bf B}$ in the above equation. However, we must consider quark condensate for hadron EDMs.
\begin{equation}
\boldsymbol{\mu}_Q=Q\left(\frac{e}{2m}\right)\boldsymbol{\sigma}, ~\mbox{and}~\mu_u=-2\mu _d,
\label{MDM}
\end{equation}
where Q is the quark charge and $<\overline{u}u>=<\overline{d}d>$ has been assumed (See {\bf \bref{condensate}} for more detail).
 It is clear from (\ref{CPeven}) that
the fermion has a magnetic dipole moment with $g = 2$
at tree level in the SM (\ref{MDM}). where
\begin{equation}
|\mu|=g\frac{e}{2m}\frac{1}{2}.
\label{g}
\end{equation}
The second term is off-diagonal and there appears the additional P-odd $\sigma^kp^k$ term in the product of the off diagonal element.
The MDM and EDM of particles are defined in the rest frame and we hereafter neglect the off-diagonal element unless it is specified. %
\footnote{This is true, especially for measuring EDM by spin precession as for most cases of neutral particles and atoms. It is not so serious for the measurements of EDMs of charged particles and neutral molecules.}

On the other hand, for an axial vector current, the general form is
\begin{equation}
J^{5\mu}=G_1(k^2)(\overline{u}_2\gamma_5u_1)P^\mu+G_2(k^2)\overline{u}_2\gamma^\mu\gamma_5 u_1+G_3(k^2)(\overline{u}_2\gamma_5u_1)k^\mu.
\label{pseudo}
\end{equation}
In this case, $G_3$ survives due to chiral symmetry breaking.
In weak interactions or higher loops in the SM or in new physics, the current includes both $J^\mu$ and
$J^{5\mu}$ in general. Thus the following CP odd effective action appears, 
\be
-i\frac{e}{4m}\overline{\psi} \gamma^5\sigma^{\mu\nu} \psi F_{\mu\nu}\nonumber\\
=\frac{e}{2m}
 \overline{\psi}\left[i\boldsymbol{\Sigma}\cdot {\bf E}-\gamma_5\boldsymbol{\Sigma}\cdot {\bf B}\right]\psi .
\label{CPodd}
\ee
This will be discussed in more detail in connection with the EDM and MDM shortly.

Also we can consider another conserved current like the vector case
\begin{equation}
a(k^2)(\gamma kk^\mu-k^2\gamma^\mu)\gamma^5,
\end{equation}
which reduces to
\begin{equation}
a(k^2){\bf k}^2\boldsymbol{\sigma}_{\perp}
\end{equation}
in the nonrelativistic limit.

This term is called an anapole term, which comes from the second term of
\begin{equation}
A_i({\bf r})=\int d^3r'\frac{J_i({\bf r}')}{|{\bf r}-{\bf r}'|}
\end{equation}
in the expansion around $r$, that is,
\begin{equation}
A_i^{(2)}({\bf r})=\left(\nabla_k\nabla_l\frac{1}{r}\right)T_{ikl}
\end{equation}
with
\begin{equation}
T_{ikl}=\frac{1}{2}\int d^3r'r_k'r_l'J_i({\bf}').
\end{equation}

 At the loop level in the SM and/or models beyond the SM, the 
following effective interaction of gauge invariant form
can be obtained:
\begin{eqnarray}
&&
 -i\overline{\psi}_i
 \left(
  A_L^{ij} P_L + A_R^{ij} P_R
 \right)
 \sigma^{\mu\nu}
 \psi_j
 F_{\mu\nu}\nonumber\\
&&\hspace*{7mm}
=
 \frac{-i}{\,2\,}
 (A_L^{ij} + A_R^{ij})
 \overline{\psi}
 \sigma^{\mu\nu}
 \psi
 F_{\mu\nu}
 +
 \frac{1}{\,2\,}
 (A_R^{ij} - A_L^{ij})
 \overline{\psi}
 \sigma^{\mu\nu} \gamma^5
 \psi
 F_{\mu\nu}\nonumber\\
&&\hspace*{7mm}
\approx
 (A_L^{ij} + A_R^{ij})
 \overline{\psi}
 \boldsymbol{\Sigma}\cdot {\bf B}\psi
   +i(A_R^{ij} - A_L^{ij})
 \overline{\psi}
 \boldsymbol{\Sigma}\cdot {\bf E}\psi.
\label{eff_dipole}.
\end{eqnarray}
Here we have neglected off diagonal parts in the second equalty.

 For the electric and magnetic dipole moments,
we take zero momentum of the photon.
 Then the imaginary part of the coefficients
of the effective interaction vanish
because of the optical theorem
(imaginary part of the forward scattering amplitude
is given by the sum of possible cuts of intermediate states).
 We find the anomalous magnetic dipole moment $a_\psi$
and electric dipole moment $d_\psi$ to be
\begin{eqnarray}
a_\psi
&=&
 \frac{g-2}{2}
=
 - \frac{2 m}{ \text{e}}\,
   \Re ( A_R^{ii} + A_L^{ii} ),\label{g-2}\\
d_\psi
&=&
 2\,\Im ( A_R^{ii} - A_L^{ii} ).
\label{EDM}
\end{eqnarray}
 Note that $A_L$ and $A_R$ must include
a fermion mass ($m_\psi$ or a fermion mass in the loop)
because the effective interaction
$\overline{\psi} \sigma^{\mu\nu} \psi$
changes the chirality which can be achieved by 
adding a mass term in the fundamental Lagrangian.
 If one of the particles in the loop is
much heavier than the others,
$A_L$ and $A_R$ are suppressed by the mass.
 Thus,
for large $A_L$ and/or $A_R$,
it is preferred that masses of particles in the loop
are similar to each other.

 The effective interaction (\ref{eff_dipole}) also
 causes a $\ell_i \to \ell_j \gamma$ decay where the decay rate is given by
\begin{eqnarray}
\Gamma( \ell_i \to \ell_j \gamma )
= \frac{ m_{\ell_i}^3 }{ 4\pi }
  \left(
   |A_L^{ij}|^2 + |A_R^{ij}|^2
  \right)\,.
\end{eqnarray}
Thus EDM and MDM have opposite parities and different in the order of magnitude.
However they appears in parallel, and have some similarities also.
One of them is their SU(6) property \cite{Georgi} and will be discussed in Appendix {\bf A}.

For an invariant electromagnetic field, EDM, MDM, anapole, and higher n-pole moments appear as
the multipole expansions of the Coulomb potential and vector potential.
These points are also discussed in Appendix {\bf B}.
 
Quarks receive additional contributions, which will be discussed for diamagnetic atoms.
Here we list the results.

A strong CP violating term connected with the $\theta$ vacuum (see Appendix {\bf G})
\begin{equation}
L_{d=4}=\frac{g_s^2}{64\pi^2}\overline{\theta}G_{\mu\nu}^a{G}_{\rho\lambda}^a\epsilon^{\mu\nu\rho\lambda}\equiv \frac{g_s^2}{32\pi^2}\overline{\theta}G^a\cdot \tilde{G}^a.
\label{theta}
\end{equation}

In new physics beyond the SM, we have other P and T violating effective actions:
the chromoelectric dipole operator (cEDM)
\begin{equation}
L_C=-\frac{i}{2}\tilde{d}_qg_s\overline{q}\sigma_{\mu\nu}\gamma_5T^aq G^{\mu\nu a}\equiv -\frac{i}{2}\tilde{d}_qg_s\overline{q}\sigma G\gamma_5q,
\label{cEDM}
\end{equation}
and the following dimension 6 operators,
\begin{equation}
L_G=-\frac{1}{6}d_Gf_{abc}G_{\mu\rho}^aG^\rho{}_{\nu}^bG_{\lambda\sigma}^c\epsilon^{\mu\nu\lambda\sigma}\equiv -\frac{1}{3}d_Gf_{abc}G^aG^b\tilde{G}^c,
\label{dim6a}
\end{equation}
the so-called Weinberg term \cite{Weinberg}, and 
\begin{equation}
L_{d=6}=\sum C^a_{ij}\overline{\psi_i}O_a\psi_i\overline{\psi_j}O_a\gamma_5\psi_j.
\label{dim6b}
\end{equation}
Here $\psi_i$ and $\psi_j$ are leptons and/or nucleons. $O_a$ are scalar, vector, and tensor gamma matrices.
We will explain the detailed physical implications in the diamagnetic atom in section 6.

In the SM, weak interactions act with matter and gauge fields in the form,
\begin{equation}
H_{weak}=\overline{\psi}P_L\Gamma_\mu \psi W^\mu\equiv J_\mu W^\mu.
\end{equation}
However, except for the top quark, fermion masses are small compared to weak bosons masses,
They are descrived as the four-fermion coupling
\begin{equation}
H_{weak}=J_\mu J^\mu.
\end{equation}
This is the case for tree diagrams. If you consider loop diagrams or new physics beyond the SM, we will encounter more general forms.
We will discuss this in the Appendix C.

%%%%%%%% sec: experimental bounds  %%%%%%%%%
\subsection{Experimental Bounds}
We have no experimental signal of the EDM yet but have upper limits.
They are as follows \cite{PDG}.
It is very impressive that recently we have a more precise upper limit of the electron EDM from molecule (YBF) than from atom (Tl).

\begin{eqnarray}
d_e~\text{from thallium atom}~d(Tl)
&=& (6.9 \pm 7.4)\times 10^{-28}\,\text{e cm}%
~\text{\cite{Regan:2002ta}}\\
d_\mu
&=& (3.7 \pm 3.4)\times 10^{-19}\,\text{e cm}%
~\text{\cite{Bennett}}\\
d_n
&<& 2.9\times 10^{-26}\,\text{e cm}\ \ (90\%C.L.)%
~\text{\cite{Baker:2006ts}}\\
d({}^{199}Hg)
&<& 3.1\times 10^{-29}\,\text{e cm}\ \ (95\%C.L.)%
~\text{\cite{Hg}}\\
d_e~\mbox{from the molecule $d$(YbF)}
&=& (-2.4\pm 5.7_{stat}\pm 1.5_{syst})\times 10^{-28}\,\text{e cm}%
~\text{\cite{YbF}}
\end{eqnarray}

For reference, we give here the muon anomalous MDM, a non-null signal of new physics beyond the SM.

The deviations of the SM predictions from the experimental result
are given by
\begin{eqnarray}
\Delta a_\mu[\tau]
&\equiv& a_\mu^{\text{exp}} - a_\mu^{\text{SM}}[\tau]
 = 14.8(8.2)\times 10^{-10},\nonumber\\
\Delta a_\mu[e^+e^-]
&\equiv& a_\mu^{\text{exp}} - a_\mu^{\text{SM}}[e^+e^-]
 = 30.3(8.1)\times 10^{-10}.
\label{g-2}
\end{eqnarray}
Here the hadronic contributions
to $a_\mu^{\text{SM}}[\tau]$ and $a_\mu^{\text{SM}}[e^+e^-]$
were calculated~\cite{Davier:2009ag} by using data of
hadronic $\tau$ decay and $e^+e^-$ annihilation to hadrons,
respectively.
 These values of $\Delta a_\mu[\tau]$ and $\Delta a_\mu[e^+e^-]$
correspond to $1.8\sigma$ and $3.7\sigma$ deviations
from the SM predictions, respectively.
EDMs and MDMs come from similar diagrams apart from CP transfomation and 
the differences of magnitudes in the MDM and EDM stem from the cancellation of diagrams and
symmetry.
%%%%%%%%  sec: Standard Model %%%%%%%%%%%%%%
\section{STANDARD MODEL}
In this section we give the EDMs of quarks, hadrons, and leptons in the SM framework.
Structure of matter multiplets in SM+(Dirac) neutrino is 
\begin{eqnarray}
\label{Q}
 &&Q = 
\left( 
 \begin{array}{ccc}
u_1 & u_2  & u_3  \\ 
d_1 & d_2  & d_3  
 \end{array}   \right)  \sim \left(3,2,\frac{1}{6}\right), \nonumber\\
&& u^c=(u_1^c~~u_2^c~~u_3^c)\sim \left(\overline{3},1,\frac{-2}{3}\right),\nonumber\\
&&d^c=(d_1^c~~d_2^c~~d_3^c)\sim \left(\overline{3},1,\frac{1}{3}\right),\\
&&L=
\left(
\begin{array}{c}
\nu\\
e
\end{array} \right)\sim \left(1,2,\frac{-1}{2}\right),\nonumber\\
&&e^c\sim (1,1,1),\nonumber\\
&&\nu^c\sim (1,1,0).\nonumber
\end{eqnarray}    

CP violation occurs as per Kobayashi-Maskawa mechanism, that is, CP phase in CKM mixing matrix for quarks or MNS matrix for leptons.
Diagrammatically, it resembles with Lepton Flavour Violation (LFV) processes but for the former it is necessary to incorporate the non-zero Jarlskog parameter.
Apart from the uncovered new phenomena in neutrino, the SM has the deficiency of baryon assymmetry. Jarlskog introduced $A_{CP}$ \cite{Jarlskog} defined by
\begin{equation}
[M_uM_u^\dagger, M_dM_d^\dagger]=iA_{CP}.
\end{equation}
Here $M_u,~M_d$ are up-type and down-type quark mass matrices.
The observables are not these matrices but those which are invariant under rebasing and rephasing, that is, eigen values and CKM mixing matrix.
By construction, $A_{CP}$ is traceless and Hermitian, and characterizes the effect of CP violation. Its explicit form is
\begin{equation}
\mbox{det}A_{CP}=(m_t^2-m_c^2)(m_t^2-m_u^2)(m_c^2-m_u^2)(m_b^2-m_s^2)(m_b^2-m_d^2)(m_s^2-m_d^2)J_{CP},
\label{Jarlskog2}
\end{equation}
where $J_{CP}$ is given by (\ref{Jarlskog}).
As we will see the detail shortly, one-loop diagram (Fig. (\ref{fig:dEDM_SM1loop}) gives zero contribution to the EDM. As for two loop diagrams (Fig. \ref{fig:dEDM_SM2loop}),
using this $J_{CP}$ and denoting by $f$ the Green function of $f$ flavoured fermion \cite{Hamzaoui}, the f quark EDM has the form 

\begin{equation}
i\sum_{jkl}\Im (V_{jk}V_{lf}V_{jf}^*V_{lk}^*)fjklf=\frac{1}{2}\Im (V_{jk}V_{lf}V_{jf}^*V_{lk}^*)f(jkl-lkj).
\label{Jarlskog3}
\end{equation}
and $J_{CP}\approx 3\times 10^{-5}$ is twice the area of unitary triangle
$V_{ud}^*V_{ub}+V_{cd}^*V_{cb}+V_{td}^*V_{tb}$.
One finds \cite{Shaposhnikov} that 
\be
n_B/n_\gamma\approx (m_t^2-m_c^2)(m_t^2-m_u^2)(m_c^2-m_u^2)(m_b^2-m_s^2)(m_b^2-m_d^2)(m_s^2-m_d^2)/T^{12}\times J_{CP}\approx 10^{-20}
\ee
This falls short of the observed baryon asymmetry $n_B/n_\gamma\approx 
10^{-10}$.
We will show the detail loop by loop in the subsequent sections.
\subsection{Quark EDM}
For definiteness,
we consider the EDM of $d$ quark.
 The advantage of $d$ in comparison with $u$
may be that $t$ can be in the loop
with the weak interaction to avoid the GIM cancellation \cite{Glashow}.

\subsubsection{One Loop}
%%%%%%%%%  SM 1-loop for EDM of d  %%%%%%%
\begin{figure}[t]
\begin{center}
\includegraphics[scale=0.6]{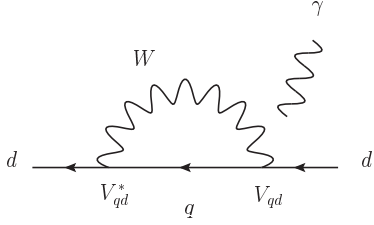}
\caption{
 The diagram for the EDM of $d$ quark
at one loop level in the SM\@.
}
\label{fig:dEDM_SM1loop}
\end{center}
\end{figure}
 In the one loop level (Fig.~\ref{fig:dEDM_SM1loop}),
elements of the CKM matrix
appear as $|V_{qd}|^2$.
 Therefore,
there is no EDM (imaginary part of coefficient) apparently.

\subsubsection{Two Loop}
%%%%%%%%%  SM 2-loop for EDM of d  %%%%%%%
\begin{figure}[t]
\begin{center}
\includegraphics[scale=0.6]{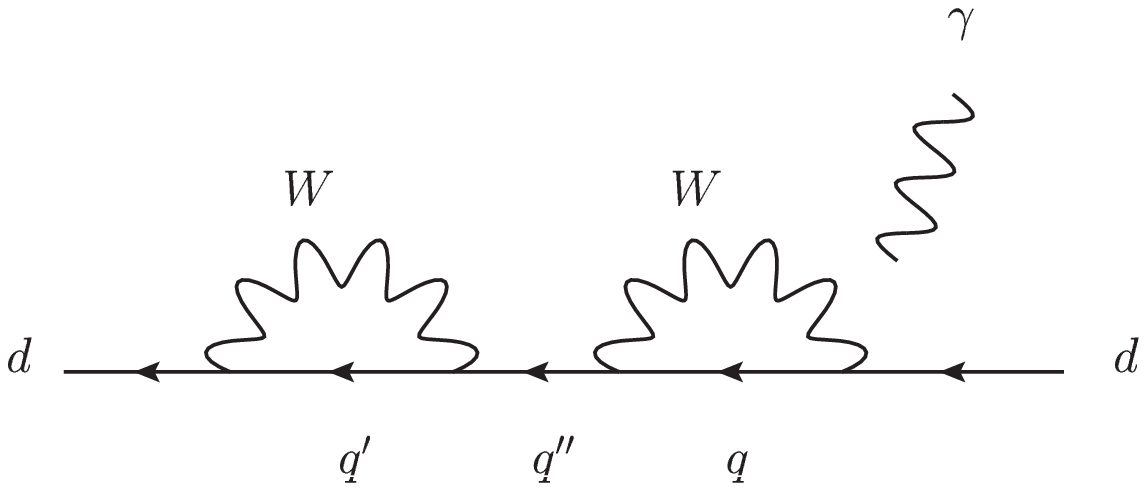}
\includegraphics[scale=0.6]{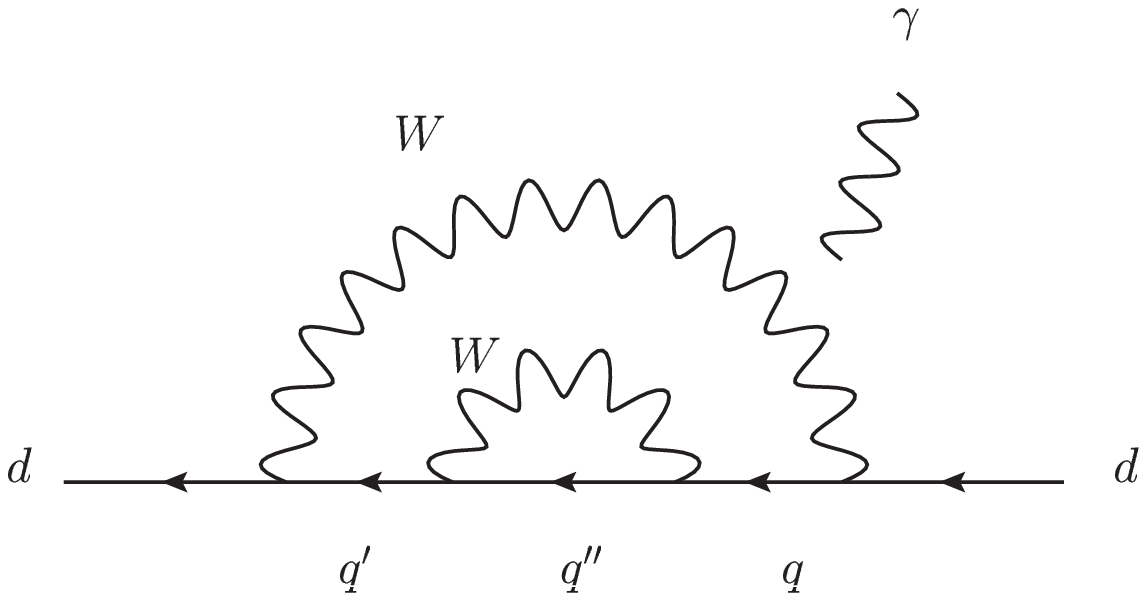}
\caption{
 Diagrams for the EDM of $d$ quark
at two loop level in the SM\@.
}
\label{fig:dEDM_SM2loop}
\end{center}
\end{figure}
 At two loop level in the SM,
two types of diagrams
have an imaginary coefficient potentially.
 The diagrams are shown in Fig.~\ref{fig:dEDM_SM2loop}.
 It is, however, shown that the imaginary part of each diagram
vanishes by the summation of contributions from all quarks
of internal lines~\cite{Shabalin:1978rs}.

 For the diagrams in Fig.~\ref{fig:dEDM_SM2loop},
it is clear that $q^{\prime\prime}$ must not be $d$ quark
because it gives $|V_{q^\prime d}|^2 |V_{qd}|^2$ of real value.
 By the same reason,
$q^\prime$ must be different from $q$.

\subsubsection{Three Loops}
%%%%%%%%%  SM 3-loop for EDM of d  %%%%%%%
\begin{figure}[t]
\begin{center}
\includegraphics[scale=0.17]{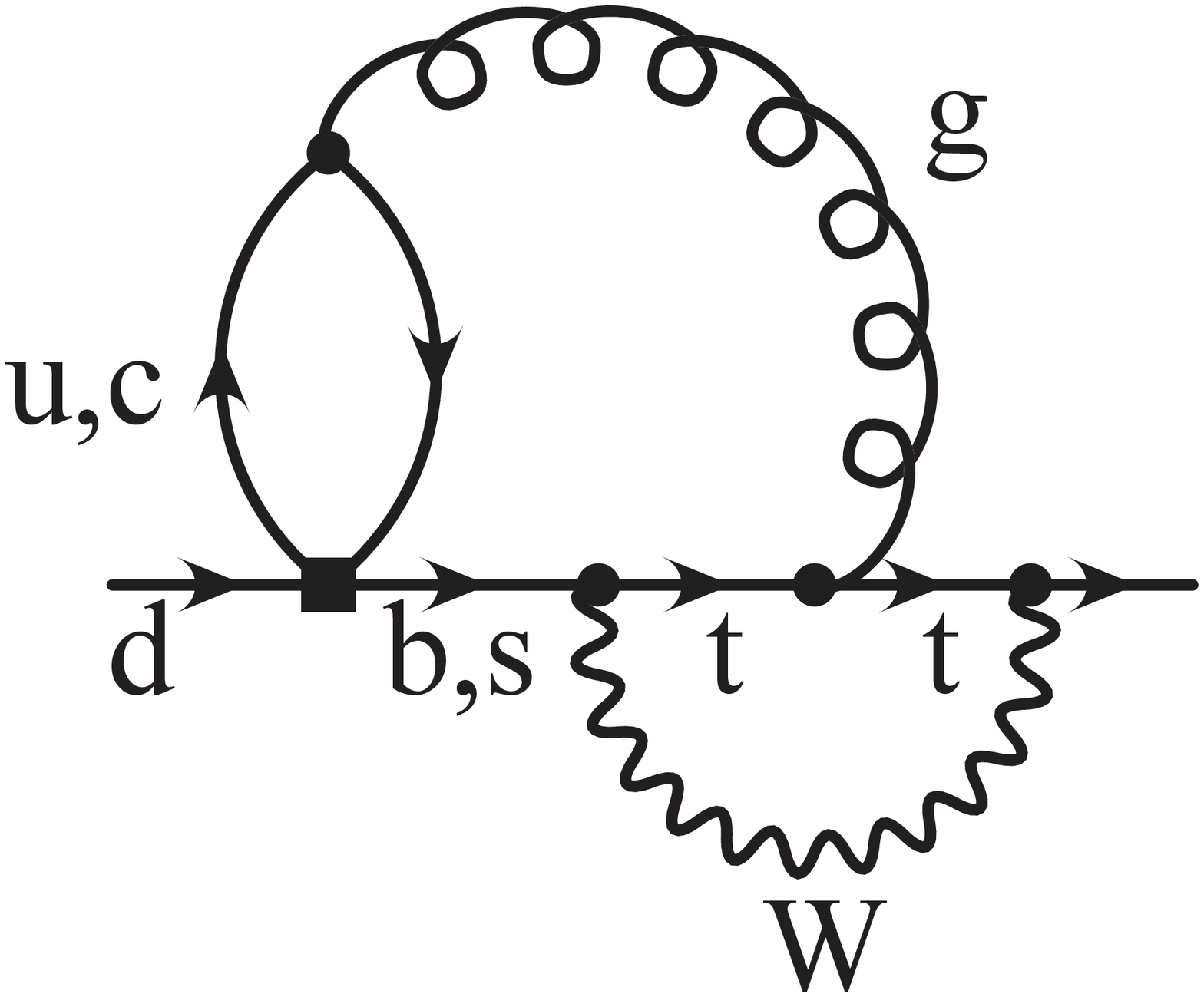} \ \ 
\includegraphics[scale=0.17]{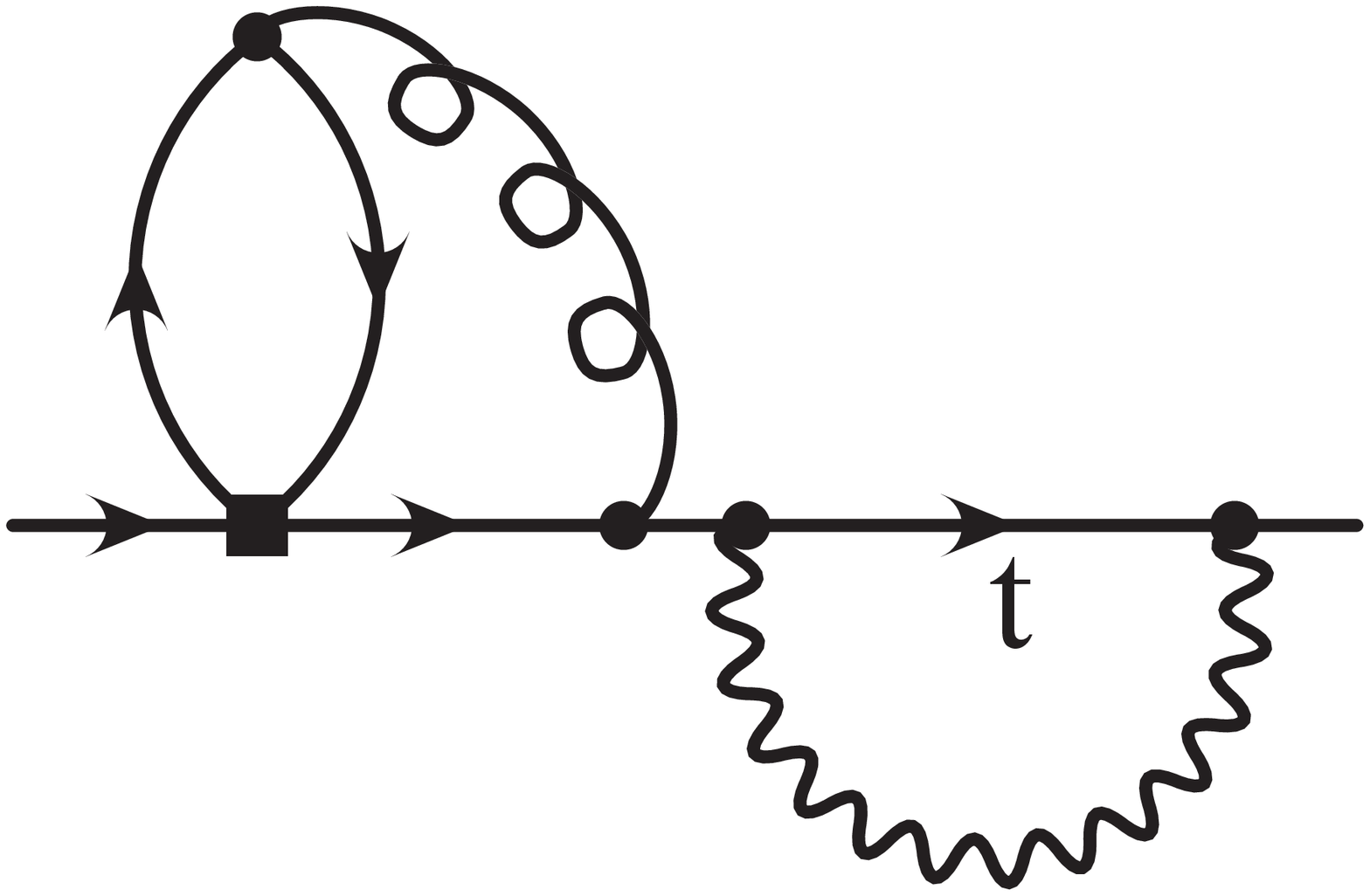} \ \ 
\includegraphics[scale=0.17]{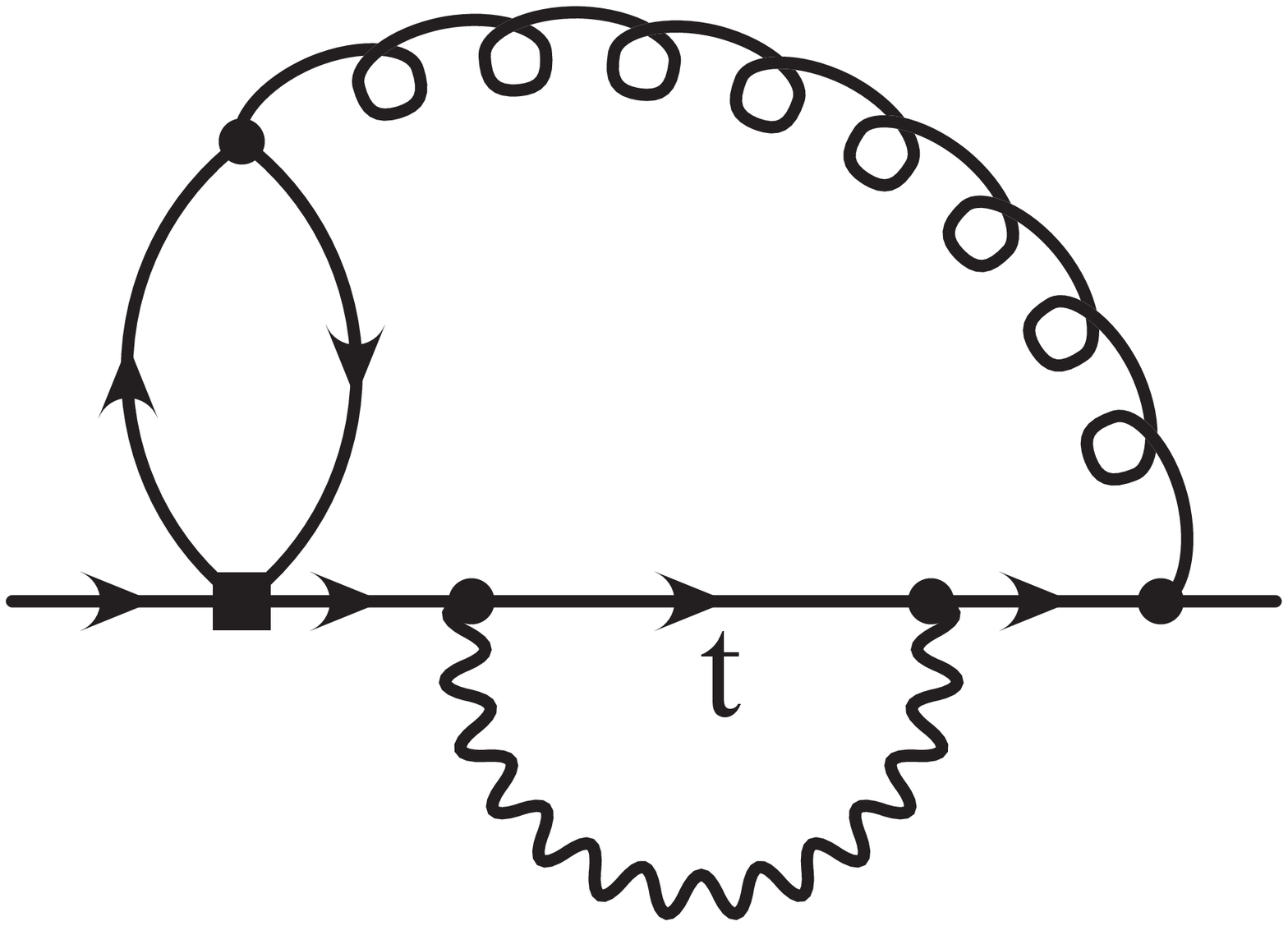}
\end{center}
%\label{fig:dEDM_SM3loop}
\caption{
Diagrams for the EDM of $d$ quark
at three loop level in the SM \protect\cite{Czarnecki:1997bu}.
}
\label{fig:dEDM_SM3loop}
\end{figure}

 Fig.~\ref{fig:dEDM_SM3loop} shows three loops diagrams
which contribute to $d$ quark EDM in the SM\@.
 The formula for the contribution is
given in \cite{Czarnecki:1997bu} as
\begin{eqnarray}
&&
\frac{d_d}{\text{e}}
\simeq
  \frac{ m_d m_c^2 \alpha_s G_F^2 J_{CP} }{ 108\pi^5 }
  \bigg\{
   \left( L_{bc}^2 - 2 L_{bc} + \frac{\pi^2}{\,3\,} \right) L_{Wb}
   + \frac{5}{\,8\,} L_{bc}^2\nonumber\\
&&\hspace*{30mm}
 {}- \left( \frac{335}{36} + \frac{2}{\,3\,} \pi^2 \right) L_{bc}
   - \frac{1231}{108}
   + \frac{7}{\,8\,} \pi^2
   + 8\zeta (3)
  \bigg\}\,,
\end{eqnarray}
where $L_{ab} \equiv \ln (m_a^2/m_b^2)$.
\label{d_d}
 It results in
\begin{eqnarray}
 d_d \simeq - 10^{-34}\,\text{e}\ \text{cm}
\end{eqnarray}
while the triple log approximation (taking only $L^3$ term) gives
\begin{eqnarray}
 d_d \simeq + 10^{-34}\,\text{e}\ \text{cm}.
\end{eqnarray}

%%%%%%%  sec: Neutron  EDM  %%%%%%%%%%%
\subsection{Neutron EDM}
\begin{figure}[t]
\begin{center}
\includegraphics[scale=0.3]{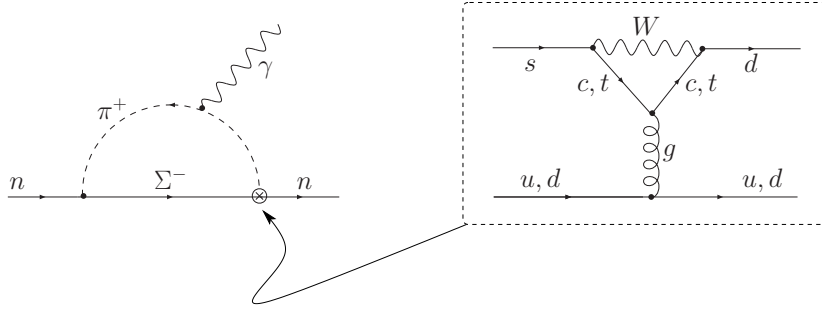}
\caption{
 Diagram for the EDM of neutron in the SM \protect\cite{Khriplovich:1981ca}.
\label{fig:nEDM_SM}}
\end{center}
\end{figure}

 The dominant contribution to the neutron EDM
in the SM comes from "two loop" diagram
in Fig.~\ref{fig:nEDM_SM}.
 The interaction on the left part of the loop is given
by the phenomenological interaction hamiltonian
\begin{eqnarray}
&&
H
= i G_F m_\pi^2
  \overline{u}_n
  ( A + B \gamma^5 )
  u_\Sigma \varphi_\pi\,,\\
&&
 A = -1.93, \ B = -0.65,
\end{eqnarray}
where $u_n$, $u_\Sigma$, and $\varphi_\pi$
stand for wave functions of neutron, $\Sigma^-$ baryon ($dds$),
and $\pi^+$, respectively.
 The interaction on the right part of the loop
is so-called "strong penguin" whose effective operator is given by
\begin{eqnarray}
{\cal H}_{\text{pen}}
= \frac{iG_F \alpha_s(\bar{m}) \Delta}{12 \sqrt{2}\,\pi }
  s_{23} s_{13} c_{23}\sin\delta
  \ln\frac{m_t^2}{m_c^2}\,
  \bar{s} \gamma_\mu (1-\gamma^5) \lambda^a d\,
  \sum_{q=u,d} \bar{q} \gamma^\mu \lambda^a q 
\end{eqnarray}
Here $\Delta\approx 1.3$ arises due to strong interaction. Note that the ${\cal H}_{\text{pen}}$ seems to be obtained
for $m_c \simeq m_t < m_W$.
 With these interactions,
the neutron EDM was estimated as
\begin{eqnarray}
 d_n=d_n^{short}+d_n^{long}\simeq 10^{-32}\, \text{e}\ \text{cm}\,.
\end{eqnarray}
Here the first is the contribution from Fig. \ref{fig:dEDM_SM3loop} ($O(\alpha_sG_F^2)~\approx~ 10^{-34}$ ecm) and the second
from Fig. \ref{fig:nEDM_SM} \cite{Khriplovich:1981ca}.

If we incorporate the rephasing invariance of strange wave function, this value is modified to $1.4\times 10^{-31}\leq |d_n|\leq 9.9\times 10^{-33}$ e~cm \cite{He}.
Recently there appeared new type of diagram which contribute to EDM in loopless diagram \cite{Mannel}.  Thus some controversies are still left even in the naive SM scheme.

Furthermore, there are new CP violating five and six dimensional operators,
\begin{equation}
{\cal L}_{CPV}=\sum_q d_q\overline{q}(\sigma F)\gamma_5 q+\sum_q \tilde{d}_q\overline{q}(\sigma G)\gamma_5 q+wGG\tilde{G}+...\end{equation}
The details of this contribution will be discussed in the section of diamagnetic atom.

\begin{figure}[t]
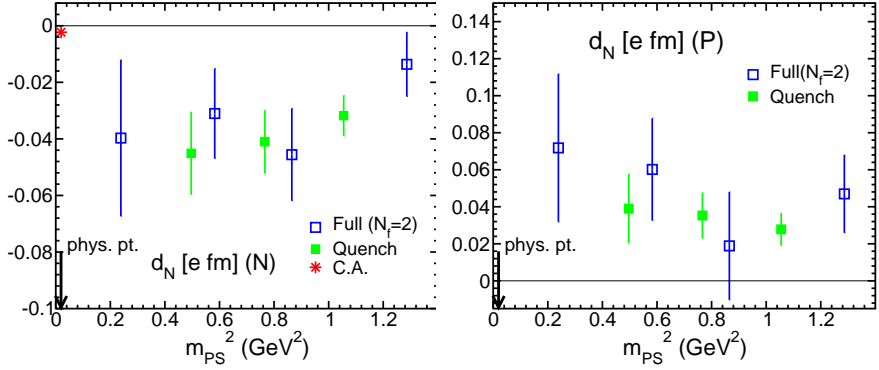

\begin{center}
\includegraphics[scale=0.4]{latticen.eps}
\includegraphics[scale=0.4]{latticep.eps}
\caption{The EDM as a function of the pseudoscalar meson mass squared $m_{PS}^2$ for neutrons (left panel) and protons (right panel). The arrow shows the physical point of the pion mass squared, $m_\pi^2=0.0195 \mbox{GeV}^2$, and the star symbol represents the the result of current algebra (C.A.) \protect\cite{Shintani}. They used the Wilson fermion which does not allow to take the chiral symmetry limit. The error bar in the diagram does not include the syastematic error due to this. More improved model is ongoing by them \cite{Shintani2}.}
\label{fig:lattice}
\end{center}
\end{figure}

The concrete and model independent calculations are expected in the lattice QCD.
The EDM of neutron is estimated by lattice calculation (see Fig. \ref{fig:lattice}).

It is also possible to guess the rough estimate of hadron EDM using SU(6).
The detailed discussion is given in Appendix {\bf A}.

%%%%%%  sec: Lepton EDM  %%%%%%%%%
\subsection{Lepton EDM}

\begin{figure}[t]
\begin{center}
\includegraphics[scale=0.6]{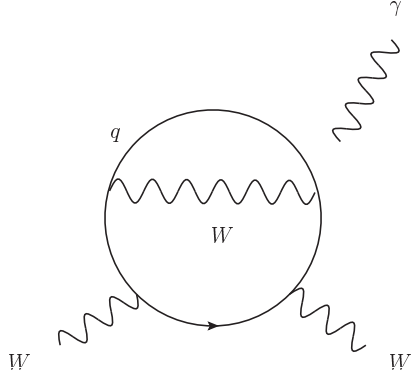}
\caption{
 Diagram for the EDM of $W$ boson
at two loop in the SM\@.
}
\label{fig:WEDM_SM2toop}
\end{center}
\end{figure}
\begin{figure}[t]
\begin{center}
\includegraphics[scale=0.6]{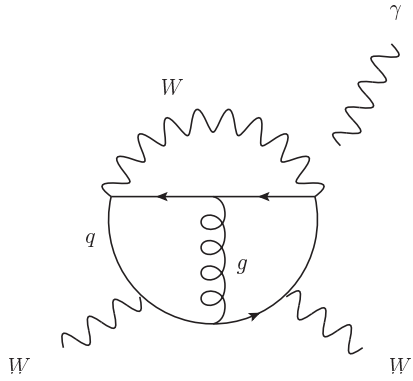}
\caption{
 Three loop diagram
which may give a nonzero contribution
to the EDM of $W$ boson.
}
\label{fig:WEDM_SM3loop}
\end{center}
\end{figure}

 For definitteness,
let us concentrate on electron.
 Similarly to quark EDM,
one and two loop diagrams do not
contribute to electron EDM\@.
 In order to avoid GIM cancelation
($\propto (m_i^2 -  m_j^2)/m_W^2$),
CKM matrix is better to be used than
Maki-Nakagawa-Sakata matrix (lepton mixing).
 Two $W$ bosons (at least) should attached
to the electron line in order to use
a quark loop.
 Then,
the electron EDM is caused by the $W$ boson EDM\@.
 It was shown that the $W$ boson EDM vanishes
at two loop level~\cite{Pospelov:1991zt}.

 The two loop diagram is shown in Fig.~\ref{fig:WEDM_SM2toop}.

$J_{CP}$ defined by (\ref{Jarlskog}) is antisymmetric
under $j\leftrightarrow l$ (\ref{Jarlskog3}) (corresponding to side line's quarks), whereas it is symmetric in Fig.6.
 Adding another loop with gluon
(see Fig.~\ref{fig:WEDM_SM3loop}),
the $W$ boson EDM in three loop was estimated as
\begin{eqnarray}
d_W
\simeq
 J_{CP}
 \left( \frac{1}{16\pi^2} \right)^2
 \left( \frac{g^2}{\,8\,} \right)^2
 \frac{\alpha_s}{4\pi}
 \frac{ \text{e} }{2m_W}
\simeq 8\times 10^{-30}\,\text{e cm}\,.
\end{eqnarray}
 The electron EDM in four loop
was estimated with $d_W$ as
\begin{eqnarray}
d_e
\simeq
 \frac{g^2}{32\pi^2}
 \frac{m_e}{m_W}\,
 d_W
\simeq 8\times 10^{-41}\,\text{e cm}\,.
\end{eqnarray}

\section{Beyond the Standard Model}
In this section we will discuss models beyond the SM.

As we have shown in the previous section, the SM predicts rather smaller values of EDMs than the experimental upper limits by roughly ten orders of magnitude.
However, we also know that CP violation in the SM is insufficient for baryon asymmetry in the real world.
Also we have many direct signals of new physics beyond the SM like neutrino
oscillations and muon g-2 etc.
Even if we stand in the SM, we have new 4- and 6-dimensional CP violating effective actions like (\ref{theta}) to (\ref{dim6b}), which have never been discussed so much in the previous section.
Then it is very natural to estimate how much such new physics or new models predict the EDMs.
In this section we concentrate on new physics beyond the SM.
As for the new 4- and 6-dimensional CP violating effective actions, we will discuss in
the sections of diamagnetic atoms and molecules.

\subsection{Minimal Supersymmetric Standard Model (MSSM)}
In the MSSM, all particles have their SUSY partners;
sfermions $\tilde{f}$ (bosons) for fermions $f$,
Higgsinos (fermions) for Higgs bosons, gauginos (fermions) for gauge bosons. Also another Higgs doublet is added to the SM for recovering chiral anomaly
free condition once broken by this doubling \cite{Martin}. Yukawa coupling is given by
\begin{equation}
W_{MSSM}=Y_u\overline{u}QH_u-Y_d\overline{d}QH_d-Y_e\overline{e}LH_d+\mu H_uH_d
\label{MSSM}
\end{equation}
SUSY is broken at O(1TeV) by soft SUSY breaking terms which retain hierarchy problem. MSSM is a minimally extended supersymmetric SM and we will consider below the constrained MSSM (cMSSM) and $\nu$MSSM including light neutrino masses.
In general soft breaking terms are
\begin{eqnarray}
L^{(1)}_{SB}&=&-(m_{\tilde{q}})^2_{ij}\tilde{Q}_{Li}^\dagger \tilde{Q}_{Lj}-(m_{\tilde{u}})^2_{ij}\tilde{u}_{Ri}^* \tilde{u}_{Rj}-(m_{\tilde{d}})^2_{ij}\tilde{d}_{Ri}^* \tilde{d}_{Rj}-(m_{\tilde{l}})^2_{ij}\tilde{L}_{Li}^\dagger \tilde{L}_{Lj}\nonumber\\
&-&(m_{\tilde{e}})^2_{ij}\tilde{e}_{Ri}^* \tilde{e}_{Rj}-\mu_1^2H_u^\dagger H_u-\mu_2^2H_d^\dagger H_d-\mu_S^2S^* S-b(H_uH_d+c.c.)\\
&-&\left(A_{uij}\tilde{u}^*_{Ri}\tilde{Q}_{Li}H_u-A_{dij}\tilde{d}^*_{Ri}\tilde{Q}_{Li}H_d-A_{eij}\tilde{e}^*_{Ri}\tilde{L}_{Li}H_d+c.c.\right)\nonumber
\label{soft1}
\end{eqnarray}
\begin{eqnarray}
L^{(2)}_{SB}=-\frac{1}{2}\left(M_3\tilde{g}^a\tilde{g}^a+M_2\tilde{W}^a\tilde{W}^a+M_1\tilde{B}^a\tilde{B}^a+c.c.\right)
\label{soft2}
\end{eqnarray}
These include many CP violating phases, in general.
\begin{figure}[t]
\begin{center}
\includegraphics[scale=0.6]{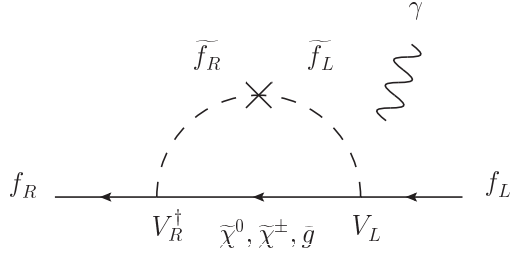}
\caption{
 Diagrams of SUSY contributions to the EDM of fermions.
 Gluino $\tilde{g}$ contributes only for quark EDM\@.
}
\label{fig:EDM_SUSY}
\end{center}
\end{figure}

%\begin{figure}[t]
%\begin{center}
%\includegraphics[scale=0.7]{Barr-Zee.eps}
%\includegraphics[scale=0.6]{diagram_EDM_SUSY-2.eps}
%\includegraphics[scale=0.6]{diagram_EDM_SUSY-3.eps}
%\caption{
% Two loop corrections to the EDM. 
%}
%\label{fig:Barr-Zee}
%\end{center}
%\end{figure}

 For the loop correction to the Higgs masses,
the problem on its quadratic divergence
is cancelled by the loop of SUSY partners
(different statistics with same coupling).

 SUSY particles contribute
to fermion EDM at one loop
shown in Fig.~\ref{fig:EDM_SUSY}.
 The neutralinos $\tilde{\chi}^0$ and
charginos $\tilde{\chi}^\pm$ are mass eigenstates,
and they are linear combinations of
Higgsinos and gauginos of $SU(2)_L$ and $U(1)_Y$.
 The $d$ quark EDM from the diagram
is estimated~\cite{Buchmuller:1982ye} as 
\begin{equation}
d_d/e=\frac{1}{2m}(-\frac{1}{3})\frac{4}{3}\frac{\alpha_s}{\pi}v\Im(V^{d\dagger}_R A_dV_L^d)_{11}\frac{\mu m}{(M^2-\mu^2)^2}\left(\frac{1}{2}+3\frac{\mu^2}{M^2-\mu^2}-\frac{\mu^2(\mu^2+2M^2)}{(M^2-\mu^2)^2}ln\frac{M^2}{\mu^2}\right).
\end{equation}
Here $v$ is the common vacuum expectation value of $H_u$ and $H_d$. $m$ and $M$ are masses of d quark and the universal squark mass (by assumtion), respectively. 
If we adopt $v\Im(V^{d\dagger}_R A_DV_L^d)\sim M^2$ with rough estimations of $M=100$ GeV and maximal mixings, we obtain
\begin{eqnarray}
d_d
\sim 10^{-22}\,\text{e cm}\,.
\end{eqnarray}
 The value is clearly in conflict with
the experimental bound on $d_n$.
 The naive estimation was done
with $M_{\tilde{q}} \simeq 100\,\text{GeV}$ and
a sizable CP-violating phase $\sin\phi \sim 1$.
 Therefore,
the contradiction can be resolved
by small phase (approximate CP symmetry)
and/or heavy masses of SUSY particles.
However, we adopt not such fine tuning but the
universal soft SUSY breaking (cMSSM).
$V_{L(R)}$ is the unitary matrix which rotates left (right)-handed weak eigen states, and therefore if $A_{ij}\propto Y_{ij}, ~V^{d\dagger}_R A_DV_L^d$ is a real, diagonal matrix and the imaginary parts of its matrix elements vanish.
Thus the small EDM leads us to relations of trilinear terms in
(\ref{MSSM}) and (\ref{soft1})
\begin{equation}
A_u=A_{u0}Y_u,~~A_d=A_{d0}Y_d,~~A_e=A_{e0}Y_e.
\end{equation}
Also we accept the universal soft SUSY breaking which is realized by gravity or gauge mediated SUSY breaking.
\begin{eqnarray}
&&M_1,M_2,M_3\sim m_{1/2},\\
&&m_Q,m_L,m_u,m_d,m_e,m_{H_u},m_{H_d}\sim m_0.
\end{eqnarray}
 CP-violating phases appear
only in flavor off-diagonal parts of matrices
(Hermitian Yukawa matrices) and
the CP violating effect is suppressed
by small mixings only due to RGE.
\begin{eqnarray}
\delta_{LL}^q&=&\frac{(m_{\tilde{q}}^2)_{ij}}{m_{\tilde{q}}^2},~~\delta_{RR}^u=\frac{(m_{\tilde{u}}^2)_{ij}}{m_{\tilde{u}}^2},~~\delta_{RR}^d=\frac{(m_{\tilde{d}}^2)_{ij}}{m_{\tilde{d}}^2},\nonumber\\
\delta_{LL}^l&=&\frac{(m_{\tilde{l}}^2)_{ij}}{m_{\tilde{l}}^2},~~\delta_{RR}^e=\frac{(m_{\tilde{e}}^2)_{ij}}{m_{\tilde{e}}^2}.
\end{eqnarray}
\begin{figure}[t]
\begin{center}
\includegraphics[scale=0.4]{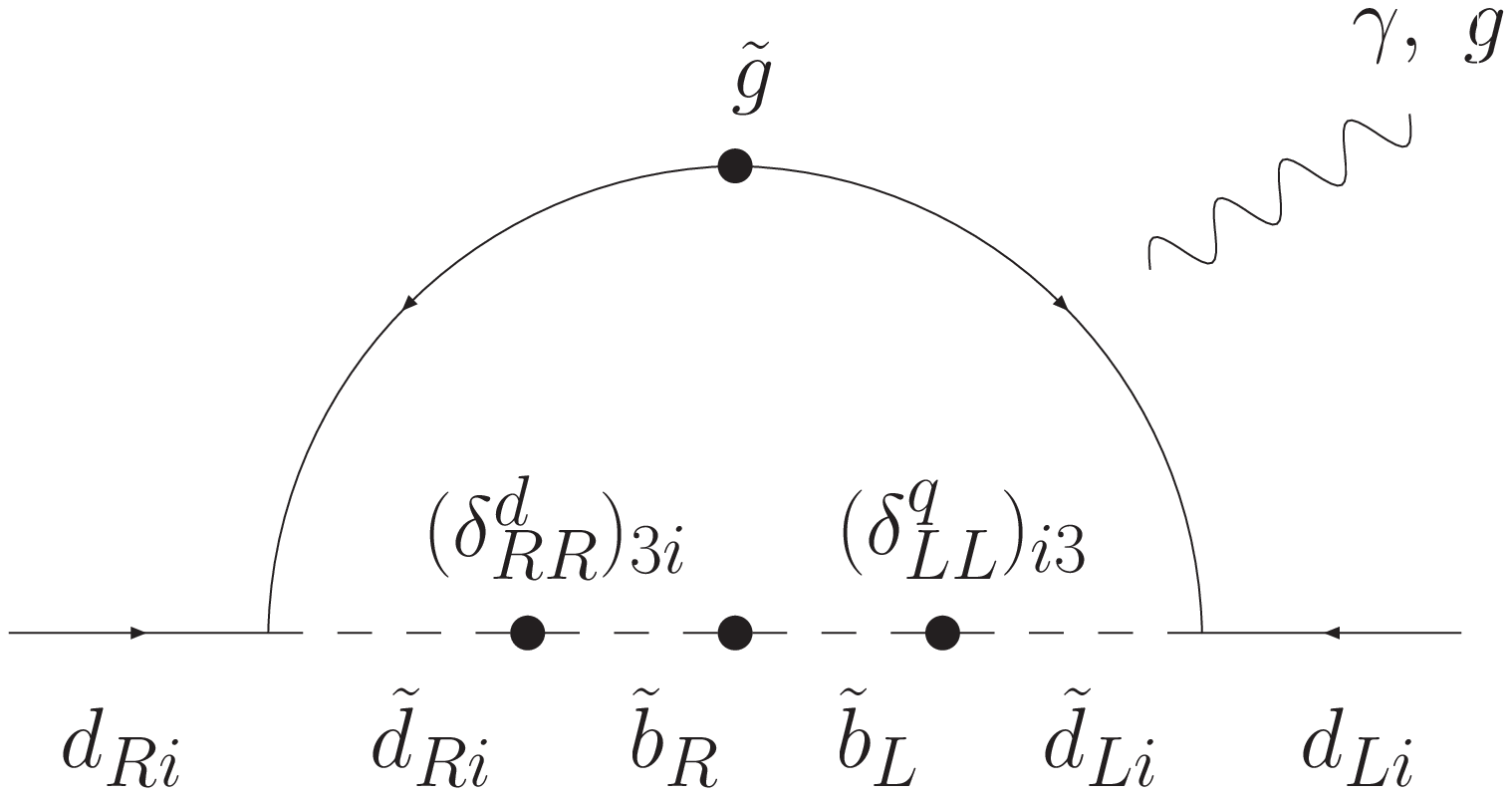}
\includegraphics[scale=0.6]{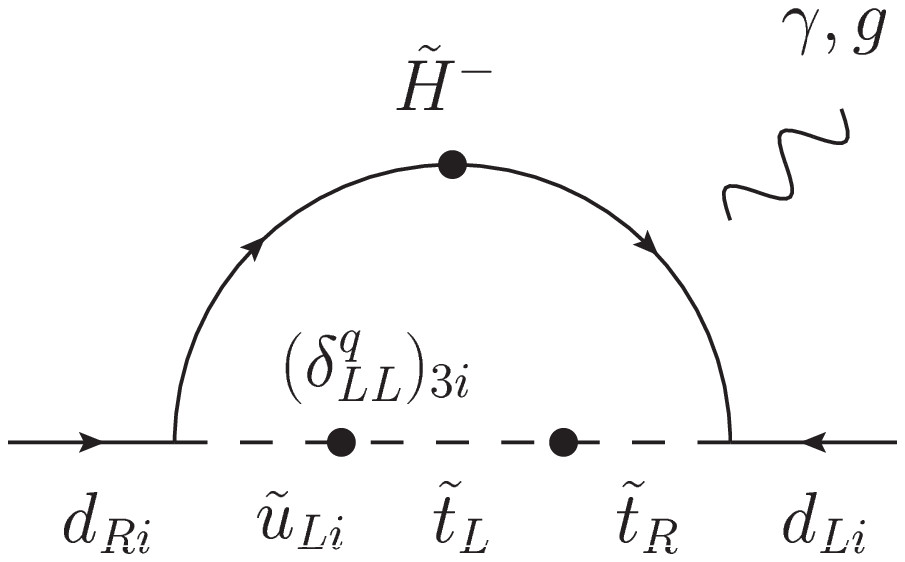}
\caption{
 Diagrams of SUSY contributions to the EDM of fermions
in case of flavored CP violation
where CP violating phase appears with
change of flavor \protect\cite{Hisano:2008hn}.
}
\label{fig:EDM_flavor}
\end{center}
\end{figure}

 The diagrams for the flavored case
are shown in Fig.~\ref{fig:EDM_flavor}.
When both left- and right-handed squarks (sleptons) have mixings,
they contribute to the EDM in the form:
\begin{equation}
J_{LR}^{(d_i)}=\Im \{\delta_{RR}^dy_d\delta_{LL}^q\}_{ii},~~J_{LR}^{(u_i)}=\Im \{\delta_{RR}^uy_u\delta_{LL}^q\}_{ii}.
\end{equation}
 It was shown (see e.g.\ \cite{Hisano:2008hn}) that
$d_d$ can be $\sim 10^{-25}\text{-}10^{-26}\,\text{e cm}$,
and then SUSY parameters are constrained
by hadronic EDM\@.

\begin{figure}[t]
\includegraphics[scale=0.5]{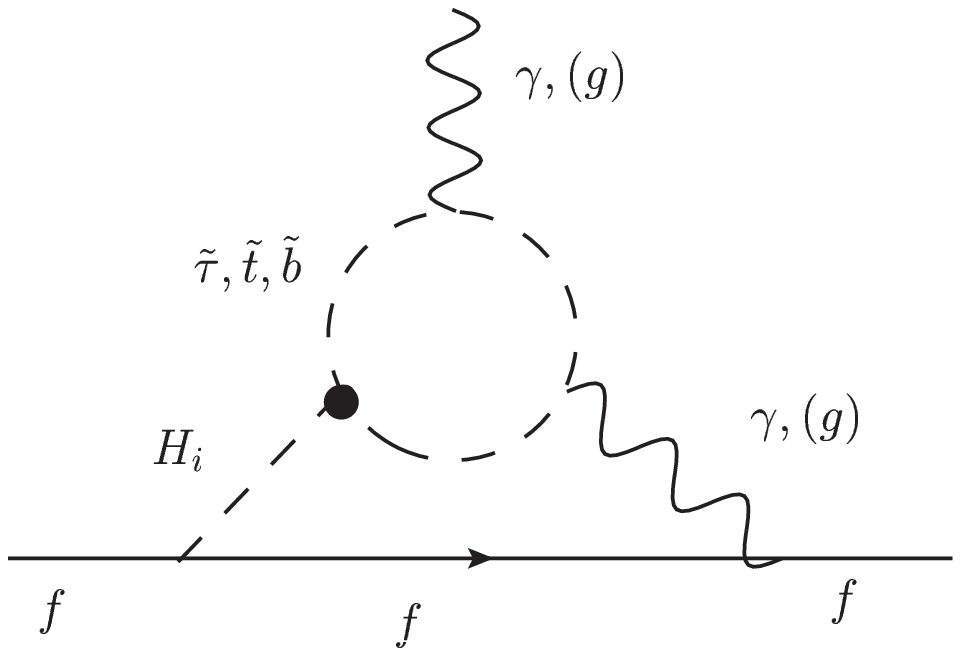}
\includegraphics[scale=0.5]{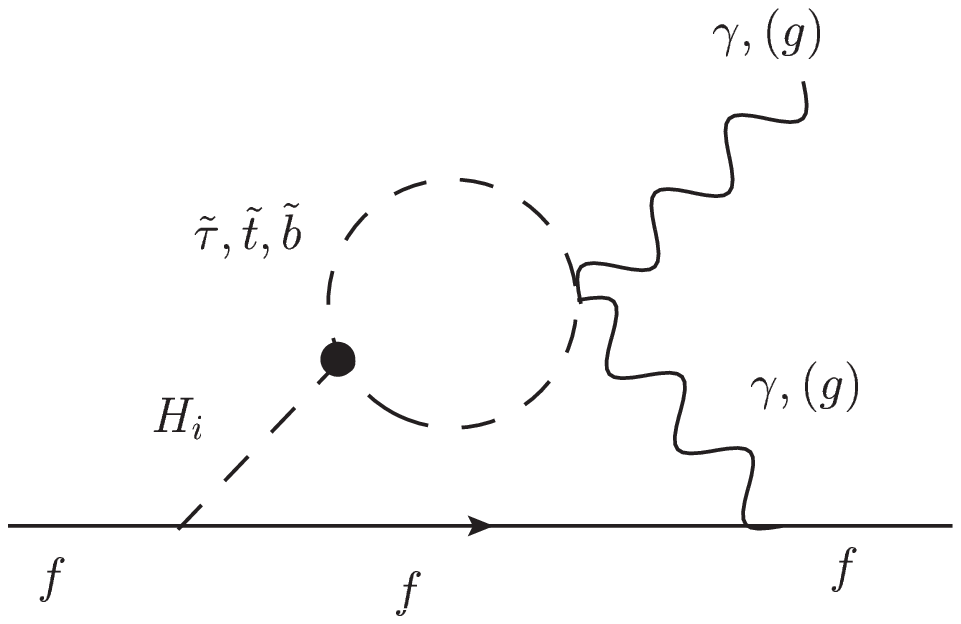}\\[3mm]
\includegraphics[scale=0.5]{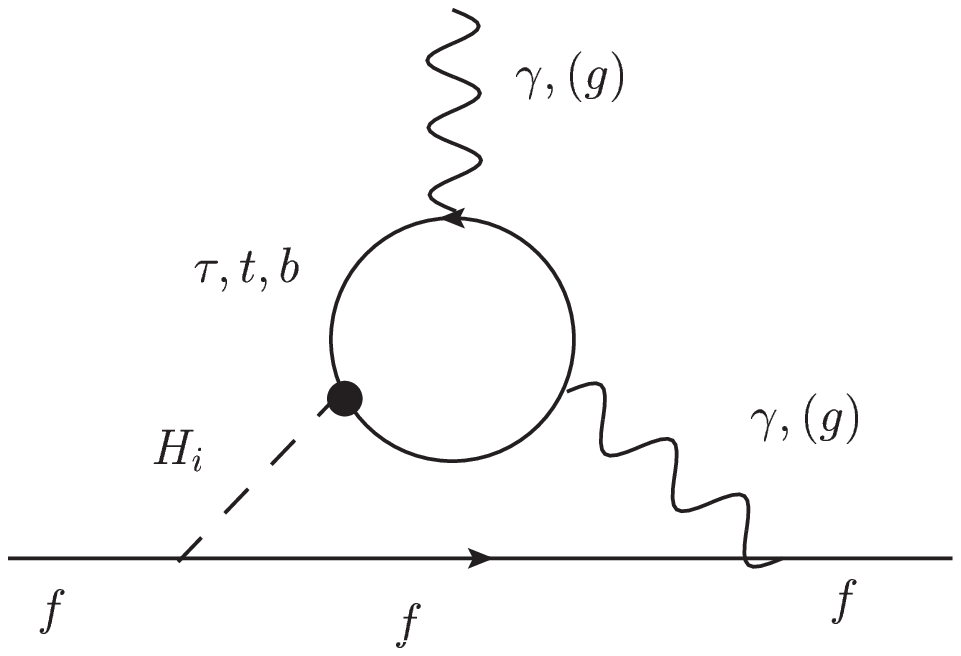}
\includegraphics[scale=0.5]{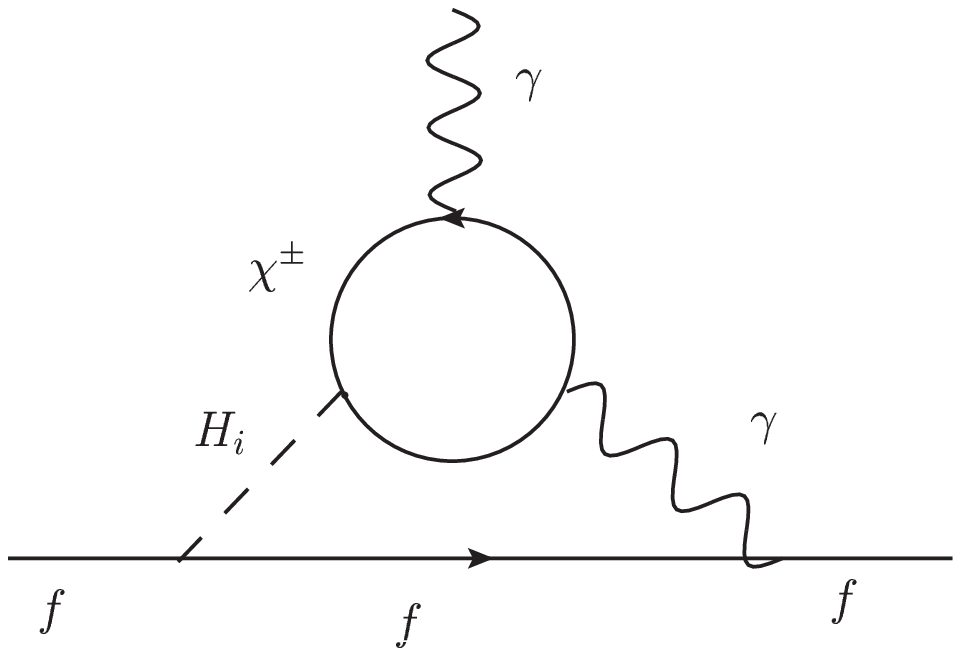}
\smallskip
\noindent
\caption{  Barr-Zee  diagrams: the  $H_i$  lines  denote all  three
neutral Higgs bosons,  including CP-violating Higgs-boson mixing, and heavy
dots   indicate   resummation   of   threshold  corrections   to   the
corresponding Yukawa couplings \protect\cite{Ellis1}.
}
\label{fig:Barr-Zee}
\end{figure}

Also there are additional diagrams called Barr-Zee diagrams that contribute to the EDM beyond the one loop level (see Fig.\ref{fig:Barr-Zee}).

Thus the MSSM gives an elegant backgroud but itself does not predict any definite relations between quarks and leptons including all observations in neutrino.

In other word, its predictions are not testified from many constraints from
various already known observations. These observations must be complicatedly related in reliable models.

It is Grand Unified Theory (GUT) which fulfiles these deficiencies \cite{F-Y}.

\subsection{Minimal Supersymmetric SU(5) GUT}

Here and hereafter, "minimal" means the minimum number of Higgs fields with renormalizable Yukawa coupling.

In SU(5) model \cite{Georgi2}, matter multiplets in (\ref{Q}) are classified to 

\begin{eqnarray}
 {\bf 5}^* = 
\left( 
 \begin{array}{c}
d_1^c \\ 
d_2^c \\
d_3^c\\
e\\
\nu\\
 \end{array}   \right)_L~~~~{\bf 10}=\left(
\begin{array}{ccccc}  
0 & u_3^c& -u_2^c& u_1& d_1\\
  &0 &u_1^c&u_2& d_2\\
 & & 0& u_3& d_3\\
 & & & 0& e^+\\
  & & & & 0\\
\end{array}
\right)_L~~~~{\bf 1}=\nu^c_L.
\end{eqnarray}
We need two Higgs ${\bf 5}^*_H$ and ${\bf 24}_H$, and
SU(5) breaks down to $SU(3)_c\times SU(2)_L\times U(1)_Y$ by
\begin{equation}
{\bf 24}_H=\mbox{diag}\left(V,V,V,-\frac{3}{2}V,-\frac{3}{2}V\right).
\end{equation}
Here ${\bf 5}$ and ${\bf 10}$ are broken to
\begin{equation}
{\bf 5}=(1,2)(1/2)+(3,1)(-1/3)),~~{\bf 10}=(1,1)(1)+(\overline{3},1)(-2/3)+(3,2)(1/6).
\end{equation}
Then, $SU(3)_c\times SU(2)_L\times U(1)_Y$ breaks down to $SU(3)\times U(1)_Q$ via
\begin{equation}
{\bf 5}^*=(0,0,0,0,v/\sqrt{2}).
\end{equation}

Yukawa coupling has the form,
\begin{equation}
W=\frac{1}{4}f^u_{ij}{\bf 10}_i{\bf 10}_j{\bf 5}_H+\sqrt{2}f^d_{ij}{\bf 10}_i{\bf 5^*}_j5^*_H+f^\nu_{ij}{\bf 5}^*_i{\bf 1}_j{\bf 5}_H+M_{ij}{\bf 1}_i{\bf 1}_j.
\end{equation}
Here the products imply
\begin{eqnarray}
{\bf 10}_i{\bf 10}_j{\bf 5}_H&=&\epsilon_{abcde}{\bf 10}_i^{ab}{\bf 10}_j^{cd}{\bf 5}_H^e\nonumber\\
{\bf 10}_i{\bf 5^*}_j{\bf 5}^*_H&=&{\bf 10}_i^{ab}{\bf 5}^{*a}_j{\bf 5}^{*b}_H~~\mbox{etc.}
\end{eqnarray}
with $a,b=1,...,5$.
Then mass matrices have the following forms
\begin{equation}
M^d=M^e=f^dv/\sqrt{2}, ~~M^u=f^uv/\sqrt{2}
\label{SU5mass}
\end{equation}
at GUT scale. This gives nice $b-\tau$ unification. The  disparity between their observed masses is supposed to be due to renormalization effect from $M_{GUT}$ to their mass shell. Unfortunately, we can not explain the disparities between the first and second families even if we take the renormalization effects since it predicts wrong relation
\begin{equation}
m_d/m_s=m_e/m_\mu,~~m_s/m_b=m_\mu/m_\tau.
\label{mass1}
\end{equation}
It also predicts too fast proton decay \cite{Rujura}. Hadronic EDM in SUSY SU(5) was discussed in \cite{Kakizaki}. 
%Compare this with (\ref{massmatrix}) in SO(10) GUT.

Flipped SU(5) changes 
\begin{equation}
u^c\leftrightarrow d^c,~~e^c\leftrightarrow \nu^c
\end{equation}
and, therefore, we obtain in place of (\ref{SU5mass})
\begin{equation}
M_u=M_\nu.
\label{SU5mass2}
\end{equation}
This does not lead to apparent pathology. Moreover, it is attractive from doublet-triplet splitting:
Higgs super potential has the form
\begin{equation}
W_H={\bf 10}\times {\bf 10}\times {\bf 5}+\overline{\bf 10}\times\overline{\bf 10}\times\overline{\bf 5}, 
\end{equation}
which give rise to triplet mass
\begin{equation}
\langle (1,1,0)_{10}\rangle(\overline{3},1;1/3)_{10}(3,1:-1/3)_5+\langle (1,1;0)_{\overline{10}}\rangle(3,1;-1/3)_{\overline{10}}(\overline{3},1;1/3)_{\overline{5}}
\end{equation}
but has no doublet mass since $5+\overline{5}$ has no partner in $10+\overline{10}$
(the missing partner mechanism). This is a solution to the doulet-triplet problem without additional adjoint Higgs.
However, flipped SU(5) drives us to
unrenormalizable heavy Majorana neutrino mass term,
\begin{equation}
{\bf 10}_i{\bf 10}_j\overline{\bf 10}_H\overline{\bf10}_H/\Lambda
\end{equation}
for the seesaw mechanism.

%\begin{equation}
%m_d/m_s=9m_e/m_\mu,~~3m_s/m_b=m_\mu/m_\tau.
%\label{mass2}
%\end{equation}
%changes up sectors and down sectors, 
%\begin{equation}
%M^u=M^\nu
%\label{SU5mass2}
%\end{equation}
%where $M^\nu$ is the Dirac neutrino and it raises no conflict.

The other approaches are to introduce unrenormalizable term \cite{Perez},
\begin{eqnarray}
W_Y&=&\epsilon_{abcde}\left(f_{1ij}{\bf 10}^{ab}_i{\bf 10}^{cd}_j{\bf 24}_{Hf}^e{\bf 5}_H^f+f{2ij}{\bf 10}^{ab}_i{\bf 10}^{cf}_j{\bf 24}_{Hf}^d{\bf 5}_H^e\right)/\Lambda\nonumber\\
&+&g_{1ij}{\bf 5}^*_{Ha}{\bf 24}a_{Hb}{\bf 10}^{bc}_i{\bf 5}^*_{jc}/\Lambda+g_{2ij}{\bf 5}^*_{Ha}{\bf 10}^{ab}_i{\bf 24}^b_{Hc}{\bf 5}^*_{jc}/\Lambda\\
&+&\Delta Y_5\frac{\langle {\bf 24}_H\rangle}{\Lambda} {\bf 5}^*_i{\bf 10}_j{\bf 5}^*_H\nonumber
\end{eqnarray}
or to add another Higgs in Yukawa coupling,
\begin{equation}
Y_{45}{\bf 5}^*_i{\bf 10}_j{\bf 45}^*_H
\end{equation}
etc.
Unfortunately in SU(5) model, right-handed heavy Majorana neutrino belongs to the singlet and we have no constraint on it.  Usually it is assumed to be diagonal but there is no reason to justify it.
The other undetermined parameters (like $m_0.M_{1/2}, A_0, tan\beta$) crucially depend on this assumption.

These points are remedied in the case of renormalizable SO(10) GUT, which is discussed in the next subsection (cEDM and parity odd nuclear interaction).

\subsection{Minimal Supersymmetric SO(10) GUT}
In the SO(10) Grand Unified Theory \cite{Fritzsch},
fermions belong to a multiplet of {\bf 16} representation as
\begin{eqnarray}
\psi
\equiv
 \left(
  u_R^r, u_R^g, u_R^b,
  d_R^r, d_R^g, d_R^b,
  e_R, \nu_R,
  u_L^r, u_L^g, u_L^b,
  d_L^r, d_L^g, d_L^b,
  e_L, \nu_L,
 \right)^T\,.
\end{eqnarray}
 Note that the right-handed neutrino $\nu_R$
is included naturally.
 
\begin{figure}[t]
\begin{center}
\includegraphics[scale=0.6]{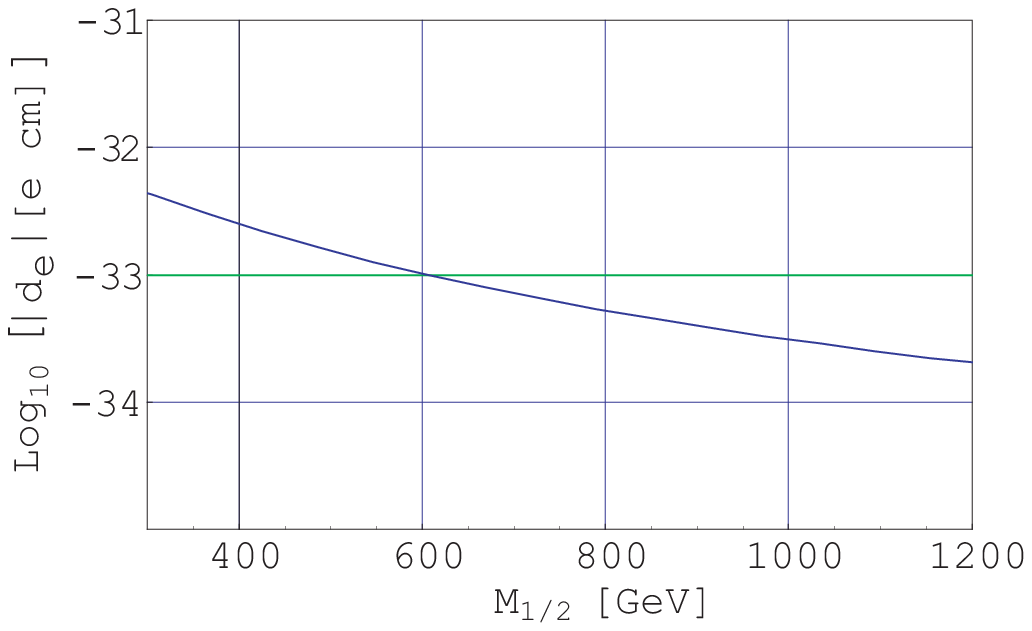}
\includegraphics[scale=0.6]{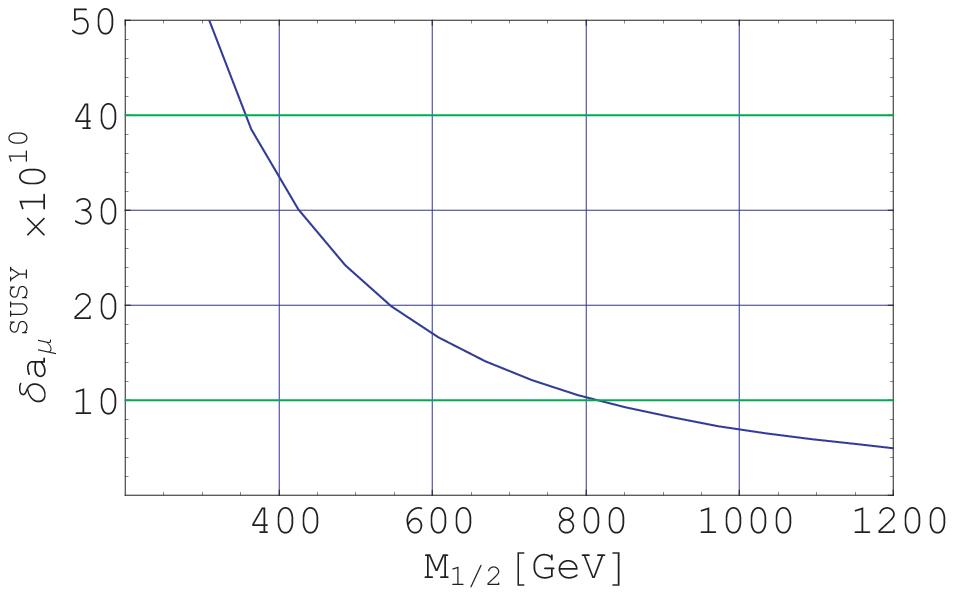}
\includegraphics[scale=0.6]{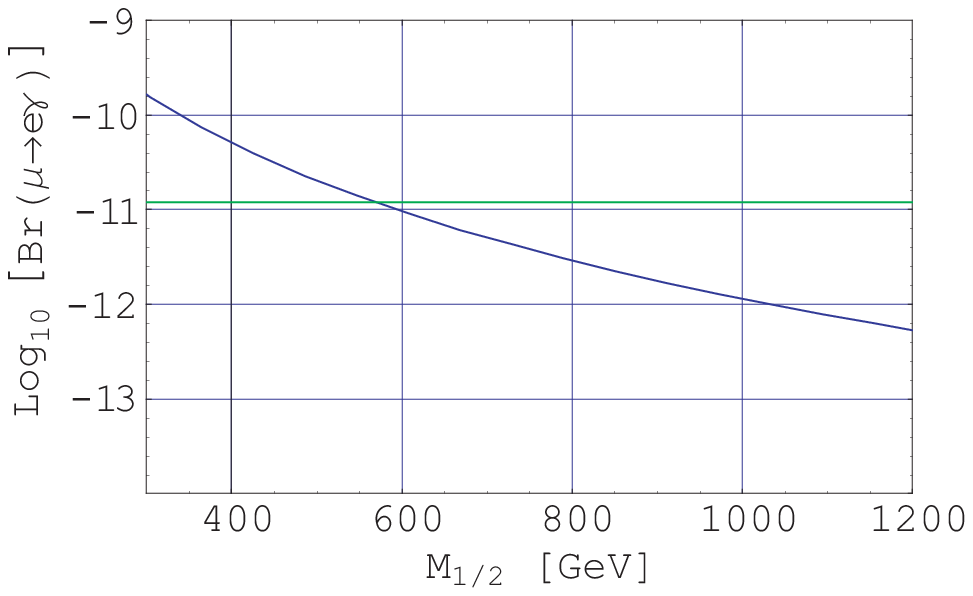}
\includegraphics[scale=0.6]{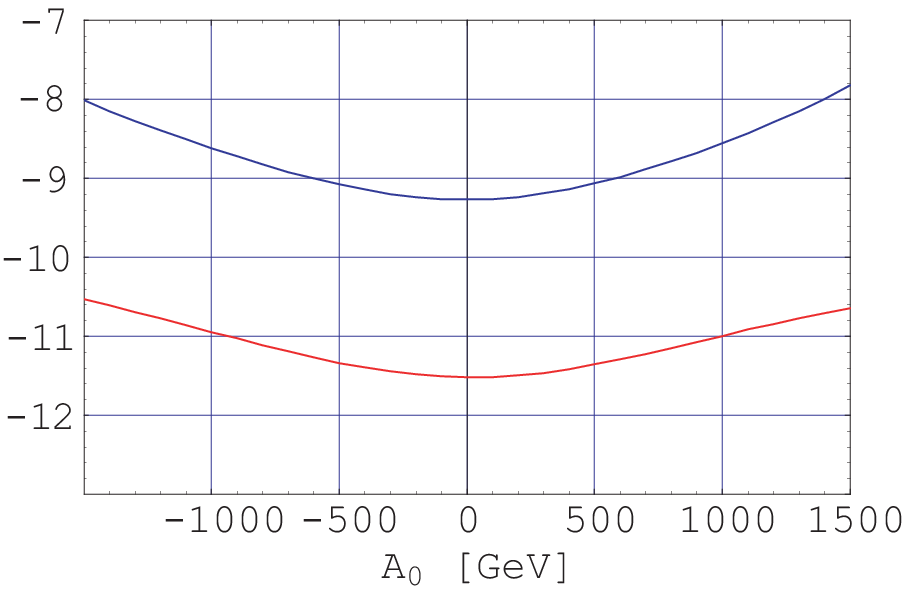}
\caption{
 The predictions for the electron EDM $|d_e|$,
the muon anomalous MDM $\delta a_\mu$ (\ref{g-2}),
and the decay branching ratio of $\mu \to e \gamma$
in the minimal SUSY SO(10)
with respect to the universal gaugino mass $M_{1/2}$. Trilinear term $A_0$ is assumed to be zero except for the last panel. The last panel is added for the reference to see the behaviour of non zero $A_0$, where
the branching ratios,  
$\mbox{Br}(\tau \rightarrow \mu \gamma)$ (top)
and 
$\mbox{Br}(\mu \rightarrow e \gamma)$ (bottom) are given
 as functions of $A_0$ (GeV) 
 for $m_0 =600$ GeV and $M_{1/2}=800$ GeV.  All are cited from \protect\cite{Okada-etal}.
}
\label{fig:eEDM_MSSO10}
\end{center}
\end{figure}

 So-called minimal renormalizable SO(10) model
includes Higgs bosons of {\bf 10} 
and $\overline{{\bf 126}}$  in Yukawa couplings.
This is because
\begin{equation}
{\bf 16}\times {\bf 16}={\bf 10}+{\bf 120}+{\bf 126}.
\end{equation}
In order to make singlet in Yukawa renormalizable coupling, therefore,
Higgs can be ${\bf 10}, {\bf 120}, \overline{{\bf 126}}$.
\footnote{If we relax the renormalizability , different SO(10) models are also possible \cite{Barr2}. However, in this case, we have much less predictivity}
One Yukawa coupling leads to the conclusion that the CKM mass matrix is unity and we needs
at least (minimal) two Higgs, ${\bf 10}+{\bf 120}$ or ${\bf 10}+\overline{{\bf 126}}$.
We select the latter set. This is because
\begin{equation}
\overline{{\bf 126}}=(6,1,1)+(\overline{10},1,3)+(10,3,1)+(15,2,2)
\end{equation}
under $SU(4)_c\times SU(2)_L\times SU(2)_R$ and the second and third terms play essential role in type I and type II seesaw,
respectively. In its SUSY version~\cite{Okada-etal},
${\bf 126}$ is necessary to be added.
Providing the Higgs VEVs, 
 $H_u = v \sin \beta$ and $H_d = v \cos \beta$ 
 with $v=174 \mbox{GeV}$, 
 the quark and lepton mass matrices can be read off as%
\begin{eqnarray}
  M_u &=& c_{10} M_{10} + c_{126} M_{126}   \nonumber \\
  M_d &=&     M_{10} +     M_{126}   \nonumber \\
  M_D &=& c_{10} M_{10} -3 c_{126} M_{126}    \\
  M_e &=&     M_{10} -3     M_{126}   \nonumber \\
  M_T &=& c_T M_{126} \nonumber \\  
  M_R &=& c_R M_{126}  \nonumber \; , 
 \label{massmatrix}
\end{eqnarray} 
where $M_u$, $M_d$, $M_D$, $M_e$, $M_T$, and $M_R$ 
 denote the up-type quark, down-type quark, 
 Dirac neutrino, charged-lepton, left-handed Majorana, and 
 right-handed Majorana neutrino mass matrices, respectively. 
Note that all the quark and lepton mass matrices 
 are characterized by only two basic mass matrices, $M_{10}$ and $M_{126}$,   
 and four complex coefficients 
 $c_{10}$, $c_{126}$, $c_T$ and $c_R$, 
 which are defined as 
 $M_{10}= Y_{10} \alpha^d v \cos\beta$, 
 $M_{126} = Y_{126} \beta^d v \cos\beta$, 
 $c_{10}= (\alpha^u/\alpha^d) \tan \beta$, 
 $c_{126}= (\beta^u/\beta^d) \tan \beta $, 
 $c_T = v_T/( \beta^d  v  \cos \beta)$) and 
 $c_R = v_R/( \beta^d  v  \cos \beta)$), respectively.  
These are the mass matrix relations required by 
 the minimal SO(10) model. 
 The model is very predictive
by virtue of the relation between quark Yukawa matrix,
lepton Yukawa matrix, and neutrino Majorana matrix.
 Fig.~\ref{fig:eEDM_MSSO10} \cite{Okada-etal} shows
the prediction for the electron EDM $|d_e|$ in the minimal SUSY SO(10)
with respect to the universal gaugino mass $M_{1/2}$.
 The muon EDM $|d_\mu|$ exists above $|d_e|$
by roughly a factor of $10^2$.
 The muon anomalous MDM $a_\mu$ and
the LFV decay branching ratio of $\mu \to e\gamma$ 
are also predicted (see Fig.~\ref{fig:eEDM_MSSO10}).  EDM anf LFV are due to the essentially same diagrams, apart from the fact that the former (latter) comes from diagonal (offdiagonal) part of sfermion mass matrix.\footnote{We have added the last panel with nonzero $A_0$ reflecting the recent discovery of Higgs-like particle around $126$ GeV by the LHC. See the last part of Discussion for more detail.}

The effective Lagrangian 
 relevant for the EDM, MDM, and the LFV processes ($\ell_i \rightarrow \ell_j \gamma$) is described in (\ref{eff_dipole}).
${R, L} = (1 \pm \gamma_5)/2 $ is  
 the chirality projection %
\begin{eqnarray}
 {\cal L}_{\mbox{eff}}= 
 - i \frac{e}{2} m_{\ell_i} \overline{\ell}_j \sigma_{\mu \nu} F^{\mu \nu} 
 \left(A_L^{j i} P_L + A_R^{j i} P_R \right) \ell_i  \; , 
\label{eff_dipole2}
\end{eqnarray}
where $P_{L,R}$ are Left-Right projection operators, 
 and  $A_{L,R}$ the photon-penguin couplings of 1-loop diagrams 
 in which chargino-sneutrino and neutralino-charged slepton 
 are propagating.  it should be noted that we have changed the normalization of $A_{L,R}$ from (\ref{eff_dipole}) by $\frac{em_l}{2}$.
The explicit formulas of $A_{L,R}$ etc. used in our analysis 
 are summarized in \cite{Okada-etal} \cite{Hisano-etal}. 
If the diagonal components of $A_{L,R}$ have imaginary parts, the EDMs of the charged leptons are given by
\begin{equation}
d_{l_i}/e=-m_{l_i}\Im (A_L^{ii}-A_R^{ii})
\end{equation}
in the new normalization.
The rate of the LFV decay of charged-leptons is given by 
\begin{eqnarray}
\Gamma (\ell_i \rightarrow \ell_j \gamma) 
= \frac{e^2}{16 \pi} m_{\ell_i}^5 
 \left( |A_L^{j i}|^2  +  |A_R^{j i}|^2  \right) \; , 
\end{eqnarray}
while the real diagonal components of $A_{L,R}$ 
 contribute to the anomalous magnetic moments of 
 the charged-leptons such as 
\begin{eqnarray}
 \delta a_{\ell_i}^{SUSY} = \frac{g_{\ell_i}-2}{2} 
  = -  m_{\ell_i}^2 
  \Re \left[ A_L^{i i}  +  A_R^{i i}  \right]  \; . 
\end{eqnarray}
In order to clarify the parameter dependence 
 of the decay amplitude, 
 we give here an approximate formula of the LFV decay rate 
 \cite{Hisano-etal}, 
\begin{eqnarray}
\Gamma (\ell_i \rightarrow \ell_j \gamma) 
 \sim  \frac{e^2}{16 \pi} m_{\ell_i}^5 
 \times  \frac{\alpha_2}{16 \pi^2} 
 \frac{| \left(\Delta  m^2_{\tilde{\ell}} \right)_{ij}|^2}{M_S^8} 
 \tan^2 \beta \; , 
 \label{LFVrough}
\end{eqnarray}
where $M_S$ is the average slepton mass at the electroweak scale, 
 and $ \left(\Delta  m^2_{\tilde{\ell}} \right)_{ij}$ 
 is the slepton mass estimated in Eq.~(\ref{leading}). 
We can see that the neutrino Dirac Yukawa coupling matrix 
 plays the crucial role in calculations of the LFV processes. 
We use the neutrino Dirac Yukawa coupling matrix of Eq.~(\ref{Ynu})
 in our numerical calculations. 
In the leading-logarithmic approximation, 
 the off-diagonal components ($i \neq j$)
 of the left-handed slepton mass matrix are estimated as 
\begin{eqnarray}
 \left(\Delta  m^2_{\tilde{\ell}} \right)_{ij}
 \sim - \frac{3 m_0^2 + A_0^2}{8 \pi^2} 
 \left( Y_{\nu}^{\dagger} L Y_{\nu} \right)_{ij} \; ,  
 \label{leading}
\end{eqnarray}
where the distinct thresholds of the right-handed 
 Majorana neutrinos are taken into account 
 by the matrix $ L = \log [M_G/M_{R_i}] \delta_{ij}$. 

Unlike the muon MDM, quark and lepton EDMs have still null observation. This is of course due to tiny CP violation and due to the cancellation of diagrams where $\gamma$ (gluon) couples with slepton (squark) and where it does with Higgsino (gluino) in Fig. \ref{fig:EDM_flavor} \cite{Nath}.

%\begin{figure}[t]
%\begin{center}
%\includegraphics[scale=0.8]{muonedm.eps}
%\includegraphics[scale=0.8]{muonedm2.eps}
%\caption{The cancellation of EDM. For quarks $\gamma$ is replaced by gluon \cit%e{Nath}.}
%\label{fig:muedm}
%\end{center}
%\end{figure}

If we consider gauge mediation scenario for SUSY breaking,
$A_0\approx 0$.
In the basis where both of the charged-lepton 
 and right-handed Majorana neutrino mass matrices 
 are diagonal with real and positive eigenvalues, 
 the neutrino Dirac Yukawa coupling matrix at the GUT scale 
 is found to be \cite{F-O}
\begin{eqnarray}
 Y_{\nu} = 
\left( 
 \begin{array}{ccc}
-0.000135 - 0.00273 i & 0.00113  + 0.0136 i  & 0.0339   + 0.0580 i  \\ 
 0.00759  + 0.0119 i  & -0.0270   - 0.00419  i  & -0.272    - 0.175   i  \\ 
-0.0280   + 0.00397 i & 0.0635   - 0.0119 i  &  0.491  - 0.526 i 
 \end{array}   \right) \; .  
\label{Ynu}
\end{eqnarray}     
Semi-leptonic LFV processes are discussed in \cite{F-K-I}.

Thus the EDM, MDM, lepton flavor violations etc. are all closely connected, which are expected to be explained universally by GUT. However, we do not adhere to any special model in this review, and will discuss more phenomenological models in the following subsections. 

These may be the remnants from GUT or may be independent of GUT.
For instance, the adjoint representation of SO(10), ${\bf 45}$ is decomposed into
\begin{equation}
{\bf 45}=(1,3,1)+(3,1,1)+(15,1,1)+(6,2,2)
\end{equation}
under $SU(4)_c\times SU(2)_L\times SU(2)_R$ and leads to Left-Right symmetric model, $g_L=g_R$.
Also
${\bf10}$ representation is decomposed into 
\begin{equation}
{\bf 10}=(1,2,2)+(6,1,1),
\end{equation}
which leads us to two Higgs $SU(2)_L$ doublets under the SM.
Also $\overline{{\bf 126}}$ is
\begin{equation}
\overline{{\bf 126}}=(6,1,1)+(10,3,1)+(\overline{10},1,3)+(15,2,2).
\end{equation}
If $(\overline{10},1,3)$ ($(10,3,1)$)  has vev, it gives type I (type II, or Higgs triplet Model) seesaw model.

In the following we consider these models independently of GUT principally. \footnote{He et al. discussed neutron EDM in those models \cite{H-M-P}.}

\subsection{Two Higgs Doublet Model}
\begin{figure}[t]
\begin{center}
\includegraphics[scale=0.5]{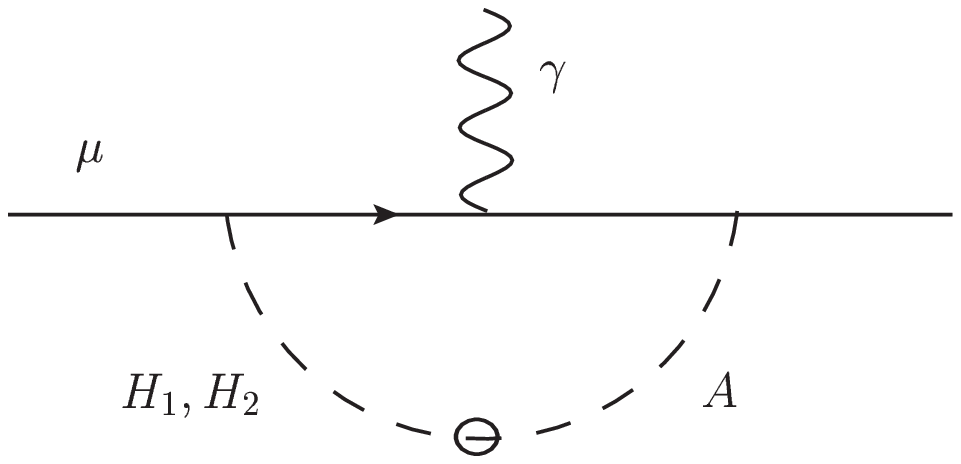}
\includegraphics[scale=0.5]{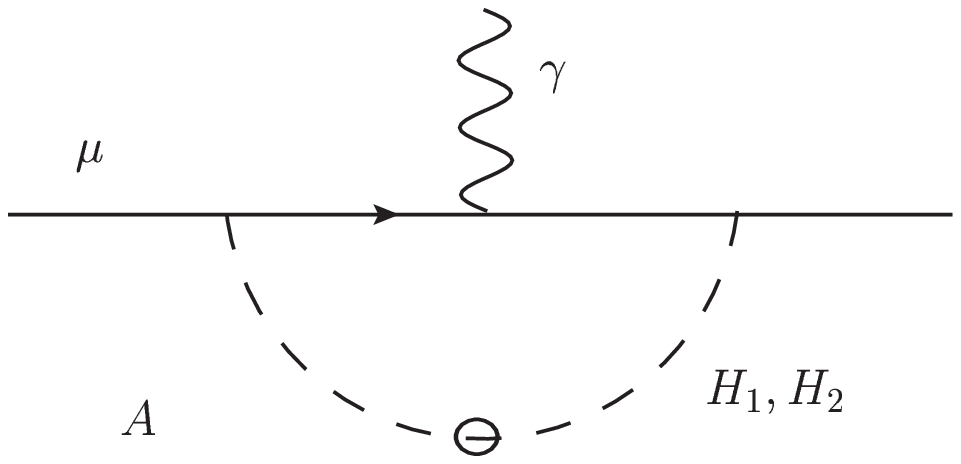}\\[3mm]
\includegraphics[scale=0.5]{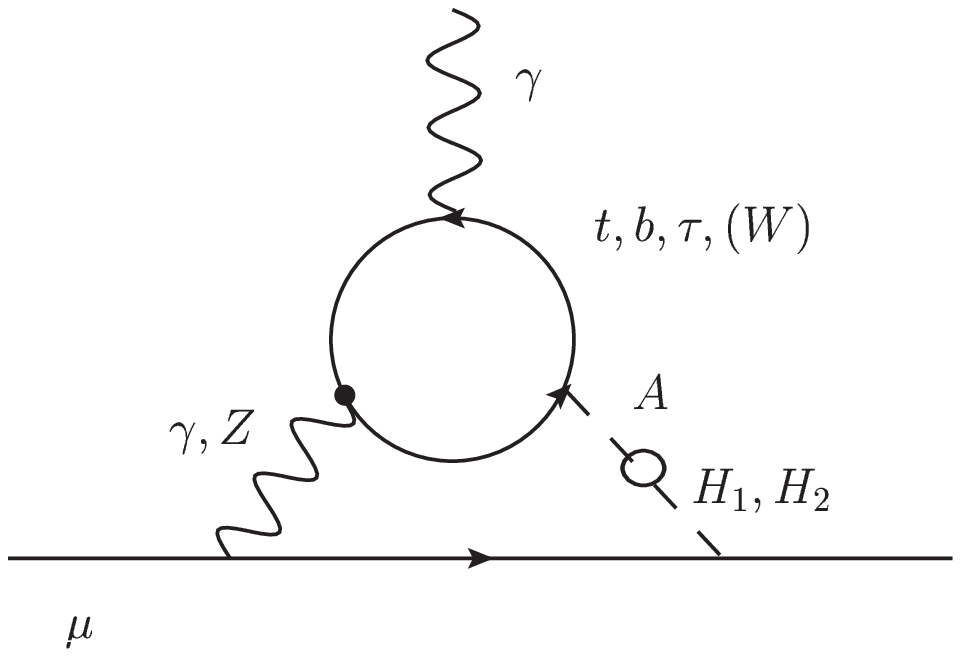}
\includegraphics[scale=0.5]{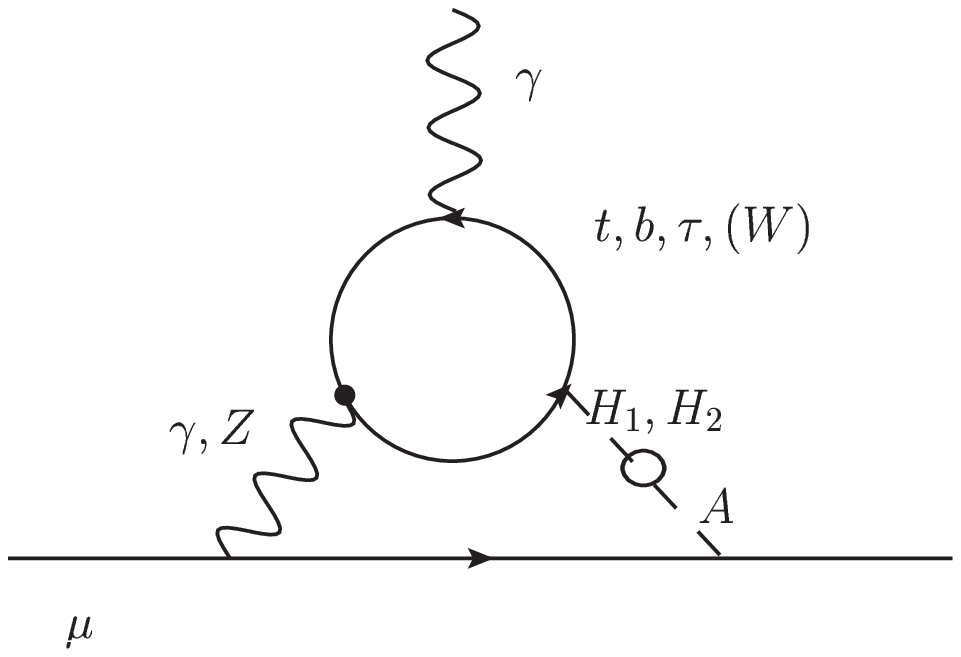}
\caption{
 Diagrams of muon EDM in two Higgs doublet model \protect\cite{Barger:1996jc}.
}

\label{fig.muEDM}
\end{center}
\end{figure}

 Most of models beyond the SM
has some new Higgs bosons.
 As the simplest extension of the Higgs sector of the SM
which has only one Higgs doublet $\phi_1$,
another Higgs doublet $\phi_2$ is introduced
in the Two Higgs Doublet Model \cite{2HDM}.
 There are several types of the model
depending on which doublet couples with which fermion:
\begin{eqnarray}
\text{type I (SM-like)}
&:&
 \text{$\phi_1$ couples with all fermions}\nonumber\\
&&
 \text{$\phi_2$ decouples with fermions}\nonumber\\
\text{type II (MSSM-like)}
&:&
 \text{$\phi_1$ couples with down-type quarks and charged leptons}\nonumber\\
&&
 \text{$\phi_2$ couples with up-type quarks}\nonumber\\
\text{type III (general)}
&:&
 \text{both of Higgs doublets couple with all fermions}\nonumber\\
etc. && \nonumber
\end{eqnarray}
 If CP-violating term exists in the Higgs potential,
e.g.\ $(\phi_1^\dagger \phi_1)(\phi_1^\dagger \phi_2)$ with an imaginary coefficient,
there appears the mixing between CP-even ($H^0$)
and CP-odd ($A^0$) neutral Higgs bosons.
 Then these Higgs bosons can contribute to the EDM
(see Fig.\ref{fig.muEDM}).
 The mixing between $H^0$ and $A^0$
provides also CP-violating electron-nucleon effective interactions
($\bar{e}i\gamma^5 e \bar{N}N$, etc.)
which will contribute to the atomic EDM\@.

Barger et al. \cite{Barger:1996jc} gave large $d_\mu$ close to the proposed experiments. It should be remarked that their values are calculated in units of $\Im Z$ where \cite{Weinberg3}
\bea
&&\la \phi_2^0\phi_1^{0*}\ra_q=\sum_n\frac{\sqrt{2}G_FZ_n}{q^2-m_{H^n}^2},\nonumber\\
&&
\la \phi_2^0\phi_1^0\ra_q=\sum_n\frac{\sqrt{2}G_F\tilde{Z}_n}{q^2-m_{H^n}^2}
\eea
and it is probable that $|\Im Z|\approx 0$. Indeed, the masses of neutral and charged Higgses and phases are tightly constrained from
 $R_b\equiv \frac{\Gamma(Z\rightarrow b\overline{b})}{\Gamma(Z\rightarrow hadrons)},~
\Gamma(b\rightarrow s\gamma),~\overline{B}^0-B$ mixing, $\rho$ parameter etc.,
and we should take those constraints into account \cite{Grant}.

\subsection{Higgs Triplet Model}
\label{subsec:HTM}
 In the Higgs Triplet Model \cite{HTM},
we introduce a SU(2) triplet $Y=2$ scalar as
\begin{eqnarray}
\Delta
\equiv
 \left(
  \begin{array}{cc}
   \Delta^+/\sqrt{2} & \Delta^{++}\\
   \Delta^0 & -\Delta^+/\sqrt{2}
  \end{array}
 \right), \ \ \
{\mathcal L}_{triplet Yukawa}
= - h_{\alpha\beta} \overline{L_\alpha^C} i\sigma^2 \Delta P_L L_\beta
  + h.c.
\end{eqnarray}
 This model generates neutrino masses without right-handed neutrinos
with the triplet vacuum expectation value $v_\Delta$
which is given by the explicit breaking of the lepton number.
 This model is very predictive
because of a clear relation
\begin{eqnarray}
m_{\alpha\beta} = \sqrt{2} v_\Delta h_{\alpha\beta},
\end{eqnarray}
where $m_{\alpha\beta}$ denotes the Majorana mass matrix for neutrinos.
 There is no new interaction with quarks
and no effect on quark EDM\@.
 Unfortunately,
this model can not give a large contribution to lepton EDM also
because of the absence of the new interaction with right-handed fermions.
 For example,
one loop diagram for electron has a factor of $|h_{\alpha e}|^2$
(similarly to Fig.~\ref{fig:dEDM_SM1loop}).

\subsection{Left-Right Symmetric Model}
Left-Right (LR) model \cite{PS} is used in a variety of ways and needed to be clarified.
If we consider it as a remnant of SO(10), $SO(10)\rightarrow SU(4)_c\times SU(2)_L\times SU(2)_R$, it satisfies at $v_{PS}$ energy scale
\begin{equation}
g_L=g_R
\end{equation}
and this Pati-Salam (PS) model is unified at $M_{GUT}$ as
\begin{equation}
\frac{M_4}{\alpha_4}=\frac{M_{2L}}{\alpha_{2L}}=\frac{M_{2LR}}{\alpha_{2R}}=\frac{M_{1/2}}{\alpha_{GUT}}.
\label{PS}
\end{equation}
Also mixing matrices for left-handed and right-handed fermions are the same.
Of course these constraints are realized at $v_{PS}$ but start to be violated as the energy goes down to the SM scale by renormalization effects.

However, if we consider a model of $SU(2)_L\times SU(2)_R\times U(1)_{B-L}$, we are free from the above constraints.
For instance, in the framework of SO(10) GUT, $v_R$ is of order of $O(10^{12})$GeV. However, if we go apart from GUT but still consider that the mixing matrix for right-handed quarks $V_R$ has a similar structure as that for left-handed quarks $V_L$, the lower limit of $M_{WR}$ is relaxed to $M_{WR}>1.6$ TeV \cite{Beal}.
Moreover we may go beyond the restriction to V-A and V+A interactions only and consider general form;
\begin{eqnarray}
L_{\mu\rightarrow e\nu\overline{\nu}}&=&-\frac{4G_F}{\sqrt{2}}\Big[g_{RR}^S(\overline{e}_R\nu_{eL})(\overline{\nu_\mu}_L\mu_R)+g_{RL}^S(\overline{e}_R\nu_{eL})(\overline{\nu_\mu}_R\mu_L)\nonumber\\
&+&g_{LR}^S(\overline{e}_L\nu_{eR})(\overline{\nu_\mu}_L\mu_R)+g_{LL}^S(\overline{e}_L\nu_{eR})(\overline{\nu_\mu}_R\mu_L)\nonumber\\
&+&g_{RR}^V(\overline{e}_R\gamma^\mu \nu_{eR})(\overline{\nu_\mu}_R\gamma_\mu\mu_R)+g_{RL}^V(\overline{e}_L\gamma^\mu\nu_{eL})(\overline{\nu_\mu}_L\gamma_\mu\mu_L)\\
&+&g_{LR}^V(\overline{e}_L\gamma^\mu \nu_{eL})(\overline{\nu_\mu}_R\gamma_\mu\mu_R)+g_{LL}^V(\overline{e}_L\gamma^\mu \nu_{eL})(\overline{\nu_\mu}_R\gamma_\mu\mu_R)\nonumber\\
&+&\frac{g_{RL}^T}{2}(\overline{e}_R\sigma^{\mu\nu}\nu_{eL})(\overline{\nu_\mu}_R\sigma_{\mu\nu}\mu_L)+\frac{g_{LR}^T}{2}(\overline{e}_L\sigma_{\mu\nu}\nu_{eR})(\overline{\nu_\mu}_L\sigma_{\mu\nu}\mu_R)+H.c.\Big]\nonumber
\end{eqnarray}
 
 The charge of $U(1)$ in the LR model has a clear meaning
as the difference between the baryon number $B$
and the lepton number $L$
in contrast to the mysterious hypercharge $Y$ in the SM\@.
 Similarly to $SU(2)_L$ doublet in the SM,
the right-handed fermions compose doublet of $SU(2)_R$ in the LR Model.
 Therefore,
the right-handed neutrinos $\nu_R$ are introduced naturally
as $SU(2)_R$ partners of right-handed charged leptons.
 After the spontaneous breaking of $SU(2)_R\otimes U(1)_{B-L}$
to $U(1)_Y$,
 the hypercharge is given by 
\begin{equation}
Y/2 = I_{3R} + (B-L)/2.
\label{Y}
\end{equation}
Since electric charge is connected by
\begin{equation}
Q=I_{3L}+Y/2,
\end{equation}
(\ref{Y}) implies the charge quantization, which is one of great achievements of \cite{PS}.

 Since we require the parity symmetry to the theory,
the gauge coupling of $SU(2)_R$ must be the same
as the one of $SU(2)_L$: $g_2 \equiv g_{2L} = g_{2R}$.
 The Higgs field which gives the Yukawa terms
is a complex bidoublet of $SU(2)_L\otimes SU(2)_R$
with $B-L = 0$.
 The bidoublet field can be expressed as
\begin{eqnarray}
\Phi
\equiv
 \begin{pmatrix}
  \phi_1^0 & \phi_2^+\\
  \phi_1^- & \phi_2^0
 \end{pmatrix},
\end{eqnarray}
which transform as $\Phi \to
\Phi^\prime = U_L \Phi U_R^\dagger$
under $SU(2)_L$ and $SU(2)_R$.
\begin{eqnarray}
\left<\Phi\right>=\begin{pmatrix}
  \kappa & 0 \\
  0 & \kappa'
 \end{pmatrix}
\end{eqnarray}
with $\kappa\neq \kappa'$  gives rise to the breaking of L-R symmetry.
However, $\Phi$ is neutral (B-L=0) and $U(1)_{B-L}$ is not broken.
So we need another Higgs. 
 Usually,
two complex triplet fields
($\Delta_L$ for $SU(2)_L$ and $\Delta_R$ for $SU(2)_R$) with $B-L = 2$
are also introduced to generate Majorana neutrino masses
(see also the Higgs Triplet Model in Sect.~\ref{subsec:HTM})
\begin{equation}
\Delta_L=(3,1,2)~~\Delta_R=(1,3,2).
\end{equation}
Then, the gauge symmetry breaking proceeds as follows:
first $\Delta_R^0$ acquires vev $v_R$, leading to $SU(2)_L\times U(1)_Y$ with
(\ref{Y}), which furthermore breaks to $U(1)_Q$ by the vev of $\Phi$.

 The triplet Yukawa coupling for $\Delta_R$
must be equal to the coupling for $\Delta_L$
because of the parity symmetry which is spontaneously broken by $v_R$, and we have
\begin{equation}
v_R\gg \kappa,~\kappa'\gg v_L.
\end{equation}

Thus the two Higgs doublets model (not of all but its measure part) and Higgs triplet model in the previous subsections are combined together in left-right model.

\begin{figure}[t]
\begin{center}
\includegraphics[scale=0.8]{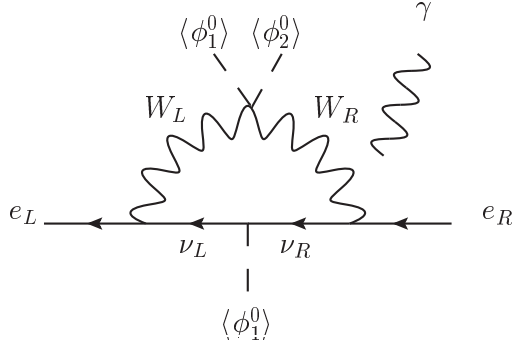}
\caption{
 One of diagrams of gauge boson contributions to the electron EDM in LR symmetric model.
 Gauge bosons contribute to quark EDM also.
}
\label{fig:eEDM_LRM}
\end{center}
\end{figure}
Their vevs $v_R$ and $v_L$ are different to each other and from 
$\kappa$ and $\kappa'$.

 Figure~\ref{fig:eEDM_LRM} shows one of diagrams
which contribute to the EDM of the electron.
 In a simple case where there is only one flavor of leptons,
the electron EDM is estimated~\cite{Bernreuther:1990jx} as
\begin{eqnarray}
|d_e|
<
 \left\{
  \begin{array}{l}
   \displaystyle
   8.2\times 10^{-27}\,
   \frac{|\text{Im}(m_D)|}{\text{MeV}}\, \text{e cm} \quad
   \text{for} \left( \frac{m_R}{m_W} \right)^2 \gg 1,
\\
   \displaystyle
   3.3\times 10^{-26}\,
   \frac{|\text{Im}(m_D)|}{\text{MeV}}\, \text{e cm} \quad
   \text{for} \left( \frac{m_R}{m_W} \right)^2 \ll 1,
  \end{array}
 \right.
\end{eqnarray}
where $m_D$ denotes the Dirac mass of the neutrino
($m_D \overline{\nu_{L}} \nu_{R}$)
and $m_R$, $m_W = 80\,\text{GeV}$ are the masses of heavy right-handed neutrino and the light $W$ boson, respectively.

 The contributions of Higgs bosons in the LR model
are not significant~\cite{Liu:1985zp}.

Another important contribution of $W_R$ in CP violation may be the neutrinoless double beta decay \cite{Takasugi}

\subsection{Fourth Family Model}

We set the quarks of the fourth family \cite{Frampton} as
\begin{equation}
(t',b')^T.
\end{equation}
The mixing angles and CP violating phases for N families are given by
\begin{equation}
N^2-(2N-1)=\frac{N(N-1)}{2}+\frac{(N-1)(N-2)}{2},
\end{equation}
where the first term is mixing angles and the second CP phases
(see Appendix {\bf D} for Majorana fermion case).

If we consider 4-generation SM (SM4) \cite{Hamzaoui}, we can construct new Jarlskog parameter in place of (\ref{Jarlskog2})
\begin{equation}
A_{(234)}=(m_{t'}^2-m_t^2)(m_{t'}^2-m_c^2)(m_t^2-m_c^2)(m_{b'}^2-m_b^2)(m_{b'}^2-m_s^2)(m_b^2-m_s^2)J_{(234)}.
\label{Jarlskog4}
\end{equation}
If we take heavy quark masses $t'$ and $b'$ in the range of 300 to 600 GeV,
$A_{(234)}/T_{EW}^{12}$ can be of order $n_B/n_\gamma$.
\begin{equation}
|V_{ud}|^2+|V_{us}|^2+|V_{ub}|^2=0.9999(10).
\end{equation}

Thus SM4 enhances CP violation and therefore the EDM also.

We may consider inside the loop only $t,~t',~b'$ heavy and identify as 
\begin{equation}
c=u=U,~~d=s=b=D.
\end{equation}
Then two loop diagram of Fig.2 vanishes for u quark but survives for d,s quarks, giving
\begin{equation}
d_d\approx \lambda^7\frac{\alpha_s}{4\pi}\frac{\alpha_W}{4\pi}\frac{1}{16\pi^2}G_fm_d\frac{m_t^2}{m_W^2}\approx 3\times 10^{-32} ecm,
\end{equation}
where $\lambda$ is the Wolfenstein parameter, $\lambda=|V_{us}|=0.22$.
This is only two orders of magnitude larger than the SM result of section (\ref{d_d}).
%For lepton EDM, there are variations even in SM4.
%That is, it depends on whether we consider light neutrinos as Dirac or Majorana.
However, if we consider the chromoelectric dipole moment of the s quark,
\begin{eqnarray}
\tilde{d}_s&\approx& \Im (V_{ts}^*V_{tb}V_{hb}^*V_{hs})\frac{\alpha_s}{4\pi}\frac{\alpha_W}{4\pi}\frac{1}{16\pi^2}G_Fm_s\frac{m_t^2}{m_s^2}\nonumber\\
&\approx& \lambda^5\frac{\alpha_s}{4\pi}\frac{\alpha_W}{4\pi}\frac{1}{16\pi^2}G_Fm_s\frac{m_t^2}{m_s^2}.
\end{eqnarray}
Using the estimation of the relation between $d_N$ and $\tilde{d}_s$ by \cite{K-K-Z}, we have \cite{Hamzaoui}
\begin{equation}
d_N\approx -\frac{1}{2}\tilde{d}_s\approx 5\times 10^{-30}~\mbox{e cm}.
\end{equation}

\subsection{Extra Dimensions}
The motivations for extra dimensions are diverse for both SUSY and non-SUSY. There are many SUSY breaking scenarios. The extra dimension makes the geometrical SUSY breaking possible. The gauge supermultiplets propagate in the bulk, and we get gaugino mass
\begin{equation}
M_a\approx \frac{\langle F\rangle}{M_5^2R_5},
\end{equation}
which is called gaugino mediation.

Even if the theory itself is CP invariant, it may be violated by extending the theory to extra dimensions. This is because the compactification of the extra dimensions does not respect the symmetry in general.
The CP phases come either from the boundary condition of extra dimensions (One of Scherk-Scwarz mechanisms \cite{Scherk}) or vev of fifth gauge field (Hosotani mechanism \cite{Hosotani}) \cite{Cosme} \cite{Grzadkowski}.
\begin{equation}
L=i\overline{\psi}\gamma^N(\partial_N-ieA_N)\psi-M\overline{\psi}\psi.
\end{equation}
Inclusion of a torsion term results in a nonminimal term
\begin{equation}
\kappa \overline{\psi}\sigma F \psi
\end{equation}
or the fermion mass term via Hosotani mechanism 
\begin{equation}
\overline{\psi}(M+i\gamma _5X_4)\psi
\end{equation}
with
\begin{equation}
X_4\equiv\int dy A_4.
\end{equation}
Here $y$ is the coordinate of the extra dimension.
The rotation of mass term gives rise to 
\begin{equation}
\kappa'\overline{\psi}\sigma F\gamma_5 \psi.
\end{equation}
The concrete constraints from the observation are given, for instance \cite{Adachi},
\begin{equation}
d_n=\frac{4}{3}d_d-\frac{1}{3}d_u=\frac{4}{3}d_d,
\end{equation}
simply because they did not consider up quark and
\begin{equation}
d(KK)\sim -2.3\times 10^{-23}(Rm_W)^2 ~\mbox{[e cm]}.
\end{equation}
Since $\frac{4}{3}d(KK)$ must be less than the experimental upper limit,
we have
\begin{equation}
\frac{1}{R}>33 m_W \simeq 2.6 \mbox{[TeV]}.
\end{equation}
\section{The EDMs of Atoms}
The origin of the difficulties of the measurement of electron EDM is due to the absence of resonance unlike the neutron.
A possible way is to perform the resonance experiment on neutral atom and to interpret the result in terms of electron EDM or hadron EDM. These object has very tiny values and let's consider the effect linear in the EDM.
In the subsequent atomic and molecular experiments we treat an internal electic field ${\bf E}_{int}$ induced by atom or molecule as well as an external elctric field ${\bf E}$. This ${\bf E}$ induces the EDM $e{\bf r}_i$ with the intrinsic $\sum_i\beta\boldsymbol{\sigma}_i\equiv \sum_i{\bf d}_e^i$. Thus the total Hamiltonian is a sum of unperturbed P,T-even terms,
\be
H_0=\sum_ic\boldsymbol{\alpha}\cdot \boldsymbol{p}_i+\beta_imc^2+V_{nucl}(r^i)+\sum_{i<j}V_C(r_{ij}),
\label{H0}
\ee
and T,P-odd term
\be
H_{PTV}=-\sum_i{\bf d}_e^i\cdot {\bf E}_{int}^i-\sum_i{\bf d}_e^i\cdot {\bf E}-e\sum_i{\bf r}_i\cdot {\bf E}.
\label{PTV}
\ee

The last term of $H_0$ is a two-body interaction and can not be solved exactly.

The first and third terms are P-odd and the second P-even.
So the first and second order energy shifts are given by
\be
E_m^1=-\sum_i\la m_0|{\bf d}_e^i|m_0\ra \cdot {\bf E}
\ee
and
\be
E_m^2=\sum_{n\neq m}\sum_i\left\{\frac{\la m_0|{\bf d}_e^i\cdot {\bf E}_{int}|n_0\ra\la n_0|e{\bf r}_i\cdot {\bf E}|m_0\ra}{E_m^0-E_n^0}+\frac{\la m_0|e{\bf r}_i\cdot {\bf E}|n_0\ra\la n_0|{\bf d}_e^i\cdot {\bf E}_{int}|m_0\ra}{E_m^0-E_n^0}\right\}.
\ee
Here $|m_0\ra$ is an eigen state of $H_0$. It should be remarked that, as will be shown in \bref{exp}, EDM appears as the coefficient of the energy shift linear in the external electric field.
So
\be
E_m=E_m^1+E_m^2\equiv -{\bf d}'\cdot {\bf E},
\ee
where
\bea
{\bf d}'=\sum_i\la m_0|{\bf d}_e^i|m_0\ra
&-&\sum_{n\neq m}\sum_i\left\{\frac{\la m_0|{\bf d}_e^i\cdot {\bf E}_{int}|n_0\ra\la n_0|e{\bf r}_i|m_0\ra}{E_m^0-E_n^0}\right.\nonumber\\
&&\left.+\frac{\la m_0|e{\bf r}_i|n_0\ra\la n_0|{\bf d}_e^i\cdot {\bf E}_{int}|m_0\ra}{E_m^0-E_n^0}\right\}.
\label{Schiffth}
\eea
However this ${\bf d}'$ vanishes as follows.
\bea
&&\la m_0|e{\bf r}^i\cdot {\bf E}|n_0\ra\la n_0|{\bf d}_e^i\cdot {\bf E}_{int}|m_0\ra
=ie\la m_0|{\bf r}_i\cdot {\bf E}|n_0\ra\la n_0|{\bf d}_e^i\cdot [{\bf p}_i,~H_0]|m_0\ra
\nonumber\\
&&=ie\la m_0|{\bf r}_i\cdot {\bf E}|n_0\ra\la n_0|{\bf d}_e^i\cdot {\bf p}_i|m_0\ra(E_m^0-E_n^0).
\eea
Using the communtation relation $[r_i,p_j]=i\delta_{ij}$ the second term of \bref{Schiffth} cancells with the first term. This is the famous Scfiff's theorem \cite{Schiff}. Since the expectation value of 
\be
\boldsymbol{\Sigma}\cdot{\bf E}_{int}=[\boldsymbol{\Sigma}\cdot\nabla,~H_0]
\label{CR}
\ee
does not contribute to a linear Stark effect, the residual EDM is becomes 
\be
V_{EDM}=-d_e(\beta-1)\boldsymbol{\Sigma}\cdot {\bf E}_{int}
\ee
and the residual energy shift is
\bea
\Delta E&=&-d_e\la m_0|(\beta-1)\boldsymbol{\Sigma}\cdot {\bf E}|m_0\ra-2d_e\sum_{n\neq m}\frac{\la m_0|{\bf r}\cdot {\bf E}|n_0\ra\la n_0|(\beta-1)\boldsymbol{\Sigma}\cdot {\bf E}_{int}|m_0\ra}{E_m-E_n}\nonumber\\
&=&-{\bf d}(atom)\cdot {\bf E}.
\label{transition}
\eea
The first term has no enhancement factor unlike the the second term and is much smaller than the second term, and
\be
{\bf d}(atom)=-2d_e\sum_{n\neq m}\sum_i\frac{\la m_0|{\bf r}^i|n_0\ra\la n_0|(\beta-1)\boldsymbol{\Sigma}^i\cdot {\bf E}_{int}|m_0\ra}{E_m-E_n}.
\ee

$d_{atom}$ has a large value when these states are almost degenerate.
However, this enhancement is reflected in quite different ways in paramagnetic atoms and diamagnetic atoms.
Though (\ref{transition}) itself is rather universal, $H_{PTV}$ is variant.
One example is P,T-odd Nucleon-electron interaction like (see Appendix {\bf C} for the detail)
\begin{equation}
+iG_{S'}\overline{N}N\overline{L}\gamma_5L+iG_{P'}\overline{N}\gamma_5N\overline{L}L.
\end{equation}
There are other CP violating effective interactions (see the last part of section IIB).
Another important interaction is due to Schiff moment.
The origin of Schiff moment itself is not unique.

As we mentioned, in the nonrelativistic Hamiltonian for a system of particles of finite size, there is no interaction energy of first order
in the EDM if there is no misalignment of charge and moment distribution.
Schiff also indicated in \cite{Schiff} how this theorem is violated by relativistic (Breit equation $O\left((v/c)^2\right)$) and the misalignment (the Schiff moment), where $v$ is the velocity of electron or nucleon.
Prior to this discovery, Salpeter \cite{Salpeter} indicated that radiative corrections of $O\left((v/c)^3\right)$ enhances hydrogen EDM.
 
Sandars pointed out that relativistic effect of electron EDM in heavy alkali atom gives large atomic EDM \cite{Sandars}.

Let us proceed to discuss in more detail for hydrogen-like atom.
The EDM has, by definition, odd parity and naively vanishes between the states with same parity. 

For nonrelativistic case, its energy levels are
\begin{equation}
E=-\frac{mZ^2\alpha^2}{2n^2}.
\end{equation}
Here $n$ is the principal quantum number, and this energy is degenerate
$n^2$-ply, $\sum_{l=0}^{n-1}(2l+1)=n^2$ (see Eq.(\ref{finestructure})).

If we consider the relativistic effects (spin effects) the degenerate energy levels are split into $n$ fine-structure components at different $j$ \cite{Landau}.
Let us obtain the relativistic terms w.r.t.  $O(v/c)$ (see Appendix {\bf E} for relativistic expansion and Appendix {\bf F} for nonrelativistic approximation for more detail).

At the first order of $v/c$, we obtain the Pauli equation
\begin{equation}
i\hbar \frac{\partial \varphi}{\partial t}=H\varphi=\left[\frac{1}{2m}\left({\bf p}-\frac{e}{c}{\bf A}\right)^2+e\Phi-\frac{e}{2mc}\boldsymbol{\sigma}\cdot{\bf B}\right]\varphi.
\end{equation}
In further approximation of $O\left((v/c)^2\right)$, we assume ${\bf B}=0~(i.e.~{\bf A}=0)$, and we get
\begin{equation}
H=\frac{{\bf p}^2}{2m}+e\Phi-\frac{{\bf p}^4}{8m^2c^2}-\frac{e\hbar}{4m^2c^2}\boldsymbol{\sigma}\cdot[{\bf E}\times {\bf p}]-\frac{e\hbar^2}{8m^2c^2}\nabla\cdot{\bf E}.
\end{equation}
If ${\bf E}$ is centrally symmetric,
\begin{equation}
{\bf E}=-\frac{{\bf r}}{r}\frac{d\Phi}{dr}.
\end{equation}
The spin-orbit interaction (the fourth term) becomes
\begin{equation}
V_{sl}=\frac{e\hbar}{4m^2c^2r}\boldsymbol{\sigma}\cdot [{\bf r}\times {\bf p}]\frac{d\Phi}{dr}=\frac{\hbar^2}{2m^2c^2r}\frac{dU}{dr}{\bf l}\cdot{\bf s}.
\end{equation}
For many electron case of atomic number $Z$,
\begin{equation}
V_{sl}=\sum \alpha_a{\bf l}_a\cdot{\bf s}_a,
\end{equation}
where
\begin{equation}
\alpha_a=\frac{\hbar^2}{2m^2c^2r_a}\frac{dU(r_a)}{dr_a},
\end{equation}
\begin{equation}
|U(r_a)|\approx \frac{Ze^2}{a_B}\approx \frac{Z^2me^4}{\hbar^2},
\end{equation}
and, therefore,
\begin{equation}
\alpha\approx Z^4\left(\frac{e^2}{\hbar c}\right)^2\frac{me^4}{\hbar^2}.
\label{ls}
\end{equation}
%\begin{equation}
%\overline{\alpha}\approx \left(\frac{Ze^2}{\hbar c}\right)^2\frac{me^4}{\hbar^2%}
%\label{ls}
%\end{equation}
For given total ${\bf L}$ and ${\bf S}$, the averaged $V_{SL}$ is
\begin{equation}
V_{SL}=A{\bf S}\cdot {\bf L},
\end{equation}
\begin{equation}
{\bf L}\cdot{\bf S}=\frac{1}{2}[J(J+1)-L(L+1)-S(S+1)].
\end{equation}
Since the value of ${\bf L}$ and {\bf S} are same for a multiplet, energy splitting is given by the Lande's interval rule,
\begin{equation}
\Delta E_{J,J+1}=AJ.
\label{Lande}
\end{equation}
Then we obtain
\begin{eqnarray}
&&1s_{1/2}\nonumber\\
&&\left(2s_{1/2},~2p_{1/2}\right),~~2p_{3/2} \\
&&\left(3s_{1/2},3p_{1/2}\right),~~\left(3p_{3/2},3d_{3/2}\right),~~3d_{5/2} .\nonumber
\label{finestructure}
\end{eqnarray}
The remaining degeneracy is removed by the hyperfine-structure components caused by the
radiative correction (Lamb shift \cite{Kroll}).
So using this hyper finesplitting, we obtain large atomic EDM \cite{Salpeter}.

To estimate atomic EDM we need two informations.
One is that of atomic wave functions and another is that of P (or T) violating 
interactions. 

%The latter has quite different aspects for paramagnetic and diamagnetic atoms.

%has, by definition, odd parity and naively vanishes between the states with same parity. 

Atomic EDM is due to those of constituents, electrons and nucleons.
For electron EDM it is very important that the atom has an unpaired electron, and electron EDM is proportional to $Z^3$ \cite{Sandars} (For the review, see \cite{G-F}\cite{P-R}.).
If there is no unpaired electron (diamagnetic atom), we can measure quark (or hadron) EDM.
For proton EDM, nucleus has an unpaired proton. In this case polarized molecule takes an important role \cite{Sandars2}.

We are dealing with many electrons system and the electron wave functions are not in general exact. In this case the expectation values of the EDM depends on the representation.

\begin{equation}
<b|{\bf r}|a>\equiv {\bf r}_{ba}=\frac{1}{E_b-E_a}<b|H_0{\bf r}-{\bf r}H_0|a>,
\label{rep1}
\end{equation}
where $H_0=-\frac{\hbar^2}{2m}\nabla^2+V$.
Inserting this into (\ref{rep1}), we obtain
\begin{eqnarray}
{\bf r}_{ba}&=&\frac{1}{E_b-E_a}<b|\nabla^2{\bf r}-{\bf r}\nabla^2|a>=-\frac{i}{m\omega_{ba}}{\bf p}_{ba}\\
&=&\frac{1}{m\omega_{ba}^2}(\nabla V)_{ba}.
\end{eqnarray}
These three representations are of course equivalent. However, if we use the approximate wave functions, these values are different in general.
So we must be careful what is the origins of discrepancies, due to different approximations or to representations \cite{Branden}.
\subsection{Relativistic Effects}
The relativistic equation of atom with CP violating interaction ($\xi$ term) is
\begin{equation}
\left[\gamma_\mu\left(\partial_\mu-\frac{ie}{\hbar c}A_\mu\right)-i\frac{mc}{\hbar}\right]u=\xi\frac{e}{4mc^2}\gamma_5\gamma_\mu\gamma_\nu F_{\mu\nu}u.
\label{modified}
\end{equation}
Electron EDM breaks the CP invariance and CP violating energy equation is
\begin{equation}
(H_0+H')u=Eu.
\end{equation}
Here $H_0$ is the Hamiltonian of the original single electron Dirac equation in the external field,
\begin{equation}
H_0=m\beta c^2+\boldsymbol{\alpha}\cdot (c{\bf p}-e{\bf A})+e\phi,
\end{equation}
which leads to \bref{H0} in the static limit and $H'$ is CP violating interaction Hamiltonian of the right-hand side of (\ref{modified}) \cite{Salpeter} \footnote{This term only contributes to the intrinsic EDM and we denote $H'$, distinguishing it from $H_{PTV}$ of \bref{PTV}.}
\begin{eqnarray}
H'&=&\xi \frac{e\hbar}{2mc}\beta\left(\boldsymbol{\Sigma}\cdot{\bf E}+i\boldsymbol{\alpha}\cdot {\bf B}\right)\nonumber\\
&\approx&-\xi \frac{e\hbar}{2mc}\beta\boldsymbol{\Sigma}\cdot\nabla \phi,
\label{H'}
\end{eqnarray}
where $\boldsymbol{\Sigma}$ is defined by \bref{Sigma} and
\begin{eqnarray}
\boldsymbol{\alpha}
\equiv \beta\boldsymbol{\gamma}=
 \begin{pmatrix}
 {\bf 0} & \boldsymbol{\sigma} \\
 \boldsymbol{\sigma} & {\bf 0}
 \end{pmatrix}. 
\end{eqnarray}
%(See also (\ref{pseudo}\bref{tensor}).)
$\xi$ is dimensionless constant which measures the EDM in units of the Bohr magneton. Thus $\xi\ll 1$ implies that the EDM is small compared with e times the Compton wavelength.
The last approximation in (\ref{H'}) comes from the relativistic suppression due to the mixing of the upper half with the lower one.  
It should be remarked that $\boldsymbol{\sigma}\cdot {\bf E}$ is T-odd (T is time reversal operator) but $\boldsymbol{\alpha}\cdot {\bf B}$ (the suppressed term) T-even.\\

We consider a hydrogen-like atom with charge Z, where ${\bf E}=\frac{Ze}{r^2}{\bf e}_r$.
\begin{equation}
H'=-\xi Z\alpha r^{-2} s_r~~\mbox{with}~s_r=\boldsymbol{\sigma}\cdot {\bf e}_r/2
\label{single}
\end{equation}
in atomic units $e=m=\hbar=1$.  The Hamiltonian of the above single electron system in the mean field approximation is generalized for many electrons system as
\cite{Sandars}
\begin{equation}
H_0'=\sum_i \left[\beta_imc^2+\boldsymbol{\alpha}_i\cdot c{\bf p}_i+e\phi_i\right]+\sum_{j\neq k}\frac{1}{2}\left[\frac{e^2}{r_{jk}}+B_{jk}\right].
\end{equation}
Here suffix indicates the quantity of i'th electron and $B_{jk}$ is the relativistic corrections. Taking the retard potential into consideration, Lagrangian can be described up to order O($(\frac{v}{c})^2$) as
\begin{equation}
\frac{e^2}{r_{jk}}+B_{jk}\equiv \frac{e^2}{r_{jk}}
\left[  1-\frac{1}{2}\left(\boldsymbol{v}_j \cdot{\boldsymbol{v}}_k+\frac{(\boldsymbol{v}_j\cdot{\bf r}_{jk})(\boldsymbol{v}_k\cdot{\bf r}_{jk})}{r_{jk}^2}\right)\right].
\label{breit1}
\end{equation}
Further higher order $\geq O\left((\frac{v}{c})^3\right)$ corrections come from photon emission (Breit interaction) \cite{Breit} and 
\bea
&&U_{jk}=\frac{e^2}{r_{jk}}-\pi\left(\frac{e\hbar}{mc}\right)^2\delta({\bf r}_{jk})-\frac{e^2}{2m^2c^2r_{jk}}\left({\bf p}_j{\bf p}_k+\frac{({\bf r}_{jk}{\bf p}_j)({\bf r}_{jk}{\bf p}_k)}{r_{jk}^2}\right)\nonumber\\
&&\frac{e^2\hbar}{4m^2c^2r_{jk}^3}\left(-(\boldsymbol{\sigma}_j+2\boldsymbol{\sigma}_k)[{\bf r}_{jk}{\bf p}_j]+(\boldsymbol{\sigma}_k+2\boldsymbol{\sigma}_j)[{\bf r}_{jk}{\bf p}_k]\right)\\
&&+\frac{1}{4}\left(\frac{e\hbar}{mc}\right)^2\left(\frac{\boldsymbol{\sigma}_j\boldsymbol{\sigma}_k}{{\bf r}_{jk}^3}-3\frac{(\boldsymbol{\sigma}_j{\bf r}_{jk})(\boldsymbol{\sigma}_k{\bf r}_{jk})}{r_{jk}^5}-\frac{8\pi}{3}\boldsymbol{\sigma}_j\boldsymbol{\sigma}_k\delta({\bf r}_{jk})\right).\nonumber
\label{breit2}
\eea
The second line corresponds to spin-orbit interaction and the third spin-spin interaction.

If we incorporate the spin of nucleus, the degeneracy of ${\bf J}$ is split (hyperfine structure),
\begin{equation}
V_{iJ}=a{\bf i}\cdot{\bf J},
\end{equation}
where ${\bf i}$ and ${\bf J}$ are the spin of nucleus and total angular momentum of electron envelope, respectively.
However, in this hyperfine splitting dominant contribution comes from magnetic dipole and electric quadrupole and does not play important role in the EDM.

Thus the linear Stark appears as relativistic effects in duplicate meanings, i.e. $1-\beta$ and $B_{jk}$ components. Many particle interaction effects are due to this relativistic effect as well as due to the other nonrelativistic excitation effects.

We proceed to the detailed calculation of single electron case (\ref{single}) \cite{Salpeter}.
The operator $s_r$ commutes with ${\bf M}^2$ and $M_z$, but has odd parity,
and
\begin{equation}
<l=j\pm\frac{1}{2},j,m|s_r|l=j\mp\frac{1}{2},j,m>=\frac{1}{2}.
\label{odd}
\end{equation}
Let the radial part of $u=r\chi_{nl}$. Then it satisfies
\begin{equation}
\left\{\left[2\mu \phi-d^2/dr^2\right]+l(l+1)r^{-2}-2\mu E_{nl}\right\}\chi_{nl}=0,
\label{Diraceq}
\end{equation}
where $\mu$ is the reduced mass of electron and set equal to unity in the subsequent equations.
It follows from (\ref{odd}) that
\begin{equation}
<n,l_+,j,m|H'|n',l_-,j,m>=-\xi Z\alpha(2j+1)^{-1}(E_n-E_n')\int dr \chi_{nl_+}\chi_{n_rl_-}
\label{odd2}
\end{equation}
with $l_\pm=j\pm\frac{1}{2}$. Therefore, naively the first-order perturbation
vanishes.  The exception is discussed in section {\bf 5.2}.
The second-order perturbation energy is obtained by use of (\ref{odd2})
\begin{equation}
\Delta E_{n,l_\pm,j}=(\xi Z\alpha/(2j+1))^2\int dr \chi_{n,l_\pm}\left[E_n+\frac{1}{2}\frac{d^2}{dr^2}+\frac{Z}{r}-\frac{1}{2}\frac{l_\mp(l_\mp+1)}{r^2}\right]\chi_{nl_\pm}.
\end{equation}
Using 
\begin{equation}
<r^{-2}>=\frac{Z^2}{n^3(l+1/2)},
\end{equation}
we obtain
\begin{equation}
\Delta E_{n,l_\pm,j}=\pm Z^4\xi^2n^{-3}(2j+1)^{-1}(l_\pm+\frac{1}{2})^{-1}\alpha^2 Ry
\label{datom}
\end{equation}
with $Ry=\frac{me^4}{2\hbar^2}=13.6$eV.
\subsection{Peculiar Property of Paramagnetic Atom}
The atomic enhancement factor defined by
\begin{equation}
K\equiv d_{atom}/d_e
\end{equation}
is given by \cite{Flambaum}
\begin{equation}
K=\sum_m\frac{4(Z\alpha)^3r_{m0}\hbar c}{(J+1)a_B^2\gamma (4\gamma^2-1)(N_0N_m)^{3/2}(E_m-E_0)}
\label{enhance}
\end{equation}
for alkali atom. Here the sum is taken over the excited state $m$,  and $N_0,N_m$ are effective principal quantum number defined in (\ref{notations}). $\gamma=\sqrt{(j+1/2)^2-Z^2\alpha^2}$ and $r_{0m}$ is electric dipole radial integral,
\begin{equation}
r_{nl,n'l'}=<n,l|r|n',l-1>=\sqrt{l}\int_0^\infty R_{n',l-1}R_{n,l}r^3 dr
\label{radialintegral}
\end{equation}
in units of $a_B=\frac{\hbar^2}{me^2}$:Bohr radius. We will derive (\ref{enhance}) (see Eq.(\ref{datom2})). 
We start with the general relativistic arguments.
For diamagnetic atoms the dominant contribution to atomic EDM comes from that of nucleus, which will be discussed later.\\
In hydrogen-like atom, states of different angular momenta $l$ with fixed principal number $n$ are degenerate in nonrelativistic approximation. The eigenfunctions with external field are the superposition of the field-free functions with different l-values, which gives the linear Stark effect.
Let us write the Dirac spinor in the form
\begin{equation}
u_\pm=r^{-1}\left(\chi_{2\pm}(r)\eta_{jl_\pm},~-i\chi_{1\pm}(r)\eta_{jl\mp}\right)^T.
\end{equation}
$\chi_i$ satisfy the following equations,
\begin{eqnarray}
\frac{d\chi_1}{dr}-\kappa\frac{\chi_1}{r}&=& \left[\frac{mc}{\hbar}\left(1-\frac{E}{mc^2}\right)-\alpha\frac{Z}{r}\right]\chi_2,\nonumber\\
\frac{d\chi_2}{dr}+\kappa\frac{\chi_2}{r}&=& \left[\frac{mc}{\hbar}\left(1+\frac{E}{mc^2}\right)+\alpha\frac{Z}{r}\right]\chi_1,
\label{Pauli2}
\end{eqnarray}
where 
\begin{equation}
\kappa=\mp (j+\frac{1}{2})~\mbox{for}~j=l\pm \frac{1}{2}.
\label{kappa}
\end{equation}
Using (\ref{H'}) and (\ref{odd}), we obtain
\begin{equation}
\left<n,j,l_+,m|H'|n,j,l_-,m\right>=-\frac{1}{2}\xi Z\alpha\int_0^\infty dr r^{-2}(\chi_{2+}\chi_{2-}-\chi_{1+}\chi_{1-}).
\label{splitting}
\end{equation}
$\chi_{2\pm}$ and $\chi_{1\pm}$ are related to each other as (\ref{Pauli2}) and we obtain
\begin{equation}
4\left<l_+|H'|l_-\right>=\xi Z\alpha^3\int_0^\infty dr(D_+\chi_+)r^{-2}(D_-\chi_-),
\end{equation}
where 
\begin{equation}
D_\pm=\frac{d}{dr}\pm \frac{j+1/2}{r}.
\end{equation}
The exact Dirac wave functions with given $n,l,j$ are \cite{Landau} \cite{Bethe}\begin{eqnarray}
\frac{\chi_2}{r}&=&-\frac{\sqrt{\Gamma(2\gamma+n_r+1)}}{\Gamma(2\gamma+1)\sqrt{n_r!}}\sqrt{\frac{1+\epsilon}{4N(N-\kappa)}}\left(\frac{2Z}{Na_B}\right)^{3/2}e^{-\frac{Zr}{Na_B}}\left(\frac{2Zr}{Na_B}\right)^{\gamma-1}\times\nonumber\\
&&\times\left[-n_rF\left(-n_r+1,2\gamma+1,\frac{2Zr}{Na_B}\right)+(N-\kappa)F\left(-n_r,2\gamma+1,\frac{2Zr}{Na_B}\right)\right]
\end{eqnarray}
and
\begin{eqnarray}
\frac{\chi_1}{r}&=&-\frac{\sqrt{\Gamma(2\gamma+n_r+1)}}{\Gamma(2\gamma+1)\sqrt{n_r!}}\sqrt{\frac{1-\epsilon}{4N(N-\kappa)}}\left(\frac{2Z}{Na_B}\right)^{3/2}e^{-\frac{Zr}{Na_B}}\left(\frac{2Zr}{Na_B}\right)^{\gamma-1}\times\nonumber\\
&&\times\left[n_rF\left(-n_r+1,2\gamma+1,\frac{2Zr}{Na_B}\right)+(N-\kappa)F\left(-n_r,2\gamma+1,\frac{2Zr}{Na_B}\right)\right].
\end{eqnarray}
Here $F$ is the confluent hypergeometric function and $n_r$ radial quantum number, the number of nodes of radial part of the wave function,
\begin{equation}
n_r=\frac{\alpha Z\epsilon}{\sqrt{1-\epsilon^2}}-\gamma,~ n=n_r+|\kappa| 
\end{equation}
with $\epsilon=\frac{E}{mc^2}$,
and
\begin{equation}
N=\sqrt{n^2-2n_r(|\kappa|-\sqrt{\kappa^2-\alpha^2Z^2})}.
\label{notations}
\end{equation}
Here we have used 
\begin{equation}
1-\epsilon^2=\frac{\alpha^2Z^2}{N^2}
\end{equation}
and normalized $\lambda r=\frac{Z}{Na_B}r$ as in the hypergeometric functions.
$N$ is called apparent principal quantum number and
\begin{equation}
E_{nl}=-\frac{mZ^2\alpha^2}{2N^2}.
\end{equation}

Using these forms,
(\ref{splitting}) finally reads, for instance \cite{Salpeter},
\begin{equation}
\left<2s_{1/2}|H'|2p_{1/2}\right>=\frac{\xi Z^3\alpha(\gamma-1)}{2\gamma(2\gamma-1)(\gamma+1)(2\gamma +1)^{1/2}}Ry,
\label{datom2}
\end{equation}
where $\gamma$ takes the value $\sqrt{1-Z^2\alpha^2}$ in this case.  For small $Z\alpha$, it is reduced to
\begin{equation}
\left<2s_{1/2}|H'|2p_{1/2}\right>=\frac{\xi Z^5\alpha^3}{8\sqrt{3}}Ry.
\end{equation}
For more general case
\begin{equation}
<j,l_+|H'|j,l_->=-\frac{4(Z\alpha)^3}{\gamma(4\gamma^2-1)(N_+N_-)^{3/2}}Ry.
\end{equation}
For heavy alkali atom, for instance, cesium, \cite{Khriplovich}
\begin{equation}
K(Cs)=d(Cs)/d_e=-\frac{16}{3}\frac{Z^3\alpha^2r(6s,6p_{1/2})}{a_B^2\gamma(4\gamma^2-1)(N_sN_p)^{3/2}}\frac{Ry}{E(6p_{1/2})-E(6s)}=118.
\label{Cs}
\end{equation}
The radial integral is experimentally known \cite{Shabanova}
\begin{equation}
r(6s,6p_{1/2})=\int_0^\infty dr r^3R_{60}(r)R_{61}(r)=5.5a_B.
\end{equation}
Eq. (\ref{Cs}) should be checked with the experimental result \cite{YbF}.
%\begin{equation}
%|d_e|=(-1.5\pm 5.5\pm 1.5)\times 10^{-26} e~cm.
%\end{equation}

For Francium (Z=87, 7s$\rightarrow 7p_{1/2}$), $K(Fr)$ is estimated as $873$.
However, as was stated in \cite{G-F}, (\ref{enhance}) is not applicable for atoms with
complex configurations and requires electrons' correlations. Such calculations are performed in, for instance, \cite{Nataraj2}
and $K(Fr)$ is modified to $895$.

There are some discrepancies on the estimate of enhancement factor of $K$ of Thallium [Xe]$4f^{14}5d^{10}6s^26p^1$ \cite{Liu} \cite{Dzuba-Flambaum} \cite{Sahoo}. The discrepancy seems to come from the starting assumptions. \cite{Dzuba-Flambaum} considered that Thallium has three valence electrons, $6s^26p^1$, whereas \cite{Sahoo} considered it having one valence electron.
If we adopt \cite{Dzuba-Flambaum},
\begin{equation}
d({}^{205}Tl)=-582(20)d_e
\end{equation}
or if we take \cite{Sahoo},
\begin{equation}
d({}^{205}Tl)=-466d_e\rightarrow d_e<1.6\times 10^{-27} \mbox{e cm}.
\end{equation}

In preparing this revised version, an interesting paper has just appeared \cite{Porsev} which asserts that this discrepancy disappears, converging to $K=-573$.

Xe has closed electron shell of $5s^2~5p^6$ but we may one elctron of $5p$ state excited to $5p^5~6s^1$, which resembles with that of Cs, [Xe]$6s^1$, whose enhancement factor was estimated to $K({}^{133}Cs)=114$ \cite{Hartley} or $120.53$ \cite{Nataraj3}.

As for ${}^{129}$Xe, the lowest excited state with a $6s$ electron has a enhancement value $K({}^{129}Xe^*)=120$ \cite{Flambaum}\cite{Player} or 111 \cite{Nataraj4}, and 
\begin{equation}
d_e<3.2\times 10^{-24} e~cm.
\end{equation}
Using these results, Ellis et al. considered the maximal EDMs of nuclei \cite{Ellis}.
\subsection{Chiral Condensate \label{condensate}}
Before discussing diamagnetic atom, we will briefly resume QCD chiral dynamics
\cite{Leutwyler}. This is because hadronic matrix elements are described in terms of quark condensates by using operator product expansion.

Let us begin with the following effective action (see Appendix {\bf G} for the implication of the effective action).
\begin{equation}
L=\overline{q}(i\gamma^\mu D_\mu-m)q-\frac{\alpha_s}{4\pi}G_{\mu\nu}\tilde{G}^{\mu\nu},
\label{effective}
\end{equation}
where
\begin{equation}
D_\mu=\partial_\mu-ig_s A_\mu^a\lambda^a.
\end{equation}
This action is invariant under $SU(3)_L\times SU(3)_R$ transformations in the limit of $m_u=m_d=m_s=0$.
That is,
\begin{equation}
Q_Lq=e^{i\alpha^a\lambda^a}q,~~Q_Rq=e^{i\beta^a\lambda^a\gamma_5}q,
\end{equation}
where $u,d,s$ quarks constitute SU(3) group,
\begin{equation}
q=(u,d,s)^T
\end{equation}
and we have the following conserved currents
\begin{equation}
j_{L,R}^{\mu a} =\overline{q}_{L,R}\lambda^a\gamma^\mu q_{L,R}.
\end{equation}
Here $\lambda^a$ are the Gell-Mann's $3\times 3$ matrices and $q_L~(q_R)$ are left-handed (right-handed) part of $q$.
\begin{eqnarray}
j^{a\mu}&=&j_L^{a\mu}+j_R^{a\mu},\\
j_5^{a\mu}&=& j_L^{a\mu}-j_R^{a\mu}.
\end{eqnarray}

So we have the conserved currents and conserved charges $Q_a$ and $Q_{5a}$.
They satisfy the algebras
\begin{equation}
[Q_a,Q_b]=if_{abc}Q_c,~~[Q_{5a},Q_b]=if_{abc}Q_{5c},~~[Q_{5a},Q_{5b}]=if_{abc}Q_c.
\end{equation}

However, this group is not exact and they are spontaneously broken to
\begin{equation}
Q_a|0\rangle =0,~~Q_{5a}|0\rangle \neq 0.
\end{equation}
Thus there appear 8 pseudo Nambu-Goldstone bosons. Pseudo implies that the original chiral symmetry ($Q_5$ transformation) is not exact.
It is broken by
\begin{equation}
H_{SB}=m_u\overline{u}u+m_d\overline{d}d+m_s\overline{s}s.
\end{equation}
This can be rewritten as
\begin{eqnarray}
H_{SB}&=&(m_u+m_d+m_s)(\overline{u}u+\overline{d}d+\overline{s}s)/3\nonumber\\
&+&(m_u-m_d)(\overline{u}u-\overline{d}d)/2\\
&+&(2m_s-m_u-m_d)(2\overline{s}s-\overline{u}u-\overline{d}d)/6.\nonumber
\label{HSB}
\end{eqnarray}
Here the first line is an SU(3) invariant, the second breaks isospin SU(2),
and the third represents the deviation of s quark mass from the SU(3) symmetry.
\begin{eqnarray}
M_\pi^2&=&(m_u+m_d)B+O(m_q^2~\mbox{ln}~ m_q),\\
M_{K^\pm}^2&=& (m_u+m_s)B+O(m_q^2~\mbox{ln}~ m_q),\\
M_{K^0}^2&=&(m_d+m_s)B+O(m_q^2~\mbox{ln}~ m_q).
\end{eqnarray}
Here $B=-\frac{2}{f_\pi^2}\langle 0|\overline{q}q|0\rangle$ with pion decay constant $f_\pi=93$MeV, and we have used the chiral limit
\begin{equation}
f_\pi=f_K,~~\langle 0|\overline{u}u|0\rangle=\langle 0|\overline{d}d|0\rangle=\langle 0|\overline{s}s|0\rangle.
\end{equation}
Adler-Bell-Jackiw axial vector singlet current anomaly \cite{Adler} and its non-Abelian version is
\begin{equation}
\partial_\mu j^{5\mu}=2i\sum_{q=u,d,s}m_q\overline{q}\gamma_5q+\frac{N_f}{8\pi^2}\left(F\tilde{F}+G^a\tilde{G}^a\right).
\label{anomaly}
\end{equation}
with the number of flavour $N_F$. 
Since
\begin{equation}
\frac{N_f}{8\pi^2}G^a\tilde{G}^a=2N_f\partial^\mu K_\mu
\end{equation}
with
\begin{equation}
K_\mu\equiv \frac{1}{16\pi^2}\epsilon_{\mu\nu\rho\sigma}A^{\nu a}\left(\partial^\rho A^{\sigma a}+\frac{1}{3}gf^{abc}A^{\rho b}A^{\sigma c}\right).
\label{Kmu}
\end{equation}

In the limit of $m_u=m_d=m_s=0$, the axial current $j^{5\mu}$ gets conserved again by
replacing $j^{5\mu}$ with
\begin{equation}
\tilde{j}_{5\mu}=j_{5\mu}-2N_f K_\mu.
\label{CVC}
\end{equation}
Thus quark condensate is essential for chiral symmetry breaking.
Nevertheless, anomalous term is crucial to the presence of $\pi^0\rightarrow 2\gamma$.
%It should be reminded that the appearance of $G\tilde{G}$ (and $F\tilde{F}$) 
%does not give rise to parity violation but does only chiral symmetry breaking.
%This is in contrast with that coming from $\theta$ vacuum (see (\ref{effective})). The latt%er breaks CP, inducing strong CP problem.
\begin{figure}[t]
\begin{center}
\includegraphics[scale=0.4]{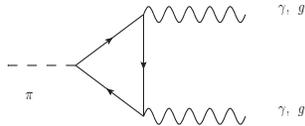}
\caption{
 Chiral anomaly induces $\pi^0\rightarrow \gamma\gamma$ via
$if_0\pi F\tilde{F}$ interaction.
}
\label{pi2gamma}
\end{center}
\end{figure}
The neutral axial vector current gives the modified PCAC relation,

\begin{equation}
\partial^\mu j_{5\mu}^0=f_\pi m_\pi^2\pi^0+\frac{\alpha}{4\pi}F\tilde{F}.
\label{aPCAC}
\end{equation}
We will come back to this problem in the next subsection. U(1) problem concerning QCD condensate is discussed in
Appendix {\bf H}.
\subsection{Peculiar Property of Diamagnetic Atom}
Diamagnetic atom has no unpaired electron, and the main contribution
of atomic EDM comes from the misalignment between charge and the EDM distribution of
nucleus. Thus hadronic part of the atomic EDM manifests itself through the Schiff moment \cite{Schiff}\cite{Sen'kov}.
 
\begin{equation}
H_{atom}=H_{electron}+H_{nucleus}+\sum_{i=1}^Z(e\Phi({\bf r}_i)-e{\bf r}_i\cdot {\bf E})-{\bf d}_{nucleus}\cdot {\bf E},
\end{equation}
where ${\bf r}_i$ are the i'th electron coordinates and ${\bf d}_{nucleus}$ is the nuclear EDM. ${\bf E}$ means the external ${\bf E}$. It should be remarked for diamagnetic atoms that the Stark term due to electron' intrinsic spin term (the first term of \bref{PTV}) is replaced by that of nucleon. $\Phi({\bf r})$ is the nuclear electrostatic potential given by
\begin{equation}
\Phi({\bf r})=\int\frac{\rho({\bf x})d^3x}{|{\bf r}-{\bf x}|},
\end{equation}
where $\rho({\bf x})$ is the charge density of nucleus.

Here it is important to notice \cite{Spevak}
\begin{equation}
-i\left[\sum_{i=1}^Z {\bf p}_i,H_{atom}\right]=-e\sum_{i=1}^Z\nabla_i\Phi({\bf r}_i)+Ze{\bf E},
\label{screen}
\end{equation}
where ${\bf p}_i$ are the momentum of atomic electrons, and the first term is the average electric field induced by
atomic electrons.
The expectation value of this commutator in the energy eigenstate vanishes and we may add
\begin{equation}
V=\la{\bf d}_{nucleus}\ra\cdot {\bf E}-\frac{1}{Z}\sum_{i=1}^Z\la{\bf d}_{nucleus}\ra\cdot \nabla_i\Phi({\bf r}_i)
\end{equation}
to $H_{atom}$ as far as we consider the expectation value.
This implies we may change 
\begin{equation}
-{\bf d}_{nucleus}\cdot {\bf E}\rightarrow -({\bf d}_{nucleus}-\langle {\bf d}_{nucleus}\rangle)\cdot {\bf E}.
\end{equation}
So the expectation value is zero. This is another statement of the Schiff theorem. From the first term of (\ref{screen}), we should consider the interaction of atomic electrons with the nucleus,
\begin{equation}
\Phi({\bf r}_i)-\frac{1}{Z}\left<{\bf d}_{nucleus}\right>\cdot\nabla_i\Phi({\bf r}_i)
\end{equation}
as the screened electrostatic potential.
Therefore, the atomic EDM reads
\begin{equation}
{\bf d}_{atom}=\sum_n\frac{\left<0|e\sum_i^Z{\bf r}_i|n\right>\left<n|e\sum_i^Z\left(\Phi({\bf r}_i)-\frac{1}{Ze}\left<{\bf d}_{nucleus}\right>\cdot\nabla_i\Phi({\bf r}_i)\right)|0\right>}{E_0-E_n}+h.c.
\label{dia}
\end{equation}
Using the charge distributions
\begin{eqnarray}
\int \rho({\bf x})d^3x&=&Z|e|,~~\int{\bf x}\rho({\bf x})d^3x=\left<{\bf d}_{nucleus}\right>,\nonumber\\
\int x^2\rho({\bf x})d^3x&=&Z|e|\left<x^2\right>_{ch},~~\int(x_kx_{k'}-\frac{1}{3}\delta_{kk'}x^2)\rho({\bf x})d^3x=Z|e|\left<Q_{kk'}\right>~~\mbox{etc}
\end{eqnarray}
\begin{equation}
\left<0_N\left|e\Phi({\bf r})-\frac{1}{Z}\left<{\bf d}_{nucleus}\right>\cdot \nabla\Phi({\bf r})\right|0_N\right>=-\frac{Ze^2}{|{\bf r}|}+4\pi e{\bf S}\cdot\nabla \delta ({\bf r})+...
\end{equation}
Here $...$ indicates electric octupole and higher pole contributions, and ${\bf S}$ is the famous Schiff moment \cite{Flambaum86},
(The detailed derivation is given in Appendix {\bf B})
%\begin{equation}
%S_k=S_k^{ch}+S_k^{nucl}.
%\end{equation}
%Here
\begin{equation}
{\bf S}^{ch}=\frac{e}{10}\sum_{p=1}^{Z}\left(r_p^2-\frac{5}{3}\langle r^2\rangle_{ch}\right){\bf r}_p.
\label{chargeS}
\end{equation}
%{\bf S}^{nucl}&=& \frac{1}{6}\sum_{j=1}^{A}{\bf d}_j(r_j^2-\langle r^2\ra_{ch})+\frac{1}{5}\su%m_{j=
%1}^A\left[ {\bf r}_j\langle {\bf r}_j\cdot {\bf d}_j\rangle -\frac{r_j^2}{3}{\bf d}_j\right]+..
%\end{eqnarray}
The $\left<Q_{kk'}\right>$ vanishes for ${}^{199}$Hg, ${}^{129}$Xe, ${}^{225}$Ra. 

There is another Schiff moment ${\bf S}^{nucl}$ due to the misalignment between the charge distribution and the EDM distribution of nucleus, whose derivation is given in Appendix {\bf I}.

Corresponding to these situations, we should consider (\ref{chargeS}) more generally
\begin{equation}
{\bf S}=\frac{1}{10}\sum_N^A\sum_ie_i\left(({\bf r}_N+{\boldsymbol \rho}_i)^2-\frac{5}{3}\langle r^2\rangle_{ch}\right)({\bf r}_N+{\boldsymbol \rho}_i).
\label{chargeS2}
\end{equation}
Here ${\bf r}_N$ is a $N$'th nucleon position and ${\boldsymbol \rho}_i$ is the position of the ith charge inside the $N$'th nucleon, and
\begin{equation}
\sum_ie_i=e_N,~~\sum_ie_i{\boldsymbol \rho}_i={\bf d}_N.
\end{equation}
Retaining the terms up to linear in ${\boldsymbol \rho}$, we have
\begin{equation}
{\bf S}={\bf S}^{ch} +{\bf S}^{nucl}
\label{totalS},
\end{equation}
where ${\bf S}^{ch}$ is given in \bref{chargeS} and
\be
{\bf S}^{nucl}=\frac{1}{6}\sum_N^A{\bf d}_N(r_N^2-\langle r^2\rangle_{ch})+\frac{1}{5}\sum_N^A\left({\bf r}_N({\bf r}_N\cdot{\bf d}_N)-\frac{1}{3}{\bf d}_Nr_N^2\right).
\ee

Usually ${\bf S}^{nucl}$ is considered to be small compared with $S^{ch}$.
The mean value of ${\bf S}^{ch}$ is nonzero only in the presence of P- and T=odd nucleon-nucleon interactions.

For the arguments of hadronic EDM, we must express hadronic CP violating interactions in terms of those of \bref{theta} and \bref{cEDM}. This will be discussed in the last part of this subsection (see \bref{cEDM2}). They are described as
\be
{\cal L}_{\pi NN}\equiv g_{\pi NN}^{(0)}\overline{N} \tau^a  N \pi^a
   +g_{\pi NN}^{(1)}\overline{N}  N \pi^0
   +g_{\pi NN}^{(2)}(\overline{N}\tau^a  N \pi^a-3 \overline{N}\tau^3N\pi^0).
\label{strongCP}
\ee
Here $g_{\pi NN}^{(i)}~(i=0,1,2)$ are CP odd coupling constants, whereas we denote the CP even strong $\pi NN$ coupling constant as $G_{\pi NN}(=13.5)$. 
The Schiff moment due to this coupling is calculated as follows.
(\ref{strongCP}) gives rise to both $S^{ch}$ and $S^{nucl}$.
\begin{figure}[t]
\begin{center}
\includegraphics[scale=0.4]{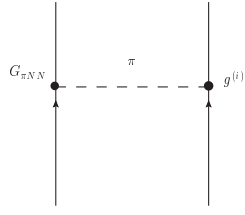}
\caption{
 One $g^{(i)}$ coupling induces effective CP-odd NN interaction, which give rise to $S^{ch}$.
}
\label{gpipi}
\end{center}
\end{figure}
P- and T-odd NN potential has the form via Fig.\ref{gpipi}.
Using, for instance
\begin{equation}
\int \frac{d^3q}{(2\pi)^3}i\boldsymbol{\sigma}_a\cdot{\bf q}\frac{e^{i{\bf q}\cdot{\bf r}}}{q^2-m_\pi^2}=\boldsymbol{\sigma}_a\cdot \nabla\frac{e^{-m_\pi r}}{4\pi r}
\end{equation}
etc., its effective potential is given by
\begin{eqnarray}
&&W({\bf r}_a-{\bf r}_b)=\frac{G_{\pi NN}m_\pi^2}{8\pi m_N}\big\{\big[g^{(0)}(\boldsymbol{\tau}_a\cdot \boldsymbol{\tau}_b)-\frac{g^{(1)}}{2}(\tau_a^z+\tau_b^z)+g^{(2)}(3\tau_a^z\tau_b^z-\boldsymbol{\tau}_a\cdot\boldsymbol{\tau}_b)\big](\boldsymbol{\sigma}_a-\boldsymbol{\sigma}_b)\nonumber\\
&&-\frac{g^{(1)}}{2}(\tau_a^z-\tau_b^z)(\boldsymbol{\sigma}_a+\boldsymbol{\sigma}_b)\big\}\cdot({\bf r}_a-{\bf r}_b)\frac{exp(-m_\pi|{\bf 
r}_a-{\bf r}_b|)}{m_\pi|{\bf r}_a-{\bf r}_b|^2}\big[1+\frac{1}{m_\pi|{\bf r}_a-{\bf r}_b|}\big].
\label{pseudoint}
\end{eqnarray}
Here we have suppressed the subscript $\pi NN$ in $g^{(i)}$.
The EDM of $j$-th nucleon ${\bf d}_j$ is generated via a diagram of Fig.\ref{nuclEDM} and is given by
\begin{figure}[t]
\begin{center}
\includegraphics[scale=0.4]{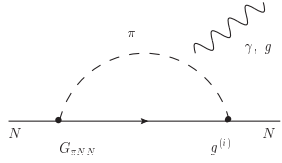}
\caption{
 One $g^{(i)}$ coupling induces ${\bf d}_N$, which gives rise to $S^{nucl}$.
}
\label{nuclEDM}
\end{center}
\end{figure}
and
\begin{equation}
{\bf d}_j=\frac{eG_{\pi NN}}{4\pi^2 m_N}\mbox{ln}\frac{m_N}{m_\pi}(g^{(0)}-g^{(2)})\boldsymbol{\sigma}_j\tau_j^z.
\label{piNd}
\end{equation}
Given T and P-odd perturbation, let us calculate the Schiff moment using diagramatic technique \cite{Nataraj} in
\begin{equation}
H=H_0+H_{res}.
\end{equation}
Here
\begin{equation}
H_0=T+V_{00}+V_{11}
\end{equation}
is unperturbative one-particle Hamiltonian and exactly solvable and
\begin{equation}
H_{res}=W+V_{22}+V_{13}+V_{31}+V_{04}+V_{40}.
\end{equation}
$W$ is the pseudoscalar interaction (\ref{pseudoint}) and $V$ the Skyrme interaction \cite{Skyrme}. Subscripts $(ij)$ refer to the final and initial numbers of quasiparticles.

Let us assume that in the 0'th order approximation, the state is $\Phi_a=|\alpha\rangle$, and define $Q$ by
\begin{equation}
Q\equiv \sum_{\beta\neq \alpha}|\beta\rangle \langle\beta|.
\end{equation}
Then perturbed wave function is given by
\begin{equation}
\Psi_a=\left(1+\frac{Q}{\epsilon_a-H_0}H_{res}+\frac{Q}{\epsilon_a-H_0}H_{res}\frac{Q}{\epsilon_a-H_0}H_{res}+...\right)\Phi_a.
\end{equation}
This is the Brillouin-Wigner expansion and $\epsilon_a$ is the single quasiparticle energy of the valence nucleon.

So in the first order perturbation of $S^z$, we obtain
\begin{equation}
\langle \Psi_a|S^z|\Psi_a\rangle=N^{-1}\langle \Phi_a|\big[1+H_{res}\left(\frac{Q}{\epsilon_a-H_0}\right)+...\big]S^z\big[1+\left(\frac{Q}{\epsilon_a-H_0}\right)H_{res}+...\big]|\Phi_a\rangle.
\end{equation}

\begin{figure}[t]
\begin{center}
\includegraphics[scale=0.8]{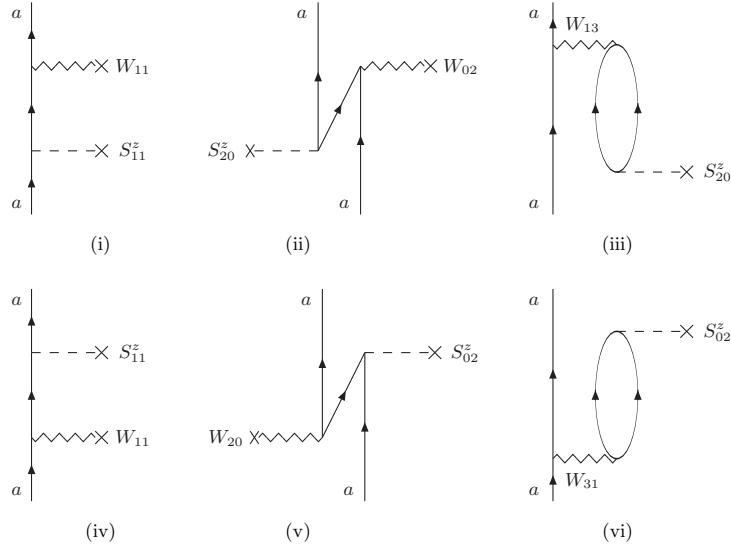}
\caption{
 First-order quasiparticle diagrams contributing to the Schiff moment \protect\cite{Jesus}
. The broken line
represents the action of the Schiff operator, the zig-zag line represents the P- and
T-odd interaction. The looped line in higher order diagram represents a generic Skyrme interaction.}
\label{fig:Goldstone}
\end{center}
\end{figure}

The first-order (in $H_{res}$) quasiparticle (Goldstone) diagram is given in Fig.\ref{fig:Goldstone} \cite{Jesus}. 

%\begin{figure}[t]
%\begin{center}
%\includegraphics[scale=0.8]{AB.eps}
%\caption{Examples of diagrams contributing to the collective
%response to the Schiff operator \cite{Jesus}.  The broken line
%represents the action of the Schiff operator (as in Fig.\
%\ref{fig:Goldstone}), the zig-zag line represents the P-- and
%T--violating interaction (also as in Fig.\ \ref{fig:Goldstone}), and
%the looped line represents a generic Skyrme interaction.}
%\label{fig:QRPA}
%\end{center}
%\end{figure}

%\begin{figure}[t]
%\includegraphics{AB2.eps}
%\caption{QRPA diagrams contributing to the Schiff moment.  The 
%filled bubble represents an infinite sum of quasiparticle bubbles,
%including all the forward and backward amplitudes.  The two B
%diagrams have partners (not shown) in by \cite {Jesus} and others which $W$ acts below $S^z$, cited from \cite{Jesus}}
%\label{fig:AB2}
%\end{figure}
Here the Goldstone diagram implies that 
\begin{eqnarray}
\parbox{35mm}{\includegraphics[scale=0.5]{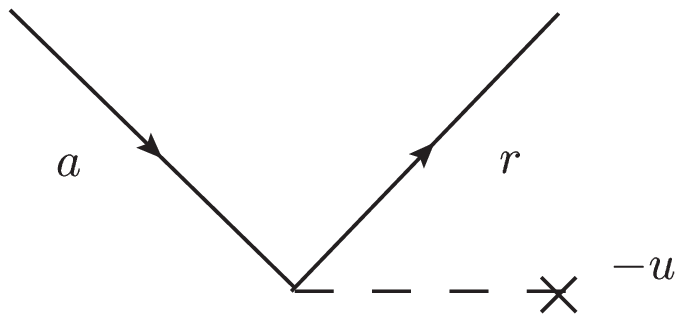}}
= \frac{|\alpha_a^r\rangle\langle r|-u|a\rangle}{\epsilon_a-\epsilon_r}.
\end{eqnarray}
Higer order quasiparticle calculations need some elaborate code and should be referred to \cite{Ban}, and
 we simply list the final results
\begin{equation}
\langle \Psi_a|S^z|\Psi_a\rangle\equiv S=(a_0+b)G_{\pi NN}g^{(0)}+a_1G_{\pi NN}g^{(1)}+(a_2-b)G_{\pi NN}g^{(2)},
\label{a_i}
\end{equation}
where the coefficients $a_i$ specify $S^{ch}$ and b does $S^{nucl}$ defined by (\ref{totalS}). The numerical results of $a_i$ and $b$ for ${}^{199}$Hg are given in Table \ref{ai}.

\begin{table}[t]
\caption{Calculated coefficients $a_i$ and $b$ for ${}^{199}$Hg. The units are e fm${}^3$. The last two references include the Skyrme interaction SkO$^{\prime}$. Five results of Ban et.al.
%\cite{Ban} 
are due to Hartee-Fock and Hartree-Fock-Bogoliubov approximations. SLy4, SIII etc. indicate several Skyrme interactions.}
%\begin{ruledtabular}
{\begin{tabular}{lcccc}
\hline 
\hline
                               &$a_0$& $a_1$ & $a_2$ & b\\
\hline
%Flambaum et al 1986 \cite{Flambaum86}        & $0.087$ & $0.087$ & $0.174$ & --%\\
Dmitriev-Sen'kov 2003 \cite{Sen'kov2}                   & $-0.0004$ & $-0.055$ & $0.009$& -- \\

%Diagram A only                 & $0.018$ & $0.074$ & $0.0$ \\
%\hline
de Jesus-Engels (averaged) \cite{Jesus}          & $0.007$ & $0.071$ & $0.018$ & --\\
Ban et al \cite{Ban} \\
SLy4(HF)                         &-0.013 & 0.006  & 0.022& -0.003 \\
SIII(HF)                         &-0.012&-0.005 & 0.016 & -0.004\\
SV(HF)                           &-0.009 & 0.0001 & 0.016 & -0.002\\
SLy(HFB)                         &-0.013 & 0.006 & 0.024 & -0.007\\
SkM${}^*$(HFB)                       &-0.041 & 0.027 & 0.069 & -0.013\\
\hline
\hline
\end{tabular}
%\end{ruledtabular}
%\caption{Calculated coefficients $a_i$ and $b$ for ${}^{199}$Hg. The units are e fm%${}^3$. The last two references include the Skyrme interaction SkO$^{\prime}$. %Five results of 
%\cite{Ban} 
%are due to Hartee-Fock and Hartree-Fock-Bogoliubov approximations. SLy4, SIII e%t al. indicate several Skyrme interactions.} 
%\label{tb:result}
\label{ai}}
\end{table}
In this Table, the first two papers considered that EDM of nucleons ${\bf d}_N$ is independent of ${\cal L}_{\pi NN}$, whereas Ban considered ${\bf d}_N$ is related as \bref{piNd}.
%In the former approach, nucleon polarization effects must be added to $a_i$ contributions %\cite{Yoshinaga}. 
The second and third papers incorporated collective modes based on the different approximation methods for nuclear structures but the results are still divergent.
 Thus it is a difficult task to precisely estimate the EDM of diamagnetic atom and to extract nucleon or quark EDMs due to lack of precise theory of nuclear structure.  Bearing this point in mind, let us consider some cases.
For ${}^{199}$Hg, numerical calculation is \cite{D-F-G-K}
\begin{equation}
d({}^{199}\mbox{Hg})=-2.8\times 10^{-17}\left(\frac{S}{e~fm^3}\right) \mbox{e cm}.
\end{equation}
In the case of ${\bf S}={\bf S}^{nucl}$, the value of the Schiff moment of $d({}^{199}Hg)$ can be presented as a sum of proton and neutron EDMs \cite{D-S}
\begin{equation}
S=s_pd_p+s_nd_n
\label{D-S}
\end{equation}
with $s_p=0.20\pm 0.02~fm^2$ and $s_n=1.895\pm 0.035~fm^2$.

Combining the experimental value \cite{Romalis} (see more up-to-date data in \cite{Hg})
\begin{equation}
d({}^{199}\mbox{Hg})<2.1\times 10^{-28} \mbox{e cm}
\end{equation}
with (\ref{D-S}), we obtain
\begin{equation}
|d_p|<3.8\times 10^{-24}~\mbox{e cm}~~|d_n|<4.0\times 10^{-25}~\mbox{e cm}.
\label{Tl}
\end{equation}

For ${}^{129}$Xe case,
numerical calculation is \cite{Dzuba}
\begin{equation}
d({}^{129}Xe)=0.38\times 10^{-17}\left(\frac{S}{e~fm^3}\right) e~cm
\end{equation}
The measurement is \cite{Vold}
\begin{equation}
d({}^{129}Xe)=(-0.3\pm 1.1)\times 10^{-26} e~cm.
\label{Xe}
\end{equation}
From (\ref{Xe}) value, \cite{Dzuba2} obtained
\begin{equation}
|d_p|\leq 4\times 10^{-21} \mbox{e cm}~~|d_n|\leq 1\times 10^{-21} \mbox{e cm}.
\label{dp}
\end{equation}
Lastly we comment on the deformed nucleus like Ra and Rn.
When atomic weight A is in 150<A<190 qand A>220, nucleus becomes deformed and has the rotation enegy levels, which enhances the Schiff moment by factor $10^2-10^3$.
For instance $a_0=5.06,~a_1=10.4,~a_2=-10.1$ for $S({}^{225}Ra)$ \cite{Bender}. 
The classification of non-spherical nucleus is similar to that for a diatomic molecule consisting of like atoms (See Chaper {\bf VI}). However, the energy levels of vivration and rotation are not so hierarchical as the molecule case.

General arguments for the CP violating four-fermion coupling are given in Appendices {\bf C} and {\bf D}.

\subsubsection{cEDM and parity odd nuclear interaction}
\begin{figure}[t]
\begin{center}
\includegraphics[scale=0.4]{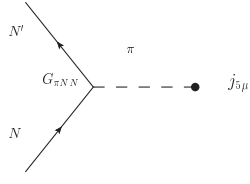}
\caption{Nambu's interpretation of pion dominance.}
\label{fig:pinn}
\end{center}
\end{figure}

In this subsection we give a very short review of chiral symmetry and its breaking in strong interactions since it has many problems.
%Basic ingredients are first
%K$\ddot{\mbox{a}}$llen-Lehman representation 
%\begin{equation}
%\langle \pi^\alpha N|O(0)|N'\rangle=i\int d^4x\frac{e^{-iqx}}{\sqrt{2\omega}}(-\Box +m_\pi^2)\langle N%|T\left(\pi^\alpha (x)O(0)\right)%|N'\rangle
%\label{Kollen}
%\end{equation}
%and its $q_\mu \rightarrow 0$ limit,
%\begin{equation}
%\langle \pi^\alpha N|O(0)|N'\rangle =im_\pi^2\int d^4 x\frac{e^{-iqx}}{\sqrt{2\omega}}\langle N|T\left%(\pi^\alpha (x)O(0)\right)|N'\rangle
%\label{Kollen2}
%\end{equation}
Let us start with the conserved axial-vector current (CAC) hypothesis \cite{Sakurai},
\begin{equation}
\partial_\mu j_5^{\mu\alpha} (x)=0.
\label{CAC1}
\end{equation}
Of course CAC requires $m_\pi=f_\pi=0$ and is not realistic. However, it makes clear to understand how to break chiral symmetry. 
(\ref{CAC1}) leads us to
\begin{equation}
\langle N'|j_{5\mu}^\alpha |N\rangle=\sqrt{\frac{m_N^2}{EE'}}F_A(t)\left[i\overline{u}'\gamma_\mu\gamma_5\frac{\tau^\alpha}{2}u+2m_N\frac{q_\mu}{q^2}\overline{u}'\gamma_5\frac{\tau^\alpha}{2}u\right],
\label{CAC2}
\end{equation}
where $F_A(0)=-g_A/g_V$.
Nambu aserted \cite{Nambu} that $1/q^2$ in the second term of (\ref{CAC2}) should be interpreted as
\begin{equation}
\frac{1}{q^2}=\lim_{m_\pi\rightarrow 0}\frac{1}{q^2-m_\pi^2}.
\label{Nambu2}
\end{equation}
This corresponds to the diagram (Fig. (\ref{fig:pinn}) ), which can be written as
\begin{equation}
G_{\pi NN}\overline{u}'\gamma_5\tau^\alpha u\frac{f_\pi q^\mu}{q^2-m_\pi^2}.
\label{pole1}
\end{equation}
Comparing (\ref{pole1}) with (\ref{CAC2}) where (\ref{Nambu2}) is inserted, we get
\begin{equation}
f_\pi G_{\pi NN}=\frac{g_A}{g_V}m_N.
\end{equation}
This is the Goldberger-Treiman's relation \cite{Goldberger}.
Here use has been made of
\begin{equation}
\langle 0|j_{5\mu}^\alpha (0)|\pi^\alpha\rangle=i\frac{f_\pi}{\sqrt{2\omega}}p_\mu.
\end{equation}
\begin{eqnarray}
\langle N'|\partial_\mu j^{\mu\alpha}_5|N\rangle&=&-i\sqrt{\frac{m_N^2}{EE'}}F_A(t)\left[-2m_N\overline{u}'\gamma_5\frac{\tau^\alpha}{2}u-2m_N\frac{q^2}{q^2-m_\pi^2}\overline{u}'\gamma_5\frac{\tau^\alpha}{2}u\right]\nonumber\\
&=&\sqrt{\frac{m_N^2}{EE'}}F_A(t)\left(\frac{m_Nm_\pi^2}{q^2-m_\pi^2}\right)i\overline{u}'\gamma_5\tau^\alpha u.
\label{PCAC1}
\end{eqnarray}
Substituting the equation of motion of $\pi$,
\begin{equation}
(\Box+m_\pi^2)\pi^\alpha=j^\alpha,
\end{equation}
into (\ref{PCAC1}), we obtain 
\begin{equation}
\langle N'|\pi^\alpha |N\rangle=-\frac{\langle N'|j^{\alpha}|N\rangle}{q^2-m_\pi^2}\approx -i\sqrt{\frac{m_N^2}{EE'}}\frac{G_{\pi NN}\overline{u}'\gamma_5\tau^\alpha u}{q^2-m_\pi^2}.
\end{equation}
Assuming the matrix elements vary little between $t=0$ and $t=m_\pi^2$, we obtain
\begin{eqnarray}
\langle N'|\partial_\mu j^{\mu\alpha}_5(0)|N\rangle&\approx &\frac{g_A}{g_V}\frac{m_N}{G_{\pi NN}}m_\pi^2\langle N'|\pi^\alpha |N\rangle\nonumber\\
&=& f_\pi m_\pi^2\langle N' |\pi^\alpha |N\rangle .
\end{eqnarray}
Thus we obtain the PCAC condition
\begin{equation}
\partial_\mu j^{\mu\alpha}_5=f_\pi m_\pi^2\pi^\alpha
\label{PCAC2}
\end{equation}
and (\ref{aPCAC}) with chiral anomaly.

%second PCAC relation
%\begin{equation}
%\partial_\mu j_{5\mu}^\alpha (x)=f_\pi m_\pi^2\pi^\alpha.
%\label{PCAC}
%\end{equation}
%Before discussing P-odd and CP-odd interaction, we discuss how $f_\pi$ and stro%ng coupling $G_{\pi NN}$ is related.
%There is a strong theorem that in parity conserving vectorlike theories such as% QCD, parity conservation is not spontaneous%ly broken \cite{Vefa}.
%\footnote{There is an argument that supersymmetric SM has a P-violation in QCD %through chiral quark-squark-gluino couplings% with non-degenerate masses \cite{O%nogi}}
%We set $O=1$ and
%\begin{equation}
%J^\alpha=\frac{ig}{2\sqrt{2}}\langle N|\gamma^\alpha(G_v+G_a\gamma^5)|N'\rangle
%\end{equation}
%Then
%\begin{equation}
%ik_\alpha J^\alpha=i(m_N-m_{N'})G_v\overline{N}N'+i(m_N+m_{N'})G_a(\overline{N}%\gamma^5N)
%\end{equation}
%On the otherhand, 
%\begin{equation}
%\langle N|k_\mu j_{5\mu}^a|N'\rangle=f_\pi m_\pi^2\langle N|\pi^a|N'\rangle=\frac{f_\pi m_\pi^2%}{k^2+m_\pi^2}G_{\pi NN}(k^2)\overline{N}\gamma_5N'
%\end{equation}
%Then we obtain
%\begin{equation}
%(m_N+m_{N'})G_a=G_{\pi NN}\sqrt{2}f_\pi
%\end{equation}
%This is the Goldberger-Treiman's relation \cite{Goldberger}.

Next, we proceed to discuss the path from the presence of the strong EDM of dimension 5 (cEDM and $\theta$ term) to the effective CP-odd $g^{(i)}$.

Let us write hadronic CP-violating operators like (\ref{theta}) and (\ref{cEDM}) etc. as $O$ and consider the following two points correlation function \footnote{The general arguments on the operator expansion of T product of two currents are given in \cite{Shifman}.}.
\begin{equation}
M^\mu\equiv \int d^4xe^{-iqx}\langle N|T\left(j_{5}^{\mu\alpha}(x)O(0)\right)|N'\rangle .
\end{equation}
From the definition of time ordered product, the right-handed side is rewritten
\begin{eqnarray}
q_\mu M_\mu&=&-i\int d^4 xe^{-iqx}\{\langle N|T\left(\partial_\mu j_5^{\mu\alpha}(x)O(0)\right)|N'\rangle \nonumber\\
&&-i\delta(x_0)\langle N|\left[j_5^{0\alpha},~O(0)\right]|N'\rangle \}.
\end{eqnarray}
Using (\ref{PCAC2}) and LSZ reduction formula \cite{LSZ} in $q\rightarrow 0$ limit, we obtain 
\begin{equation}
q_\mu M^\mu=-f_\pi\langle \pi^\alpha N|O(0)|N'\rangle-i\langle N|[Q_5^\alpha(0),O(0)]|N'\rangle
\end{equation}
or equivalently
\begin{eqnarray}
&&\lim_{q\rightarrow 0}\sqrt{2\omega}\langle \pi^\alpha N|O(0)|N'\rangle=-\frac{i}{f_\pi}\langle N|[Q_5^\alpha(0),O(0)]|N'\rangle\nonumber\\
&&-\lim_{q\rightarrow 0}\frac{q_\mu}{f_\pi}\int d^4 xe^{iqx}\langle N|T(j_5^{\mu\alpha}(x)O(0))|N'\rangle .
\label{cEDM4}
\end{eqnarray}

Substituting the concrete form of $O(0)$ as (\ref{cEDM}) into the above equation and using
\be
[Q_5^\alpha(0),q(0)]=it^\alpha\gamma_5q(0)
\ee
with the generators of group of flavour $t^\alpha$, we obtain \cite{Baluni}, \cite{Crewther}, \cite{Pospelov1}
\begin{eqnarray}
&&\mbox{RHS of (\ref{cEDM4})}=\frac{1}{f_\pi}\langle N|\tilde{d}_u(g_s\overline{u}G\sigma u-m_0^2\overline{u}u)-\tilde{d}_d(g_s\overline{d}G\sigma d-m_0^2\overline{d}d)|N\rangle\nonumber\\
&&+\frac{m_*}{f_\pi}\left[2\overline{\theta}+m_0^2\left(\frac{\tilde{d}_u}{m_u}+\frac{\tilde{d}_d}{m_d}+\frac{\tilde{d}_s}{m_s}\right)\right]\langle N|\overline{u}u-\overline{d}d|N\rangle
\label{cEDM5}
\end{eqnarray}
with
\begin{equation}
m_*=\frac{m_um_dm_s}{m_um_d+m_um_s+m_dm_s}\approx \frac{m_um_d}{m_u+m_d},~~m_0^2=\frac{\langle 0|g_s\overline{q}G\sigma q|0\rangle}{\langle \overline{q}q\rangle}.
\end{equation}
$m_0^2$ is estimated as
\be
m_0^2\approx 0.8 \mbox{GeV}^2
\ee
from QCD sum rule \cite{Belyaev}.  Here quantum corrections are also included. If we use the Peccei-Quinn mechanism \cite{PQ}, the second term of (\ref{cEDM5}) vanishes in the following way \cite{Bigi}.

\begin{equation}
L=\frac{\alpha_s}{8\pi}aG\tilde{G},
\end{equation}
where $a$ is axion field and $G\tilde{G}=\frac{1}{2}\epsilon^{\mu\nu\alpha\beta}G_{\mu\nu}^bG_{\alpha\beta}^b$.
When there exists cEDM, axion potential becomes
\begin{equation}
V_{eff}(a)=K_1 a+\frac{1}{2}Ka^2.
\label{Veffect}
\end{equation}
Here
\begin{eqnarray}
K&\equiv& -i\lim_{k\rightarrow 0}\int d^4x e^{ikx}\langle 0|T\left(\frac{\alpha_s}{8\pi}G\tilde{G}(x) \frac{\alpha_s}{8\pi}G\tilde{G}(0)\right)|0\rangle ,\\
K_1 &\equiv& -i\lim_{k\rightarrow 0}\int d^4x e^{ikx}\langle 0|T\left(\frac{\alpha_s}{8\pi}G\tilde{G}(x)\sum i\frac{\tilde{d}_q}{2}g_s\overline{q}G\sigma\gamma_5q(0)\right)|0\rangle .
\end{eqnarray}
Eq.\bref{Veffect} is obtained by considering
\be
L_{CPV}=\frac{\alpha_s}{8\pi}aG\tilde{G}+\frac{i}{2} \tilde{d}_qg_s\overline{q}G\sigma\gamma_5q
\ee
and performing path integral.

Next let us consider $\partial_\nu j^{\nu \beta}_5(0)$ as an $O(0)$
and use
\be
\partial_\nu j^{\nu \beta}_5=i\overline{q}\{t^\beta,M\}\gamma_5q,
\ee
where $M$ is the mass matrix of quarks. Then we easily obtain \cite{Eides}
\begin{eqnarray}
K&=&m_*\langle 0|\overline{q}q|0\rangle ,\nonumber\\
K_1&=&\frac{1}{2}m_*\sum_{q=u,d,s}\frac{\tilde{d}_q}{m_q}\langle 0|g_s\overline{q}G\sigma q|0\rangle .
\end{eqnarray}
So
\begin{equation}
\frac{\partial V_{eff}}{\partial a}=K_1+Ka=0
\end{equation}
leads to vanishing of the second term of (\ref{cEDM5}).

Finally we obtain \cite{Pospelov1}
\begin{eqnarray}
\label{cEDM2}
g_{\pi NN}^{(0)}&=&\frac{\tilde{d}_u+\tilde{d}_d}{f_\pi}\langle N|H_u-H_d|N\rangle \nonumber\\
g_{\pi NN}^{(1)}&=&\frac{\tilde{d}_u-\tilde{d}_d}{f_\pi}\langle N|H_u+H_d|N\rangle .
\end{eqnarray}
Here
\begin{equation}
H_q=g_s\overline{q}G\sigma q-m_0^2\overline{q}q.
\end{equation}
Thus $g_{\pi NN}^{(2)}$ vanishes if we impose Peccei-Quinn symmetry. In the absence of Peccei-Quinn symmetry, there appears 
$g_{\pi NN}^{(2)}$ \cite{Hisano2}.
The contribution of mixing of $\eta$ with $\pi$ was also considered in \cite{Hisano2}.
\section{The EDMs of Molecules}
In this section we consider heteronuclear diatomic molecule which has permanent dipole moment.
Polar paramagnetic molecules have stronger enhancement factors than paramagnetic 
atoms.
Diamagnetic molecules are more sensitive to nuclear P,T violation than diamagnetic atoms.

There are many advantageous points in molecule \cite{Bingel}.
Firstly, the polar molecule is polarized by a modest laboratory electric field $E_{lab}$ but has a vast internal electric field $E_{int}$. This implies the hugely enhanced stark effect and small fake magnetic field of $\frac{{\bf v}\times{\bf E}_{lab}}{c}$ in comparison with atomic case.
Secondly, there appears very small energy interval between nuclear rotation levels of opposite parity, which is roughly $10^{-3}$Ry as will be discussed. Also g-factor can be very small etc.
 
In general the electric dipole moment ${\bf D}$ is defined by
\begin{equation}
{\bf D}=e\left(\sum_i Z_i{\bf R}_i-\sum_j {\bf r}_j\right),
\end{equation}
where ${\bf R}_i$ and ${\bf r}_j$ are coordinates of nucleons and electrons composing molecule.
For hetero-nuclear molecule
\begin{equation}
{\bf D}_a=<a|{\bf D}|a>\neq 0
\end{equation}
and ${\bf D}$ has the permanent electric dipole moment.

So the behaviours of heteronuclear molecule and homonuclear molecule are different.

First we begin with diatomic molecule with total spin ${\bf S}=0$ case.

We first give the general rules of diatomic molecule.

In diatomic molecule, the field has axial symmetry along the two nuclei. Hence the projection of ${\bf L}$ (total orbital angular momentum of electrons) on this axis which is denoted by $\Lambda$ is conserved.

The motion of molecule is composed of the orbital motion of electrons, vibrations and rotation of nucleus,
They interact complicatedly but their interactions are approximated as independent motions as the 0'th approximation (Born-Oppenheimer approximation)
\begin{equation}
\psi=\psi_{e}\psi_v\psi_r
\end{equation}
and total energy is, therefore,
\begin{equation}
E=E_e+E_v+E_r.
\end{equation}
They are electronic energy ($\approx Ry$), and vibration and rotation energies of nucleus, respectively.
Let us consider the nuclear motions of diatomic molecule.
First we begin with the case of total spin (mainly of electrons) ${\bf S}=0$.
$E_v$ is considered as a harmonic oscillator and its energy are estimated from
\begin{equation}
M\omega_N^2a^2\approx \frac{\hbar^2}{ma^2}\equiv E_e.
\end{equation}
Here $a$ is the distance between nucleus. $M$ and $m$ are the reduced mass of nuclei and electron mass, respectively.  Therefore
\begin{equation}
E_v=\hbar\omega_N\approx \left(\frac{m}{M}\right)^{1/2}E_e.
\end{equation}
Whereas the rotation energy is
\begin{equation}
E_r=B({\bf K}-{\bf L})^2, 
\label{Erot}
\end{equation}
where ${\bf K}$ and ${\bf L}$ are total angular momentum of molecule and
electron angular momentum, respectively, and
\begin{equation}
B(r)=\frac{\hbar^2}{2Mr^2}=\frac{\hbar^2}{2I}.
\end{equation}

\begin{figure}[t]
\begin{center}
\includegraphics[scale=0.3]{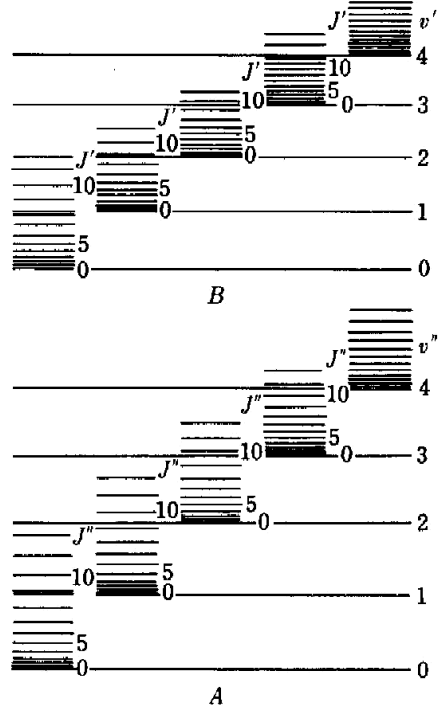}
\caption{The vibration ($v',v''$) and rotation ($J',J''$) terms of electron states A and B \protect\cite{Bingel}.
}
\label{fig:molecule}
\end{center}
\end{figure}

${\bf K}$ and the axial component of {\bf L} are conserved.
\begin{equation}
E_r=\frac{\hbar^2}{2I}l(l+1)\approx \frac{m}{M}JRy.
\end{equation}
This is much less than the atomic energy interval in general.
We show in Fig. \ref{fig:molecule} the typical spectroscopies.

$\Lambda$-doubling:\\
In (\ref{Erot}) ${\bf K}^2$ and ${\bf L}^2$ terms depend on $|\Lambda|$, and ${\bf K}\cdot {\bf L}\propto B^{2\Lambda}\approx (m/M)^{2\Lambda}$ \cite{L-L}.

Hence off diagonal parts are neglected, and $+\Lambda$ and $-\Lambda$ states are degenerate.

When we take the relativistic effect into consideration we have another
coupling of Spin of electrons ${\bf S}$ (usually nucleon spin can be neglected) with orbital angular momentum of electrons ${\bf L}_e$
and of nucleons ${\bf L}_N$. The most important energy shift is $A(r){\bf L}_e\cdot \bf S$.

Selection rule in the electric dipole transition:
\begin{eqnarray}
&&|J'-J|\leq 1\leq J+J'\\
&&+\rightarrow -, ~~-\rightarrow +
\end{eqnarray}
To obtain molecular spectra, we must consider the interactions among the above three terms; electron term $E_e$, nuclear vibration $E_v$, and rotation $E_r$.

The interaction between $E_e$ and $E_r$ is especially important.

First we consider $E_e$ for static nucleus. Unlike atomic case, conserved are not total orbital angular moment ${\bf L}$ and spin ${\bf S}$ of electrons but their projection to molecular axis
\begin{equation}
{\bf J}_z\equiv \Lambda+\Sigma =\Omega
\end{equation}
which takes the values over $\Lambda+\Sigma,\Lambda+\Sigma-1,....,\Lambda -\Sigma$. These states are described as ${}^{2\Sigma+1}\Lambda_\Omega$. For example, 
${}^2\Pi_{1/2},~{}^2\Pi_{3/2}$ for the states with $\Lambda=1,\Sigma=1/2$.

%For diatomic molecule for a and b cases,
%\begin{eqnarray}
%&&S'-S=0,\\
%&&\Lambda'-\Lambda=0,~\pm 1\\
%&&\Sigma^+\rightarrow \Sigma^+,~~\Sigma^+\rightarrow \Sigma^+ ~~\mbox{for}~\Lam%bda=0.
%\end{eqnarray}

\begin{table}[pt]
\caption{Angular momenta of diatomic molecule and their projections to molecular axis. ${\bf n}$ is the unit vector of molecular axis.}
{\begin{tabular}{c|c|c}
\hline \hline
 angular momentum  & notation & z component  \\ \hline
electron spin & ${\bf S}$ & $\Sigma$ \\ 
electron orbital angular momentum & ${\bf L}$ & $\Lambda$ \\
nucleus orbital angular momentum & ${\bf N}$ & 0 \\
total angular momentum (without spin) & ${\bf K}=\Lambda {\bf n}+{\bf N}$ & $\Lambda$ \\
total angular momentum (with spin) &${\bf J}={\bf \Omega}+{\bf N}$ & $\Omega$ \\
\hline \hline
\end{tabular}}
\end{table}

For atomic fine structure is given by (\ref{Lande}), whereas the fine structure for diatomic molecule 
\begin{equation}
\Delta E =\frac{dA\Lambda\Sigma}{d\Sigma}=A\Lambda=const.
\end{equation}
We call this relativistic interactions spin-axis interaction, which is composed of spin-orbit, spin-spin interactions, as well as the spin and orbital interactions with the rotation of molecule.
Corresponding to the relative magnitudes of these interactions, we can classify molecule energy levels as follows \cite{L-L} \cite{Hund}.
We define the magnitudes of interactions as follows.

$LA$: the coupling of orbital angular momentum with the axis (the electric interaction between the two atoms in the molecule).

$SA$: the coupling of spin angular momentum with the axis.

$\Delta E_r$: the intervals between rotational levels.

%If, $LA\gg SA,\Delta E_r$ they a
If the distances between terms with different $\Lambda$ are larger than both the intervals of fine structures ($2S+1$) and
rotational structures, they are further classified into 
%\begin{description}
\begin{itemize}
\item{Hund's case a}   ~~$LA\gg SA\gg \Delta E_r$

In this case, $\Lambda,~\Sigma,~\Omega$ are well defined and electron state is expressed as ${}^{2\Sigma+1}\Lambda_{\Omega}$.
For the $a\rightarrow a$ transition,
\begin{equation}
\Sigma'-\Sigma=0,~~\Omega'-\Omega=0,~\pm 1
\end{equation}
\begin{equation}
\nu_L=\Delta T\gg \nu_s=\frac{\partial T}{\partial \Sigma}=A\Sigma\gg \nu_J=B_v(2J+1)
\end{equation}
\begin{equation}
U_J(r)=U(r)+A(r)\Omega+B(r)\overline{({\bf J}-{\bf L}-{\bf S})^2}.
\end{equation}
Here the third term is a perturbation.
%${\bf L}\cdot{\bf A}\approx {\bf S}\cdot{\bf A}>{\bf L}\cdot{\bf S}$

${\bf L_e}$ and ${\bf S}$ precess around the internuclear axis $z$ implying that $\Lambda$ and $\Sigma $ are conserved quantum numbers. 
The total energy is described as
\begin{equation}
E=E_e+A_e\Omega+\hbar \omega(v+1/2)+B_e\{J(J+1)-2\Omega^2\}.
\end{equation}
%The total energy is described as
%\begin{equation}
%E^T(R_0)=E_s(R_0)+A\Lambda\Sigma
%\end{equation}
%and $\Sigma$ runs from $-S$ to $S$ and for a given $\Lambda$ the energy are spl%it into (2S+1) multiplets.

In this review we are interested in the transition between parity odd rotation levels of the same electron term.

\item{Hund's case b}~~$LA\gg \Delta E_r\gg SA$

$\Sigma$ is not defined.
Here the effect of the rotation of the molecule predominates over the multiple splitting and total angular momentum ${\bf J}$ and the sum ${\bf K}={\bf L}+{\bf N}$ are conserved. In this case, ${\bf S}$ is almost free from molecule (the vector ${\bf K}+{\bf S}$) precessing around {\bf J}, and $\Sigma$ is not conserved)

\begin{equation}
\nu_K\gg \nu_s,
\end{equation}
%${\bf L}\cdot{\bf A}> {\bf S}\cdot{\bf A}>{\bf L}\cdot{\bf S}$
\begin{equation}
|K'-K|\leq 1\leq K+K',
\end{equation}
\begin{equation}
H_0=H_e+B{\bf K}^2
\end{equation}
with ${\bf K}=\Lambda \hat{z}+{\bf N}$ and ${\bf J}={\bf K}+{\bf S}$.

\begin{equation}
U_K(r)=U(r)+B(r)K(K+1)+A(r)\Lambda\frac{(J-S)(J+S+1)}{2K(K+1)}
\end{equation}
with
\begin{equation}
K=\Lambda,\Lambda+1,....
\end{equation}
Here the third term is perturbation.
The total energy is
\be
E=U_e+\hbar\omega_e\left(\frac{1}{2}\right)+B_eK(K+1)+A_e\Lambda\frac{(J-S)(J+S+1)}{2K(K+1)}.
\ee

\item{Hund's case c}~~$SA\gg LA\gg \Delta E_r$

Only $\Omega$ is well defined.
This is the case where the coupling of ${\bf L}$ with the axis is small compared with the spin-orbit coupling.
%${\bf L}\cdot{\bf A}\approx {\bf S}\cdot{\bf A}<{\bf L}\cdot{\bf S}$

\begin{equation}
H_0=H_e+H_{ls}+B{\bf J}^2.
\end{equation}
\item{Hund's case d}~~$\Delta E_r\gg LA\gg SA$

This is the case where the coupling of ${\bf L}$ with the axis is small in comparison with the intervals in $E_r$.
%${\bf L}\cdot{\bf A}>{\bf L}\cdot{\bf S},~{\bf S}\cdot{\bf K}>{\bf S}\cdot{\bf A}$
\begin{equation}
H_0=H_e+B{\bf N}^2-B(J^+l^-+J^-l^+).
\end{equation}
\item{Hund's case e}~~$SA \gg \Delta E_r\gg LA$.
\end{itemize}
\subsection{Paramagnetic Molecule}
As we will show, there are a variety of paramagnetic atoms, for instnce, HgF, YbF, TlO whose
electrons configurations are ${}^{70}$Yb=[Xe]$4f^{14}6s^2$, ${}^{80}$Hg=[Xe]$4f^{14}5d^{10}6s^2$, ${}^{81}$Tl=[Xe]$4f^{14}5d^{10}6s^26p^1$. The selection rules of transitions are
\begin{eqnarray}
&&S'-S=0,\\
&&\Lambda'-\Lambda=0,~\pm 1\\
\label{Lambda}
&&\Sigma^+\rightarrow \Sigma^+,~~\Sigma^+\rightarrow \Sigma^+ ~~\mbox{for}~\Lambda=0.
\end{eqnarray}

For BiS molecule \cite{S-F}, electron configuration of Bi is [Xe]$4f^{14}5d^{10}6s^26p^3$ and
Bi${}^{++}$ has one unpaired electron. The electric field of S leads to a mixing of parity odd states:

\begin{equation}
|\Omega\rangle=|1/2\rangle=a|s_{1/2},\Omega\rangle+b|p_{1/2},\Omega\rangle+c|p_{3/2},\Omega\rangle.
\end{equation}
Here $\Omega=J_z=1/2$.
So
\begin{equation}
\langle\frac{1}{2}|\frac{1}{2}\rangle=-2ab\frac{4(Z\alpha)^2Z|e|d_e}{\gamma(4\gamma^2-1)a_B^2(N_sN_{p1/2})^{3/2}}.
\end{equation}

For total J, angular momentum of nuclei takes two values, $N_1=J+1/2$ and $N_2=J+1/2$, so the characteristic energy splitting between P-odd states is
\begin{equation} 
\Delta E_r=BN_2(N_2+1)-BN_1(N_1+1)=2B(J+1/2),
\end{equation}
which is, for BiS, four to six orders of magnitude smaller than the case of heavy atom. 
\begin{equation}
{\bf d}=\frac{2\omega d_M\langle\omega|H_d|\omega\rangle}{\Delta E_{J,\eta}}\frac{{\bf J}}{J(J+1)}
\end{equation}
and
\begin{equation}
K=\frac{d}{d_e}=3\times 10^7\frac{(-1)^{J+1/2}\eta}{(J+1/2)(J+1)}.
\end{equation}
%\end{description}

The effective electric field on the valence electron is proportional to $KE_{ind}$ for polar paramagnetic molecule.
So it is very advantageous to measure molecular EDM.

Recently the most stringent upper limit of $d_e$ was reported by using YbF \cite{YbF}. 
Yb belongs to the rare-earth elements and its electron configuration is
[Xe]+$4f^{14}6s^2$ and Yb${}^+$ ion constitutes paramagnetic molecule.
f electrons' interaction with the axis of molecule is weakened by the deep position of the f electrons and classified as Hund's c class. Their result is
\begin{equation}
d_e=(-2.4\pm 5.7_{stat}\pm 1.5_{sys})\times 10^{-28} \mbox{e cm}
\end{equation}
which sets the upper limit
\begin{equation}
|d_e|< 10.5\times 10^{-28}~\mbox{e cm}.
\end{equation}

The other experiment using ThO \cite{Vutha} is also very interesting
since a modest laboratory electric field $E_{lab}\leq 100$ V/cm fully polarizes a ThO whose internal electric field $E_{mol}$ is 100GV/cm. (The electron configuration of Thorium is Th=[Rn]$6d^27s^2$.)
This gives another advantage for polar molecules. Furthermore, the triplet state ${}^3\Delta_1$ of ThO gives the merit of g-factor cancellation (see Eq. (\ref{gfactor2})).  Also for the other molecules we can expect g-factor cancellation, where g-factors are defined by the ratio of spin rotation energy $H_E$ and $\mu_BB_z$. Here $H_E$ is given by
\begin{equation}
H_E=\beta{\bf J}^2+\Delta{\bf S}'\cdot {\bf J}-D{\bf n}\cdot {\bf E}.
\label{gfactor}
\end{equation}
Here $\Delta$ is the $\Omega$-doubling constant and ${\bf S}'$ is the effective spin and ${\bf S}'={\bf S}$ for Hund's case b. The detail of meanings of right-hand side is given in \cite{Kozlov}.
The expectation value of $H_E$ crosses zero at a specific value of electric field and the molecule becomes insensitive to magnetic field at that point.

One of the problems for molecular EDM is the difficulty of laser cooling compared with atomic case. This may be solved by first cooling composite atoms and next combining them by the Feshbach resonance \cite{Wille} and optical trap methods \cite{Takahashi} \cite{Jochim}.
The theoretical problem is to calculate matrix elements in Dirac-Coulomb + higher order approximation (see Appendix {\bf \bref{Das}}).

\subsection{Diamagnetic Molecule}
We will consider TlF as an example of diamagnetic molecule.
In searching for molecular EDM, we have two tasks.
One is to derive $d_{mol}$ from CP-odd elementary N-N and/or N-e interactions.
Another is to deduce $d_p$ and $d_n$ from the observed $d_{mol}$.

The electron configuration of Tl atom is [Xe]$4f^{14}5d^{10}6s^26p^1$
and Tl${}^+$ has a closed electron shell.
Tl${}^+$ forms also incomplete shell $6s6p$ instead of $6s^2$ \cite{Khriplovich},
\begin{equation}
|\Omega\rangle=|6s,\Omega\rangle+\beta \left(-\frac{2\Omega}{\sqrt{3}}|6p_{1/2},\Omega\rangle+\sqrt{\frac{2}{3}}|6p_{3/2},\Omega\rangle\right)
\label{spmixing}
\end{equation}
with
\begin{equation}
\beta=\frac{2}{\sqrt{3}}\frac{Ry}{E_{6s}-E_{6p}}\frac{a^2 r (6s,6p)}{r_1^2}=0.27.
\end{equation}
Here $\Omega=\pm 1/2$, and $r (6s,6p)$ is the radial integral defined by (\ref{radialintegral}) whose value is $2.3$. 
Using (\ref{weakCP2}) \cite{Khriplovich}
\begin{equation}
\langle s_{1/2}|H|p_{1/2}\rangle=\frac{Gm^2\alpha^2}{\sqrt{2}\pi}\frac{Z^2R}{(N_sN_p)^{3/2}}Ry\{\gamma(Zk_{1p}+Nk_{1n})-4j\frac{2+\gamma}{3}\langle k_{2p}\sum_p\boldsymbol{\sigma}_p+k_{2n}\sum_n\boldsymbol{\sigma}_n\rangle\},
\end{equation}
where $R$ is the relativistic factor
\begin{equation}
R=\frac{4}{\Gamma^2(2\gamma+1)}\left(\frac{a_B}{2Zr_0A^{1/3}}\right)^{2-2\gamma}
\end{equation}
with $r_0=1.2$fm.
%\begin{eqnarray}
%&&\mbox{case a: spi-axis interaction}~ >~\mbox{ energy differences between the %rotation levels}\nonumber\\
%&&\mbox{case b: spi-axis interaction}~ < ~\mbox{energy differences between the %rotation levels}
%\end{eqnarray}
%If the interactions of orbital angular momentum with axis is comprable with the% other effects, they are further classified into
%\begin{eqnarray}
%&&\mbox{case c: spin-orbit interaction}~ > ~\mbox{coupling of orbital angular m%omentum with the axis} \nonumber\\
%&&\mbox{case d: the intervals of rotation energy}~ > ~\mbox{coupling of orbital% angular momentum with the axis} \nonumber\\
%\end{eqnarray}
As for the nuclear matrix element,
\begin{equation}
\langle k_{2p}\sum_p\boldsymbol{\sigma}_p+k_{2n}\sum_n\boldsymbol{\sigma}_n\rangle\approx k_{2p}\frac{{\bf I}}{I}
\end{equation}
since a valence proton in Tl atom is $s_{1/2}$.
Reference \cite{K-L} goes further to get 
\begin{equation}
S(Tl)=-\frac{2\pi}{3}(r_q^2-r_d^2)d_p,
\end{equation}
where $r_q,~r_d$ are defined by (\ref{r^2}).
From the experimental limit \cite{Cho},
\begin{equation}
S_{exp}(Tl)\leq 0.8\times 10^{-8} e fm^3,
\end{equation}
we obtain 
\begin{equation}
d_p\leq 10^{-22} \mbox{e cm}.
\end{equation}
See (\ref{Tl}) and (\ref{dp}) for diamagnetic atom. 
The numerical calculations were estimated along the following line of thoughts \cite{Hinds}:
assuming Born-Oppenheimer approximation, total wave function of TlF is described as 
\begin{equation}
\Psi=\psi_n({\bf r_n})\psi_e({\bf r}_i)\psi_R({\bf r}_N,{\bf I}).
\end{equation}
Here $\psi_n({\bf r_n})$ describes the motion of Tl nucleus,$\psi_e({\bf r}_i)$ does F nucleus and electrons with respect to the center of mass of Tl nucleus, and $\psi_R({\bf r}_N,{\bf I})$ the spin and motion of Tl nucleus.

Let us integrate over $\psi_e$ and take (\ref{expansion2}) into account. We obtain
\begin{equation}
\langle H_{edm}\rangle =D\langle \psi_R\psi_n|{\bf a}\cdot\sum_n\left(\frac{q_n}{Z}\boldsymbol{\sigma}-\frac{{\bf d}_n}{D}\right)\left(\int_{r_i=0}^{r_n}\psi_e^*\sum_i\frac{Y_{10}^i(\Theta,\Phi)}{r_i^2}\psi_ed^3r_i\right)|\psi_R\psi_n\rangle,
\label{TlF1}
\end{equation}
where
\begin{equation}
D\boldsymbol{\sigma}=\langle\psi_n|\sum_n{\bf d}_n|\psi_n\rangle .
\end{equation}
For the present approximation (\ref{potential}), $\psi_e$ is given by
\begin{equation}
\psi_e=\Pi _i\psi_i(r_i)=\Pi_i\sum_la_l^ir_i^lY_{lm}^i(\theta_i,\phi_i)
\end{equation}
and so on.

Anyhow, analytical studies are restricted and we may need more elaborate numerical calculations as was done in the case of atomic structures or much more than that case.  However, it is certain that unknown but very fruitful frontiers are expanding in front of this field. Many experiments are preparing or ongoing. In these situations, theoretical developments are strongly awaited.

%Asahi group aims to $d({}^{129}Xe)\leq 10^{-29} e~cm$

%GUT gives the
%constraints between them \cite{Falk}.

\section{Summaries and Discussion}

We have explored the EDMs of quarks, leptons, hadrons, atoms, and molecules.
First we studied the SM predictions on the EDM and showed that those are far from the present experimental upper limits.
We have direct signals of new physics beyond the SM from neutrino oscillations and muon g-2, and many indirect ones like baryon asymmetry, DM etc.

Among them, the CP deficiency in baryon asymmetry $\eta\equiv \frac{n_B}{n_\gamma}\approx 10^{-10}$ is especially important for searching for new physics.
Namely, we can not reproduce $\eta$ via CKM CP violating phase only even if we incorporate CP violation due to a $\theta$ term and other radiative corrections in the SM framework like $G\tilde {G}$ etc.

In order to estimate the deviation of phenomena from the SM, we have tried to estimate them first in the SM precisely, including the effects of the above mentioned extra terms. 

Next we have explored many theories beyond the SM by focusing on the EDM of elementary particles.

The MSSM and two Higgs doublet model, for instance, give rather large values of EDMs. However, those values are mainly due to the ambiguities of the theories themselves. It is important to see whether such values are checked to be consistent with the other phenomena or not. We think those points are still very insufficient. More predictive models like the renormalizable minimal SO(10) GUT discussed in Section {\bf 4.3} often give more stringent values which are still several orders smaller than the present upper limits.

However, the situation is not so pessimistic. Some hope comes from unprecedented collaborations with atomic and molecular physics and elementary particles mainly via brilliant developments of laser physics. Most impressive is the new upper limit of the electron EDM from polarized molecule YbF.
As for paramagnetic atoms, theoretical calculations have been developed and seems to be convergent.
Whereas, for diamagnetic atoms there are still large discrepancies (Table I).
Lattice QCD is very promising but it is not convergent in the limit of $m_\pi=0$ (Fig.5).
However, it is certain that these situations have been improved rapidly. The large parts of such progress have been and will be done by the collaboration of a wide field of physics and chemistry. The mutual close relationships among particle, atomic, and molecular physics require the wide range of studies over these regions.

We hope that this review gives some help for these difficult tasks.

This review is restricted in theoretical part and we have not discussed many excellent ideas on the experimental side.
The latter is very attractive but is beyond the scope of this review simply due to the author's ability. We only briefly explain the mechanism and a list of experiments though it is not exhaustive. 

The procedure for the EDM measurement is as follows.
First a static external electric field {\bf E} is applied parallel to magnetic field {\bf B}.
The energy splitting is measured as a spin precession frequency $\nu_{\uparrow\uparrow}$. Next we change {\bf E} unti-parallel to {\bf B} whose precession frequency is denoted by $\nu_{\uparrow\downarrow}$. Namely,
\begin{eqnarray}
h\nu_{\uparrow\uparrow}&=&2\boldsymbol{\mu}\cdot{\bf B}+2{\bf d}\cdot{\bf E}\nonumber\\h\nu_{\uparrow\downarrow}&=&2\boldsymbol{\mu}\cdot{\bf B}-2{\bf d}\cdot{\bf E}
\label{exp}
\end{eqnarray}
and
\begin{equation}
h\Delta \nu=4{\bf d}\cdot{\bf E}.
\label{exp2}
\end{equation}
Its sensitivity is given by
\begin{equation}
\delta d=\frac{h}{2e}\frac{1}{K}\frac{1}{E}\frac{1}{\sqrt{N\tau T}}.
\end{equation}
Here N: number of sample, $\tau$: coherence time, and T: measuring time.
K is an enhance factor for paramagnetic atoms and molecules given in (\ref{enhance}) and $K\propto Z^3\alpha^2$. E is a magnitude of an internal electric field.
So experiments try to get larger values of $K,~ E, ~N, ~\tau,~ T$.
\begin{table}[pt]
\caption{A list of ongoing and planned experiments searching for EDM. Superscript * indicates estimated sensitivity.}
\scalebox{0.8}[0.7]{
{\begin{tabular}{c|c|c}
\hline \hline
Species  & Group name & Features \\ 
\hline
muon \\
\hline
$d_\mu$ & FNAL &  $10^{-21}$ e cm$^*$ (2015) \\
 & J-PARC &  $10^{-24}$ e cm$^*$  (2015) with spin frozen technique \\
 & PSI &3-4 orders below current limit${}^*$ (spin frozen technique)\\
\hline
neutron  (all $10^{-28}$ e cm$^*$) \\
\hline
%$de_n$ & Doyle et al. (f) & \\
$d_n$ & ILL (Grenoble) & $|d_n|<2.9\times 10^{-26}$ ecm (90\% C.L.) \cite{Baker:2006ts}\\
 &ILL  & Squids as magnetometer \cite{Grinten}\\
 & PSI (Zurich) SNS(Oak Ridge) & Hg co-magnetometer and Cs gradiometer \cite{Bodek}\\
  & SNS & ${}^3$He co-magnetometer and SQUIDS \cite{Beck}\\
  & KEK-RCNP (Japan) & ${}^{129}$Xe co-magnetometer  \cite{Masuda} \\
\hline
deuteron \\
\hline
$d_D$ & KVI/BNL/COSY &  $10^{-29}$ e cm${}^*$ \\
\hline
paramagnetic Atom \\
\hline
Cs &  Amherst College  & d(Cs)=$(-1.8\pm 6.7\pm 1.8)\times 10^{-24}$ ecm  \cite{Cs}\\
 & LBNL & highly improved magnetic shielding \cite{Cs2}\\
Tl& Berkeley & $d_e<1.6\times 10^{-27} (90\% C.L.)$ e cm \cite{Regan:2002ta}\\
Fr& CYRIC(Tohoku Univ.)  & K(Fr)=895 EDM measurement starts on 2014 \cite{Tohoku}.\\
Ra & KVI (Groningen)  & magneto-optical trap \cite{Jungmann}\\
\hline
diamagnetic atom \\
\hline
${}^{199}$Hg & Seattle & $d({}^{199}Hg)<3.1\times 10^{-29}$ (95\% C.L.)\cite{Hg}\\
Ra & Argonne/KVI & large enhancement $d(Ra)/d(Hg)\approx 10^{2-3}$ \\
Xe &@nEDM Collaboration & polarized liquid Xe droplets\\
   & Tokyo Institute of Technology & artificial feedback mechanism \cite{Asahi}\\   & Princeton & liquid cell\\
      & Univ. Mainz & $d({}^{129}Xe) \approx 10^{-30}$ e cm$^*$\\
Rn/Xe & Michgan & $d({}^{129}Xe)=(+0.7\pm 3.3)\times 10^{-27}$ ecm \cite{Chupp} \\
Rn & Rn EDM Collaboration & octupole enhancement of 400-600 \\
\hline
paramagnetic molecule \\
\hline
YbF & Hinds (Imperial College) et al.& the most sringent limit of $d_e$\cite{YbF}\\
ThO & ACME Collaboration &   g-factor cancellation at ${}^3\Delta_1$ \cite{Vutha}\\
PbO & DeMille (Yale) et al. &  g-factor cancellation at metastable ${}^3\Sigma^+_1$ \cite{PbO}\\
PbF &  Shafer-Ray (Oklahoma) et al. &  g-factor cancellation at ${}^2\Pi_{1/2}$ \cite{PbF}\\
HfF${}^+$ & Cornell group & trapped molecular ions in rotating electric field \cite{HfF} \\
HgF/BaF &   & same electron configuration as YbF \\
RaF & KVI  &  high $W_a$ parameter \cite{RaF} \\
FrSr & Aoki (Tokyo) et al.   & ultra cold molecule/3D optical lattice \cite{FrSr}\\
\hline
diamagnetic molecule \\
\hline
TlF & Hinds (Yale) et al. & the measured $\Delta \nu=(1.4\pm 2.4)\times 10^{-4}$ Hz\cite{TlF}\\
YbHg & Takahashi (Kyoto) et al.& ultra cold molecule/3D optical lattice \\
%\hline
%solid\\
%\hline
%GGG, Gd_2Ga_5O_{12}, Gd_3Fe_5O_{12}, PbTiO_3, Gd_3Ga_5O_{12}, solid He, liquid %Xe & & \\
\hline \hline
\end{tabular}}
}
\end{table}
We only list up ongoing and planned experiments (see Table 3). 
We have still more species, solids like
$GGG,~ Gd_2Ga_5O_{12},~ Gd_3Fe_5O_{12}, ~PbTiO_3,~ Gd_3Ga_5O_{12}$, solid He, liquid Xe (see Table nEDM Collaboration). Please refer to the corresponding sections for the terminologies in the experimental features. A few comments are in order. + signature at PbO molecule implies the parity under the mirror reflction (reflection under arbitrary plane including molecule axis) (see Eq.(\ref{Lambda})).
As for g-factor cancellation in molecular EDMs, ThO and the others' cancellation mechanisms are different: the former is due to
\begin{equation}
\mu=(\Lambda+g\Sigma)\mu_B\approx 0 ~\mbox{for}~\Lambda=2
\label{gfactor2}
\end{equation}
and the latters are due to Eq.(\ref{gfactor}).

This table is far from being exhaustive but reflects some prospect from a theoretical physicist.

For more detail see, for instance, ECT* Workshop: Violations of Discrete Symmetries in Atoms and Nuclei. Nov 15- 19, 2010 \cite{ECT}. 

We have not discussed about the EDMs of charged particles and ions. These are also very important and we have added short explanation on the storage ring in Appendix {\bf \bref{storage}}.

\begin{figure}[t]
\begin{center}
\includegraphics[scale=0.8]{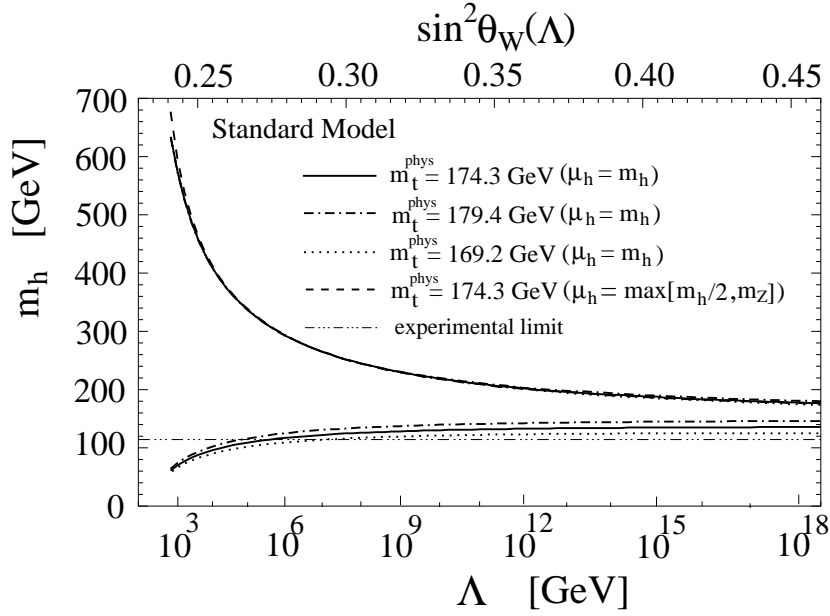}
\caption{The catastrophic RGE behaviours of quartic coupling constant at the energy scale $\Lambda$. The upper curve is the Landau pole where the coupling blows up at $\Lambda$. and the lower curve is the point that the coupling becomes negative (vacuum is unstabilized). This diagram is cited from \protect\cite{Tobe}.
}
\label{fig:bound}
\end{center}
\end{figure}

Finally we will give some comments on the recent results by the Cern Large Hadronic Collider (LHC).
On July 2012 the LHC groups announced the discovery of Higgs-like particle around $126$ GeV \cite{ATLAS} \cite{CMS}.
This is not only the discovery of the last unknown particle in the SM but also gives the serious impact to the new physics beyond the SM, especially SUSY GUT.
In this review we have estimated EDM value in the framework of SUSY GUT and given large EDM value relative to that of the SM. 
We briefly explain why $126$ GeV Higgs mass is serious for GUT and especially SUSY GUT. 
First we explain the reason why $126$ GeV is so important.
The RGE of the Higgs quartic coupling is given by \cite{Machacek}
\be
16\pi^2 \frac{d\lambda}{d\mbox{ln}\mu}=12\lambda^2-\left(\frac{9}{5}g_1^2+9g_2^2\right)\lambda+\frac{9}{4}\left(\frac{3}{25}g_1^4+\frac{2}{5}g_1^2g_2^2+g_2^4\right)+12Y_t^2\lambda-12Y_t^4.
\ee
Here the Higgs self coupling is
\begin{equation}
V=\frac{1}{2}\lambda|\Phi^\dagger\Phi|^2
\end{equation}
with 
\begin{equation}
m_h^2=\lambda v^2,~~\mbox{and}~~<\Phi>=\frac{v}{\sqrt{2}}.
\end{equation}
If the SM is assumed to be valid to the energy scale $\Lambda_{cut}=M_{Pl}=2.44\times 10^{18}$ GeV, it goes from the perturbative bound and vacuum stability bound  \cite{Hambye} \cite{Tobe} that
\be
129\mbox{GeV}\geq m_H\geq 170\mbox{GeV}
\ee
as depicted in Fig. \ref{fig:bound}.
That is, if $m_H$ was below $129$ GeV, the coupling becomes negative and vacuum unstabilized. Whereas if $m_H$ was above $170$ GeV, the coupling blows up.
$m_H=126$ GeV (very near to $129$ DeV) implies that there exists some phase transition around the GUT scale.

As for the SUSY implication, for tree level Higgs mass satisfies the inequality
\be
m_h < M_Z|\mbox{cos}(2\beta)|,
\ee
with $M_Z=91.2$GeV, which is obviouly wrong. One loop correction to $m_h$ in CMSSM is \cite{Harber2}
\be
m_h^2\approx M_Z^2\mbox{cos}^22\beta+\frac{3}{4\pi^2}\frac{m_t^4}{v^2}\left[\mbox{ln}\frac{M_S^2}{m_t^2}+\frac{X_t^2}{M_S^2}\left(1-\frac{X_t^2}{12M_S^2}\right)\right],
\label{higgs1}
\ee
where
\be
M_S=\sqrt{m_{\tilde{t}_1}m_{\tilde{t}_2}},~~X_t=A_t-\mu\mbox{cot}\beta,~~v=174\mbox{GeV}
\ee
with the trilinear Higgs-stop coupling constant $A_t$. So $126$ GeV indicates heavy stop masses and/or large $X_t$ (left-right mixing).
Experimental serach of SUSY particles also gives large sfermion masses, larger than 1 TeV \cite{ATLASsusy} \cite{CMSsusy}.
One loop EDM due to the MSSM (Fig.~\ref{fig:EDM_SUSY}) is proportional to $O(M_S^{-2})$ and heavy $M_S$ reduces EDM.
In Fig.~\ref{fig:eEDM_MSSO10} we have assumed $A_0=0$ since $A_0$ appears at two loop correction in the gauge mediation SUSY breaking and is suppressed.
In order to preserve rather small $M_S$ and still give large loop correction in \bref{higgs1}, we may take rather large $A_0$. So we have added the last fourth panel with nonzero $A_0$ in Fig.~\ref{fig:eEDM_MSSO10}. More detailed explanations will be given in the review paper of GUT prepared by us \cite{fukuGUT}.

\subsection*{Acknowledgements}
%%%%%%%%%%%%%%%%%%%%%%%%%%%%%%%%%%%%%%%%%%%%%%%%%%%%%%%%%%%%%%%
The author thanks to the all members of FPUA (Collaboration (Chair N.Sasao) :Fundamental Physics Using Atom) which drives the author to prepare this manuscript. Among them, the author expresses special thanks to K. Asahi, Y.Sakemi, T.Sato, T.Aoki, H.Nataraj, Y.Abe, Y.Takahashi, K.Enomoto, K.Matsuyanagi, M.Kajita, and W.Naylor for their stimulating comments and discussions. Sincere thanks are also due to V.V.Flambaum, E.A. Hinds, A.Ritz, P. Schmidt-Wellenburg, Y.Masuda, J.Pretz, D.P.DeMille, E.D.Commins, E.Shintani and A.J. Silenko.  The author is indebted to K.Asahi for many comments in the last version. He is also grateful to the help of H.Sugiyama at the early stage of this work. 
The work is supported in part by the Grant-in-Aid for Scientific Research from the Ministry of Education, Science and Culture of Japan (No.020540282 and No. 21104004). 

\appendix

\section{SU(6) and Dipole moments}
Both magnetic dipole moment and electric dipole moment are proportional to $eQ\boldsymbol{\sigma}$.

So we can obtain the information of the ratio of $d_p/d_n=\mu_p/\mu_n$ from SU(6) in the light quark (u,d,s) base \cite{Georgi} if CP violation in the EDM does not affect SU(3) symmetry.
They are both represented as
\begin{equation}
\langle 56|35|56\rangle
\end{equation}
Here baryons belong to 56-representation since irreducible representation of $qqq=56$ and we use that dipole moments are the generator of SU(6).
\begin{eqnarray}
|p,1/2\rangle&=&\frac{\sqrt{2}}{6}\left\{ |uud\rangle\left( 2|++-\rangle -|+-+\rangle -|-++\rangle\right)\right.\nonumber\\
&&\left.+|udu\rangle\left(2|+-+\rangle -|-++\rangle -|++-\rangle\right)\right.\\
&&\left. +|duu\rangle\left(2|-++\rangle -|++-\rangle -|+-+\rangle\right)\right\},\nonumber 
\end{eqnarray}
\begin{eqnarray}
Q\sigma_3|p,1/2\rangle &=&\frac{\sqrt{2}}{6}\left\{\frac{2}{3}|uud\rangle\left( 2|++-\rangle -|+-+\rangle +|-++\rangle \right)\right.\nonumber\\
&&\left. +\frac{2}{3}|uud\rangle\left( 2|++-\rangle +|+-+\rangle -|-++\rangle \right)\right.\\
&&\left. -\frac{1}{3}|uud\rangle\left( -2|++-\rangle -|+-+\rangle -|-++\rangle \right)\right.\nonumber\\
&&\left. +\mbox{cyclic permutations}\right\},\nonumber
\end{eqnarray}

\begin{eqnarray}
&& \langle p,1/2|Q\sigma_3|p,1/2\rangle\nonumber\\
&&=3\frac{2}{36}\left(\frac{2}{3}(4+1-1)+\frac{2}{3}(4-1+1)-\frac{1}{3}(-4+1+1)\right)=1.
\end{eqnarray}
The corresponding neutron dipole moments are given by the exchange of $u\leftrightarrow d$, resulting to $\frac{2}{3}\leftrightarrow -\frac{1}{3}$. Therefore,
\begin{eqnarray}
&& \langle n,1/2|Q\sigma_3|n,1/2\rangle\nonumber\\
&&=3\frac{2}{36}\left(-\frac{1}{3}(4+1-1)-\frac{1}{3}(4-1+1)+\frac{2}{3}(-4+1+1)\right)=-\frac{2}{3},
\end{eqnarray}
and
\begin{equation}
\frac{d_p}{d_n}=\frac{\mu_p}{\mu_n}=-\frac{3}{2}.
\end{equation}
The experimental values of MDM of proton and neutron are \cite{PDG}
\begin{equation}
\mu_p=2.792847356\pm 0.000000023,~~\mu_n=-1.91304273\pm 0.00000045
\end{equation}
and the coincidence with SU(6) prediction is good up to quantum corrections.
For the EDM, compare with the result of lattice calculations Fig.\ref{fig:lattice}.

\section{Multipole expansions}
We will study the multipole expansions of electromagnetic potential $A_\mu=(\phi, {\bf A})$ due to the charged system of finite size.
The electric and magnetic fields are defined by
\begin{equation}
\bf{\bf E}=-\frac{1}{c}\frac{\partial {\bf A}}{\partial t}-grad \phi,~~{\bf H}=rot {\bf A}.
\label{EM}
\end{equation}
Let us assume (as in the experimental environment) that the electromagnetic field is static, that is, the field is time independent. In such case, ${\bf E}~({\bf H})$ is determined only by ${\bf A}~(\phi)$.
Let us consider a stational motion of chaged particles where $e_a$ charged particles are located at ${\bf r}_a$ and study how the obserber at ${\bf R}$ feels vector potential $A_\mu$.
\begin{equation}
\phi({\bf R})=\sum_a \frac{e_a({\bf r}_a)}{|{\bf R}-{\bf r}_a|},~~{\bf A}({\bf R})=\sum_a \frac{e_a{\bf v}_a({\bf r}_a)}{|{\bf R}-{\bf r}_a|}.
\label{potential}
\end{equation}
Here we have neglected the retardation effect of fast particles.
If we included it, charge distribution has a retarded time dependence and we should replace the arguments as,
\begin{equation}
t\rightarrow t-\frac{|{\bf R}-{\bf r}_a|}{c},~~|{\bf R}-{\bf r}_a|\rightarrow |{\bf R}-{\bf r}_a|-\frac{{\bf v}\cdot ({\bf R}-{\bf r}_a)}{c}.
\label{potential2}
\end{equation}

If the scale ${\bf R} \gg {\bf r}_a$, (\ref{potential}) is expanded around ${\bf R}$,
\begin{eqnarray}
\phi&=&\frac{\sum_a e_a}{R}-\sum e_a ({\bf r}_a\cdot \nabla)\frac{1}{R}+\sum e_ae_br^i_ar^j_b\partial_i\partial_j\frac{1}{R}+....\nonumber\\
&=& \phi^{(0)}+\phi^{(1)}+\phi^{(2)}+....
\label{expansion1}
\end{eqnarray}
Then $\phi^{(l)}$ is given by 
\begin{equation}
\phi^{(l)}=\frac{1}{R^{l+1}}\sum_{m=-l}^{m=l}\sqrt{\frac{4\pi}{2l+1}}Q_{lm}^{(e)}Y_{lm}^*(\Theta, \Phi),
\label{expansion2}
\end{equation}
where
\begin{equation}
Q_{lm}^{(e)}=\sum_ae_ar_a^l\sqrt{\frac{4\pi}{2l+1}}Y_{lm}(\theta_a,\varphi_a).
\label{expansion3}
\end{equation}
gives electric $2^l$-pole moment. The superscript $(e)$ indicates electric moment distinguishing magnetic counterpart (see (\ref{magnetic})). Its continuous representation is
\begin{equation}
Q_{lm}^{(e)}=\sqrt{\frac{4\pi}{2l+1}}\int d^3r\rho({\bf r})r^lY_{lm}(\frac{{\bf r}}{r}).
\label{expansion3c}
\end{equation}
This comes from
\begin{equation}
\frac{1}{|{\bf R}-{\bf r}|}=\sum_{l=0}^{\infty}\sum_{m=-l}^l\frac{r^l}{R^{l+1}}\frac{4\pi}{2l+1}Y_{lm}^*(\Theta,\Phi)Y_{lm}(\theta,\phi).
\end{equation}
First few normalized spherical harmonics $Y_{lm}$ are
\begin{eqnarray}
Y_{00}&=& 1/\sqrt{4\pi},\nonumber\\
Y_{10}&=&i\sqrt{3/(4\pi)}\mbox{cos}\theta,~~Y_{1\pm 1}=\mp i\sqrt{3/(8\pi)}\mbox{sin}\theta e^{\pm i\phi},\nonumber\\
Y_{20}&=&\sqrt{5/(16\pi)}(1-3\mbox{cos}^2\theta),\\
Y_{2,\pm 1}&=&\pm\sqrt{15/(8\pi)}\mbox{cos}\theta\mbox{sin}\theta e^{\pm i\phi},~~Y_{2,\pm 2}=-\sqrt{15/(32\pi)}\mbox{sin}^2\theta e^{\pm 2i\phi}~~\mbox{etc.}\nonumber
\end{eqnarray}

For instance $Q_{1m}^{(e)}$ constitute electric dipole moment
\begin{equation}
Q_{10}^{(e)}=id_z,~~Q_{1\pm 1}^{(e)}=\mp \frac{i}{\sqrt{2}}(d_x\pm id_y).
\label{expansion4}
\end{equation}

Analogously, vector potential ${\bf A}$ is expanded as
\begin{equation}
A_i({\bf R})=\int \frac{J_i({\bf r})}{|{\bf R}-{\bf r}|}d^3r=A_i^{(0)}+A_i^{(1)}+A_i^{(2)}+...
\label{expansion 3}
\end{equation}
For instance $A_i^{(2)}$ is
\begin{equation}
A_i^{(2)}=\left(\nabla_k\nabla_l\frac{1}{R}\right)T_{ikl},
\label{expansion4}
\end{equation}
where 
\begin{equation}
T_{ikl}=\frac{1}{2}\int d^3rr_kr_lJ_i(r).
\label{expansion5}
\end{equation}
The identity 
\begin{equation}
\int d^3r\nabla_m(r_ir_kr_lJ_m)=0
\end{equation}
leads us to
\begin{equation}
\int d^3r(r_kr_lJ_i+r_ir_lJ_k+r_ir_kJ_l)=0,
\label{expansion6}
\end{equation}
where use has been made
\begin{equation}
\partial_mJ_m=0.
\end{equation}
This identity gives 10 constrains. Since $T_{ikl}$ has $6\times 3$ freedoms, $18-10=8$ physical freedoms remain.
We will show that five of eight freedoms constitute the $M2$ moment and the remaining three the anapole moment.
It goes from subtracting (\ref{expansion6}) from (\ref{expansion5}) that
\begin{equation}
T_{ikl}=-\frac{1}{3}\epsilon_{ikm}\epsilon_{mnr}\int d^3r r_lr_nJ_r=-\frac{1}{3}\epsilon_{ikm}\int d^3r r_lM_m,
\label{expansion7}
\end{equation}
where
\begin{equation}
M_m=\epsilon_{mnr}r_nJ_r.
\end{equation}
Dividing the $T_{ikl}$ of (\ref{expansion7}) into symmetric and antisymmetric parts w.r.t. $l,m$, we obtain
\begin{equation}
\mbox{Symmetric part of}~(\ref{expansion7})=-\frac{1}{6}\epsilon_{ikm}M_{lm}
\end{equation}
with
\begin{equation}
M_{lm}=\int d^3r(r_l\epsilon_{mnr}+r_m\epsilon_{lnr})r_nJ_r.
\end{equation}
This gives magnetic quadrupole moment. Whereas,
\begin{equation}
\mbox{Anti-symmetric part of}~(\ref{expansion7})=\frac{1}{6}\int d^3r \left[\delta_{il}\left(r_k(r_mJ^m)-r^2J_k\right)+\delta_{kl}\left(J_ir^2-r_i(r_mJ^m)\right)\right].
\label{expansion8}
\end{equation}
Here we use the identity obtained from contracting (\ref{expansion6}) w.r.t. $k$ and $l$
\begin{equation}
\int d^3r(r^2J_i+2r_ir_mJ^m)=0
\end{equation}
Then the anti-symmetric part becomes
\begin{equation}
\mbox{Anti-symmetric part of}~(\ref{expansion7})=\frac{1}{4\pi}(\delta_{il}a_k-\delta_{kl}a_i),
\label{anapole1}
\end{equation}
where
\begin{equation}
a_i=-\pi\int d^3r r^2J_i
\end{equation}
is called anapole moment.

General expression for magnetic photon corresponding to electric counterpart (\ref{expansion3c}) is
\begin{equation}
Q_{lm}^{(m)}=\frac{1}{l+1}\sqrt{\frac{4\pi}{2l+1}}\int d^3r [{\bf r}\times {\bf l}]\cdot\nabla(r^lY_{lm})
\label{magnetic}
\end{equation}
and called $2^l$-pole magnetic moment (For relativistic case $l$ is replaced by $j=|{\bf l}+{\bf s}|$).

%For electric field
%\begin{equation}
%{\bf E}({\bf r})=-\nabla \phi=-e\nabla \frac{1}{|{\bf r}-{\bf r}'|}
%\end{equation}
%and
%\begin{eqnarray}
%\nabla_l\phi^{(2)}&=&r'_mr'_n\nabla_l\nabla_m\nabla_n\frac{1}{r}\nonumber\\
%&=&r'_mr'_n\left[\left(\nabla_l\nabla_m\nabla_n-\frac{1}{5}(\delta_{lm}\nabla_n+\delta_{mn}\nabla_l+\delta_{nl}\nabla_m)\Del%ta\right)\frac{1}{r}\right.\\
%&+&\left.\frac{1}{5}(\delta_{lm}\nabla_n+\delta_{mn}\nabla_l
%+\delta_{nl}\nabla_m)\Delta\frac{1}{r}\right]\nonumber
%\end{eqnarray}

%$\spadesuit$ Landau field theory section 45
%\begin{equation}
%\overline{{\bf a}}=\nabla \frac{1}{r}\times \overline{{\bf m}}
%\end{equation}
%Here $\overline{{\bf m}}$ is time averaging of magnetic moment
%\begin{equation}
%\overline{{\bf m}}\equiv \frac{1}{2c}\sum e{\bf r}\times {\bf v}=\frac{e}{2mc}{\bf l}
%\end{equation}
%Therefore
%\begin{equation}
%{\bf H}=\frac{|e|}{2mc}\nabla\times[\nabla\frac{1}{r}\times {\bf l}]
%\end{equation}
%K-(13.49)
%\begin{eqnarray}
%&&H_M=\frac{eM}{4I(2I-1)}\left[I_mI_n+I_nI_M-\frac{2}{3}\delta_{mn}I(I+1)\right]\nonumber\\
%&&\times\frac{1}{2}\alpha_k\nabla_l(\epsilon_{klm}\nabla_n+\epsilon_{kln}\nabla_m)\frac{1}{r}.
%\end{eqnarray}

\section{C,P,T-transformations of Fermi coupling}
We consider the four fermions (current-current) coupling.
Here it is concerned with the transformation property of fermion but not with detailed dynamics, we consider it as $\overline{N}\hat{O}N\overline{L}\hat{O}'L$, where $N,L$ are spinors, and $\hat{O}$ and $\hat{O}'$ are combinations of gamma matrices. The most general forms are
\begin{eqnarray}
&&G_S\overline{N}N\overline{L}L+G_P\overline{N}\gamma_5N\overline{L}\gamma_5L\nonumber\\
&&+G_V\overline{N}\gamma_\mu N\overline{L}\gamma^\mu L+G_A\overline{N}\gamma_\mu\gamma_5N\overline{L}\gamma^\mu\gamma_5L+G_T\overline{N}\sigma_{\mu\nu}N\overline{L}\sigma^{\mu\nu}L\nonumber\\
&&+G_{V'}\overline{N}\gamma_\mu N\overline{L}\gamma^{\mu}\gamma_5L+G_{A'}\overline{N}\gamma_\mu \gamma_5N\overline{L}\gamma^{\mu}L\label{jj4}\\
&&+iG_{S'}\overline{N}N\overline{L}\gamma_5L+iG_{P'}\overline{N}\gamma_5N\overline{L}L+iG_{T'}\epsilon^{\mu\nu\rho\sigma}\overline{N}\sigma_{\mu\nu}N\overline{L}\sigma_{\rho\sigma}L.\nonumber
%\label{jj4}
\end{eqnarray}
The first two lines costitute Lorentz scalars, the third line P-odd and the fourth line P,T-odd terms. Imaginary $i$ in the last line comes from the Hermiticity of action. 
The last term of the fourth line are also expressed as
\begin{equation}
\overline{N}\sigma_{\mu\nu}N\overline{L}\sigma^{\mu\nu}\gamma_5L~\text{or}~\overline{N}\sigma_{\mu\nu}\gamma_5N\overline{L}\sigma^{\mu\nu}L
\end{equation}
since
\begin{equation}
\epsilon^{\mu\nu\rho\sigma}\gamma_\mu=-i\gamma_5\gamma^\nu\gamma^\rho\gamma^\sigma.
\end{equation}
C,P,T conjugations are defined by
\begin{eqnarray}
&&C\psi(t,{\bf r})=\gamma^2\psi^*(t,{\bf r}),\\
&&P\psi(t,{\bf r})=i\gamma^0\psi(t,-{\bf r}),\\
&&T\psi(t,{\bf r})=i\gamma^3\gamma^1\psi^*(-t,{\bf r}).
\end{eqnarray}
The fourth line of (\ref{jj4}) is T-odd since
\begin{equation}
\overline{N}N\rightarrow -\overline{N}N,~~\overline{L}\gamma_5 L\rightarrow \overline{L}\gamma_5 L~~~etc.
\end{equation}
under T-transformation.
\section{CP phases in general L-R model and generation number}
Type I (canonical) seesaw is composed of $N$ left-handed and $N$ right-handed neutrino. Right-handed neutrino has heavy Majorana mass term.
\begin{equation}
L_{Yukawa}=-\overline{\nu_R}M_D\nu_L-\frac{1}{2}\{\overline{(\nu_L)^c}M_L\nu_L+\overline{(\nu_R)^c}M_R\nu_R\}+h.c.
\label{seesaw}
\end{equation}
This is described in terms of mass eigenvectors,
\begin{equation}
L_M=(\overline{N^{(1)}},\overline{N^{(2)}})
 \left(
  \begin{array}{cc}
   U^{(1)}
   & U^{(2)}\\
   V^{(1)*}
   & V^{(2)*}
     \end{array}
 \right)^T
\left(
  \begin{array}{cc}
   M_L 
   & M_D^T\\
   M_D
   & M_R
     \end{array}
 \right)
\left(
  \begin{array}{cc}
   U^{(1)}
   & U^{(2)}\\
   V^{(1)*}
   & V^{(2)*}
     \end{array}
 \right)
\left(
  \begin{array}{c}
N^{(1)}\\
N^{(2)}
\end{array}
\right) +h.c.
\end{equation}
\begin{equation}
\nu_{lL}=\sum_{j=1}^{2N}U_{lj}N_{jL}, ~~\nu_{lR}=\sum_{j=1}^{2N}V_{jl}N_{jR},
\end{equation}
where $l=1,...,N$.

So $N\times 2N$ unitary maqtrices $U$ and $V$ are decomposed into $N\times N$
matrices,
\begin{equation}
U=\left(U^{(1)},U^{(2)}\right)^T,~~V=\left(V^{(1)},V^{(2)}\right)^T.
\end{equation}
In the SM $m_{\nu_l}=0$ and there is no mixing in neutrino sector.

For Dirac neutrino case, $U^{(2)}=V^{(2)}=0$. As we mentioned above, there exist $(N-1)(N-2)/2$ phases in this case.

For Majorana neutrino case, $U^{(2)}=V^{(1)}=0$ when there exist both left-handed (L-type $N_1,...,N_N$) and right- handed neutrino (R-type $N_{N+1},...,N_{2N}$). In this case $2N\times 2N$ unitary

$V^{(2)}=0$ is added  for only left-handed Majorana neutrino case.

\begin{equation}
H_W=\frac{G}{\sqrt{2}}\left[j_{La}^\dagger j_L^a+\lambda j_{Ra}^\dagger j_R^a+\kappa\left(j_{La}^\dagger j_R^a+j_{Ra}^\dagger j_L^a\right)\right],
\end{equation}
where
\begin{eqnarray}
j_{La}&=&\sum_l \overline{l}(x)\gamma_a(1-\gamma_5)\nu_{lL}(x),\label{currentL}\\
j_{Ra}&=&\sum_l \overline{l}(x)\gamma_a(1+\gamma_5)\nu_{lR}(x)
\label{currentR}
\end{eqnarray}
with $l=e,\mu,...$
Mass matrices has rebasing and rephasing symmetries, which does not change physics.

We start with $N$ generation of quarks (Dirac fermions). $N\times N$ unitary matrix has $N^2$ real numbers.  Of these, $2N-1$ is absorved by rephasing of 2N left-handed and right-handed quarks. An orthogonal $N\times N$ orthogonal matrix has $N(N-1)/2$ Euler angles. The remaining $(N-1)(N-2)/2$ is the number of phase parameters.
Kobayashi-Maskawa predicted that there must at least three generations to incorporate CP phase in mass matrix \cite{K-M}.
If we relax this arguments to include Majorana neutrino we can use only rephasing of charged lepton in MNS mixing matrix whose freedom is $N$.
Therefore, the number of phases in MNS matrix is $N^2-N(N-1)/2-N=N(N-1)/2$.

If we furthermore generalize the above arguments to include heavy right-handed neutrino \cite{Takasugi},
\begin{equation}
\nu_{lL}=\sum_{j=1}^{2N}U_{lj}N_{jL}, ~~\nu_{lR}=\sum_{j=1}^{2N}V_{jl}N_{jR},
\end{equation}
where $l=1,...,N$.

So $N\times 2N$ unitary maqtrices $U$ and $V$ are decomposed into $N\times N$
matrices,
\begin{equation}
U=\left(U^{(1)},U^{(2)}\right)^T,~~V=\left(V^{(1)},V^{(2)}\right)^T.
\end{equation}
In the SM $m_{\nu_l}=0$ and there is no mixing in neutrino sector.

For Dirac neutrino case, $U^{(2)}=V^{(2)}=0$. As we mentioned above, there exist $(N-1)(N-2)/2$ phases in this case.

\section{Expansion in power of $1/c$}
Relativistic equation of fermion in an external electromagnetic foeld is
\begin{equation}
\left(\gamma (p-eA)+m\right)\psi=0.
\label{Dirac}
\end{equation}
Let us study the relativistic effect as the deviation from the Schroedinger prescription which is obtained by expanding (\ref{Dirac}) in power of $1/c$.
For that purpose we must exclude $mc^2$ from energy, which implies to replace
$\psi$ to $\psi'$
\begin{equation}
\psi=\psi'e^{-imc^2t/\hbar}
\end{equation}
and 
\begin{equation}
\left(i\hbar\frac{\partial}{\partial t}+mc^2\right)\psi'=\left[c{\boldsymbol{\alpha}}\cdot\left({\bf p}-\frac{e}{c}{\bf A}\right)+\beta mc^2+e\Phi\right]\psi'.
\label{Dirac2}
\end{equation}
Substituting 
\begin{equation}
\psi'=
 \left(
  \begin{array}{c}
    \phi'\\
    \chi'
  \end{array}
 \right)
\end{equation}
into (\ref{Dirac2}), Dirac spinor is reduced to two component Weyl spinors
]\begin{eqnarray}
&&\left(i\hbar\frac{\partial}{\partial t}-e\Phi\right)\phi'=c\boldsymbol{\sigma}\cdot\left({\bf p}-\frac{e}{c}{\bf A}\right)\chi',\\
&&\left(i\hbar\frac{\partial}{\partial t}-e\phi+2mc^2\right)\chi'=c\boldsymbol{\sigma}\cdot\left({\bf p}-\frac{e}{c}{\bf A}\right)\chi'.
\label{Pauli}
\end{eqnarray}
Retaining only the term $2mc^2\chi'$ in the second equation, we obtain
\begin{equation}
\chi=\frac{1}{2mc}\boldsymbol{\sigma}\cdot \left({\bf p}-\frac{e}{c}{\bf A}\right)\phi.
\label{NR}
\end{equation}
Substituting this into the first equation, we finally obtain the famous Pauli equation,
\begin{equation}
i\hbar\frac{\partial\phi}{\partial t}=\left[\frac{1}{2m}\left({\bf p}-\frac{e}{c}{\bf A}\right)^2+e\Phi-\frac{e}{2mc}\boldsymbol{\sigma}\cdot {\bf H}\right]\phi.
\end{equation}
The current density is
\begin{equation}
{\bf j}=c\psi^*{\boldsymbol{\alpha}}\psi=c(\phi^*\boldsymbol{\sigma}\chi+\chi^*\boldsymbol{\sigma}\phi).
\end{equation}
Substituting $\chi$ of (\ref{NR}) into it, we obtain
\begin{equation}
{\bf j}=\frac{i\hbar}{2m}(\phi\nabla \phi^*-\phi^*\nabla \phi)-\frac{e}{mc}{\bf A}\phi^*\phi+\frac{\hbar}{2m}\nabla \times (\phi^*\boldsymbol{\sigma}\phi).
\label{anapole2}
\end{equation}
In the presence of the EDM (see (\ref{CPodd})), ${\bf j}$ includes pseudo-vector part,
\begin{equation}
{\bf j}_d=id_N\nabla\times(\psi^*\boldsymbol{\gamma}\psi),
\end{equation}
which in two components approximation is reduced to
\begin{equation}
{\bf j}_d=\frac{{\bf d}}{2m}\nabla\times [\phi'\boldsymbol{\sigma}\times({\bf p'+p})\phi],
\end{equation}
where ${\bf p}$ and ${\bf p}'$ are the momenta of $\phi$ and ${\phi}'$, respectively.
\section{Nonrelativistic approximation}

In the heavy nucleon limit, nucleon bilinear forms are approximated as
\begin{eqnarray}
\overline{N}(x)\gamma_0N(x)&=&\delta({\bf r}),~~\overline{N}(x)\boldsymbol{\gamma}N(x)=0,\nonumber\\
\overline{N}(x)\gamma_0\gamma_5N(x)&=&0,~~\overline{N}(x)\boldsymbol{\gamma}\gamma_5N(x)=-\boldsymbol{\sigma}_N\delta({\bf r}).
\label{heavy}
\end{eqnarray}
We are interested in P-odd and T-odd weak interaction in the Fermi coupling between electron and nucleons. In the heavy nucleon limit (\ref{heavy})
these interactions are limited in the following forms,
\begin{equation}
H=\frac{G}{\sqrt{2}}\left(k_1\overline{N}N\overline{e}i\gamma_5e+k_2\frac{1}{2}\epsilon^{\kappa\lambda\mu\nu}\overline{N}\sigma_{\kappa\lambda}N\overline{e}\sigma_{\mu\nu}e\right).
\label{weakCP}
\end{equation}
In the nonrelativistic (heavy nucleon mass) limit it reduces to 
\begin{equation}
H=i\frac{G}{\sqrt{2}}\delta({\bf r})(k_1\gamma_0\gamma_5+4k_2\boldsymbol{\sigma}\cdot\boldsymbol{\gamma}).
\end{equation}
Here you should consider $H$ is sandwiched by electron wave functions.
For the case of a nucleus of charge $Z$ and mass number $A$, it gives \cite{Khriplovich}
\begin{equation}
H=i\frac{G}{\sqrt{2}}\delta({\bf r})\left[(Zk_{1p}+Nk_{1n})\gamma_0\gamma_5+4\left(k_{2p}\sum_p\boldsymbol{\sigma}_p+k_{2n}\sum_n\boldsymbol{\sigma}_n\right)\cdot\boldsymbol{\gamma}\right],
\label{weakCP2}
\end{equation}
\begin{eqnarray}
\langle s_{1/2}|&H&|p_{1/2}\rangle=g\frac{Z^2R}{(N_sN_p)^{3/2}}Ry\left[\gamma(Zk_{1p}+Nk_{1n})-8{\bf j}\cdot\frac{2+\gamma}{3}\langle k_{2p}\sum_p\boldsymbol{\sigma}_p+k_{2n}\sum_n\boldsymbol{\sigma}_n\rangle\right.\nonumber\\
&+& \left. 8{\bf j}\cdot(1-\gamma)\langle k_{2p}\sum_p\left({\bf n}_p(\boldsymbol{\sigma}_p\cdot{\bf n}_p)-\frac{1}{3}\boldsymbol{\sigma}_p\right)+k_{2n}\sum_n\left({\bf n}_n(\boldsymbol{\sigma}_n\cdot{\bf n})-\frac{1}{3}\boldsymbol{\sigma}_n\right)\rangle\right],
\end{eqnarray}
The above arguments can be applied for both paramagnetic and diamagnetic atoms.
Let us apply the above arguments to Cs, Tl, and Xe${}^*$ atoms \cite{Khriplovich}, corresponding to the arguments in Section {\bf 5.2} in the presence of (\ref{weakCP}),
The wave function for Cs is described as
\begin{eqnarray}
|\overline{6s_{1/2},F}\rangle&=&|6s_{1/2},F\rangle-3.7\times 10^{-11}\left[ 0.41k_{1p}+0.59k_{1n}\right.\nonumber\\
&+& \left. 0.74\times 10^{-2}\left(F(F+1)-\frac{33}{2}\right)k_{2p}\right]|6p_{1/2},F\rangle ,
\end{eqnarray}
and, therefore,
\begin{eqnarray}
d(Cs)&=&e\langle \overline{6s_{1/2},F}|z|\overline{6s_{1/2},F}\rangle=-ea_B\times 1.34\times 10^{-10}\nonumber\\
&\times&\left[0.41k_{1p}+0.59k_{1n}+0.74\times 10^{-2}\left(F(F+1)-\frac{33}{2}\right)k_{2p}\right].
\end{eqnarray}
Here $F$ is the total angular momentum of the atom.
The observed value \cite{Wood} is
\begin{equation}
d(Cs)=(-1.8\pm 6.7\pm 1.8)\times 10^{-24}~e~cm.
\end{equation}
For Tl
\begin{equation}
d(\mbox{Tl})=ea_B\cdot 0.96\times 10^{-9}(0.4k_{1p}+0.6k_{1n}-2\cdot 10^{-3}k_{2p}).
\end{equation}
For Xe{}$^*$
\begin{equation}
d(\mbox{Xe}^*)=-1.3\cdot 10^{-10}ea_B(0.41k_{1p}+0.59k_{1n}).
\end{equation}
%For diamagnetic atom, $d(Xe)$ is \cite{Khriplovich} and
%\begin{equation}
%d({}^{129}Xe)=-\frac{8}{9\pi}S\frac{Z^2}{a_B^2}(R_{1/2}+2R_{3/2})\sum_n\frac{r(5p,ns)}{(N_%{5p}N_{ns})^{3/2}}\frac{Ry}{E(ns)-E(5p)}.
%\end{equation}
\section{Strong CP violation}
In the QCD world, the true vacuum is described by the $\theta$ vacuum,
\begin{equation}
|\theta\rangle\equiv \sum_ne^{-in\theta}|n\rangle,~~(n=\mbox{integer}).
\end{equation}
\begin{eqnarray}
\langle \theta'|e^{-iHt}|\theta\rangle&=&\sum_{n,m}e^{im\theta '}e^{-in\theta}\langle m|e^{-iHt}|n\rangle\nonumber\\
\label{theta1}
&=&\sum_{m,n}e^{-i(n-m)\theta}e^{im(\theta'-\theta)}\int [dA]_{n-m}e^{i\int  L d^4x}\\
&=&\sum_\nu e^{-i\nu \theta}\int [dA]_\nu e^{i\int  Ld^4x}\nonumber
%\label{theta1}
\end{eqnarray}
Using that $A_n$ gives
\begin{equation}
n=\frac{1}{16\pi^2}\int d^4 x Tr\left(G_{\mu\nu}\tilde{G}^{\mu\nu}\right)
\label{theta2}
\end{equation}
and substituting (\ref{theta2}) into (\ref{theta1}) we obtain
\begin{equation}
\langle \theta'|e^{-iHt}|\theta\rangle=\sum_\nu \int [dA]_\nu e^{i\int  L_{eff}~d^4x}
\end{equation}
with
\begin{equation}
L_{eff}=L+\frac{\theta}{16\pi^2}\int d^4 x Tr\left(G_{\mu\nu}\tilde{G}^{\mu\nu}\right)
\end{equation}

\section{U(1) problem}
$\theta$ term in (\ref{effective}) comes from the fact that the vacuum in QCD is $|\theta\rangle$, whereas $G\tilde{G}$ term in
(\ref{anomaly}) does from quantum anomaly, occurring irrelevant to Abelian and non Abelian.
In this appendix we will show that these two terms are closely related and lead us to solve U(1) problem.

In the following discussions we consider mass zero quark limit, and $N_f=3$, up, down, strange quarks.
Chiral invariant action has originally $U_L(3)\otimes U_R(3)$ symmetry.
If, as we have considered, QCD vacuum is quark condensate
\begin{equation}
\langle\overline{u}u\rangle=\langle\overline{d}d\rangle=\langle\overline{s}s\rangle,
\end{equation}
action symmetry is reduced to the flavor symmetry U(3) and generates $3^2$ NG bosons.
They are $\pi^\pm,~\pi^0,~K^\pm,~K^0,~\overline{K}^0,~\eta_8$, and $\eta_0$.
Here the first eight particles constitute octet and the last a singlet.
The observed mass eigen states, $\eta$ and $\eta'$ particles, are the linear combinations of $\eta_8$ and $\eta_0$, and their masses are $m_\eta=550$MeV, $m_{\eta'}=958$ MeV.
Weinberg showed \cite{Weinberg2} that the observed $m_{\eta'}$ is too heavy for predicted NG boson,
\begin{equation}
m_{\eta'}\leq \sqrt{3}m_\pi.
\label{u1}
\end{equation}
This is one of the U(1) problems.
Another is concerned with $\eta\rightarrow \pi^+\pi^-\pi^0$ process.

Let us explain these problems \cite{Chen}:
The octet axial vector currents satisfy
\begin{equation}
\partial^\mu J_{5\mu}^a=f_am_a^2\phi^a~~(a=1,...,8)
\label{nopole}
\end{equation}
and
\begin{eqnarray}
\delta_{ab}m_a^2f_a^2&=&i\frac{m_b^2-k^2}{m_b^2}ik_\nu\int d^4x e^{-ikx}\langle 0|T\left(\partial^\mu J_{5\mu}^a(0)\partial^\nu J_{5\nu}^b(x)\right)|0\rangle\nonumber\\
&=&i\frac{m_b^2-k^2}{m_b^2}\left\{ik_\nu\int d^4x e^{-ikx}\langle 0|T\left(\partial^\mu J_{5\mu}^a(0)J_{5\nu}^b(x)\right)|0\rangle\right.\\
&+&\left.\int d^4x e^{-ikx}\langle 0|\delta(x_0)\left[\partial^\mu J_{5\mu}^a(0),~J_{50}^b(x)\right]|0\rangle\right\} .\nonumber
\label{CCR}
\end{eqnarray}
In the low energy limit, if there is no massless pole, this reduces to

\begin{equation}
\delta_{ab}m_a^2f_a=i\int d^4x\langle 0|\delta(x^0)[J_{50}^b(x),\partial^\mu J_{5\mu}^a(x)]|0\rangle
\label{nopole2}
\end{equation}
%\frac{i(m_b^2-k^2)}{f_bm_b^2}\int d^4x e^{-ikx}\la0|T(\partial^\mu J_{5\mu}^a (0)\partial%\nu J_{5\nu} ^b(x)|0\rangle.
%\label{nopole2}
%\end{equation}

Whereas, isosinglet axial vector current constitute ABJ anomaly (\ref{anomaly}).
%\begin{equation}
%\partial^\mu J_{5\mu}=2i(m_u\overline{u}\gamma_\mu\gamma_5 u+m_d\overline{d}\gamma_\%mu\gamma_5 d)+\frac{\alpha_s}{4\pi}G\tilde{G}.
%\label{ABJ}
%\end{equation}
The isosinglet can be described as a sum of SU(3) octet and singlet,
\begin{equation}
J_{5\mu}=\frac{1}{\sqrt{3}}J_{5\mu}^{(8)}+\sqrt{\frac{2}{3}}J_{5\mu}^{(0)}
\label{isosinglet}
\end{equation}
with
\begin{eqnarray}
J_{5\mu}^{(8)}&=&\frac{1}{\sqrt{3}}(\overline{u}\gamma_\mu\gamma_5 u+m_d\overline{d}\gamma_\mu\gamma_5 d-2\overline{s}\gamma_\mu\gamma_5 s).\\
J_{5\mu}^{(0)}&=&\sqrt{\frac{2}{3}}(\overline{u}\gamma_\mu\gamma_5 u+m_d\overline{d}\gamma_\mu\gamma_5 d+\overline{s}\gamma_\mu\gamma_5 s)
\end{eqnarray}
%\begin{equation}
%\delta_{ab}m_a^2f_a=\frac{i(m_b^2-k^2)}{f_bm_b^2}\int d^4x e^{-ikx}\la0|T(\partial^\%mu J_{5\mu}^a (0)\partial^\nu J_{5\nu} ^b(x)|0\rangle
%\end{equation}
%In (\ref{ABJ},
%\begin{equation}
%\frac{\alpha_s}{4\pi}G\tilde{G}=\partial^\mu K_\mu
%\end{equation}
%and by replacing $J_{5\mu}$ by
%\begin{equation}
%\tilde{J}_{5\mu}\equiv J_{5\mu}-K_\mu
%\end{equation}
Taking (\ref{anomaly}), (\ref{Kmu}), and (\ref{CVC}) into considerations,
we obtain the same equation for isosinglet case as (\ref{CCR}) by replacing $J_{5\mu}$ with $\tilde{J}_{5\mu}$,
\begin{eqnarray}
m_0^2f_0^2&=&i\frac{m_0^2-k^2}{m_0^2}\left\{ik_\nu\int d^4x e^{-ikx}\langle 0|T\left(\partial^\mu\tilde{J}_\mu^5(0)\tilde{J}_\nu^5(x)\right)|0\rangle\right.\nonumber\\
&+&\left.\int d^4x e^{-ikx}\langle 0|\delta(x_0)\left[\partial^\mu\tilde{J}_\mu^5(0),~\tilde{J}_0^5(x)\right]|0\rangle\right\}.
\label{CCR2}
\end{eqnarray}
Here $f_0$ is the isoscalar meson decay constant.
If there is no zero mass pole like the octet cases, this relation is same as the octet case (\ref{nopole2}) except for $J_{5\mu}$ replaced by $\tilde{J}_{5\mu}$ , and we obtain
\begin{equation}
m_0^2f_0^2=m_\pi^2f_\pi^2.
\label{singletoctet}
\end{equation}
So, if SU(3) is good symmetry, it goes from (\ref{isosinglet}) and (\ref{singletoctet}) that
\begin{equation}
f_0\geq \frac{1}{\sqrt{3}}f_\pi,
\end{equation}
which directly leads to (\ref{u1}).
However if any massless particle couples to $\tilde{J}_\mu^5$, then first term of (\ref{CCR}) does not vanish and we can evade (\ref{singletoctet}) \cite{Kogut}.
't Hooft showed that this is indeed the case if we take $\theta$ vacuum into consideration correctly \cite{t'Hoft2}.
Also Witten proposed a solution compatible with quark condensate \cite{Witten}:
\begin{equation}
m_{\eta^0}^2=\frac{4N_f^2}{f_{\eta^0}}\left(\frac{\partial^2E_\theta}{\partial\theta^2}\right)_{\theta=0},
\end{equation}
where
\begin{equation}
\left(\frac{\partial^2E_\theta}{\partial\theta^2}\right)_{\theta=0}=\frac{1}{N_c^2}\left(\frac{1}{16\pi^2}\right)^2\int d^4x\langle T\left(Tr(G(x)\tilde{G}(x))Tr(G(0)\tilde{G}(0))\right)\rangle
\end{equation}
\begin{equation}
\langle \pi^+\pi^-\pi^0|\eta\rangle=\frac{m_u-m_d}{F_\pi m_q}\lim_{k\rightarrow o}\langle \pi^+\pi^-|\partial^\mu J_{5\mu}(k)|\eta\rangle.
\end{equation}
The right-hand side vanishes due to momentum conservation. However, it is experimentally observed as $\Gamma (\eta\rightarrow \pi^+\pi^-\pi^0)\approx 200$eV.
This process is occurred via SU(2) violating operator \cite{Sutherland}
\begin{equation}
{\cal L}=\frac{1}{2}(m_u-m_d)(\overline{u}u-\overline{d}d)
\end{equation}
and 
\begin{equation}
\langle 3\pi |{\cal L} |\eta\rangle\rightarrow \frac{(m_u-m_d)A}{\sqrt{2}f_\pi^2},
\end{equation}
where
\begin{eqnarray}
A&\equiv& \langle \pi\pi |(m_u\overline{u}\gamma_5 u+m_d\overline{d}\gamma_5 d)|\eta\rangle\nonumber\\
&=&\frac{1}{2i}\langle \pi\pi |\partial^\mu \tilde{J}_\mu^5(0)|\eta\rangle.
\end{eqnarray}
So this process is suppressed by axial vector current conservation even if $m_u\neq m_d$.
This is another U(1) problem.
%\begin{equation}
%L_{QCD}\rightarrow L_{QCD}+2N_f\varphi \partial^\mu K_\mu
%\end{equation}
%under the chiral transformation
%\begin{equation}
%\psi'=e^{i\gamma_5\varphi}\psi.
%\end{equation}
\section{Schiff moment}
There are several origins for the Schiff moment. Here we discuss the Schiff moment induced by the nuclear EDM when the charge and the EDM distributions ($\rho_q$ and $\rho_d$, respectively) in the nucleus are different \cite{K-L}.

The interaction of the electron with the dipole moment of finite size nucleus is
\bea
V_s&=&\int d^3r'[\rho_d({\bf r}')-\rho_q({\bf r}')]{\bf d}_N\cdot \nabla'\frac{-e}{|{\bf r}-{\bf r}'|}\\
&=&\frac{1}{2}e\int d^3r'[\rho_d({\bf r}')-\rho_q({\bf r}')]d_{N,l}r_m'r_n'\nabla_l\nabla_m\nabla_n\frac{1}{r}.
\eea
Here we may assume \cite{Bohr}:

$\rho_q$ is spherically symmetric.

${\bf d_N}$ coincides with the EDM of a valence nucleon, ${\bf d_N}=d_{p,n}\boldsymbol{\sigma}$.

$\rho_d$ is due to the valence nucleon.

Then
\begin{equation}
V_s=\frac{1}{2}ed_{p,n}\int d^3r'4\pi r'^2\left[\rho_d({\bf r}')\langle \sigma_ln_mn_n\rangle-\rho_q(r')\frac{1}{3}\delta_{mn}\langle \sigma_l\rangle\right]\nabla_l\nabla_m\nabla_n\frac{1}{r},
\end{equation}
where ${\bf n}={\bf r}'/r'$. 
Let us divide $
\nabla_l\nabla_m\nabla_n$ as
\begin{eqnarray}
&&\left[\nabla_l\nabla_m\nabla_n-\frac{1}{5}(\delta_{lm}\nabla_n+\delta_{mn}\nabla_l+\delta_{nl}\nabla_m)\Delta\right]\nonumber\\
&&+\frac{1}{5}(\delta_{lm}\nabla_n+\delta_{mn}\nabla_l+\delta_{nl}\nabla_m)\Delta .
\end{eqnarray}
The first term corresponds to the electron interaction with the $2^3$-pole moment of the nucleus.

\begin{eqnarray}
\mbox{The second term}&=&\left[\rho_d\langle\sigma_ln_mn_n\rangle-\rho_q\frac{1}{3}\delta_{mn}\langle\sigma_l\rangle\right]\frac{1}{5}(\delta_{lm}\nabla_n+\delta_{mn}\nabla_l+\delta_{nl}\nabla_m)\nonumber\\
&=&-\left[\frac{1}{3}\rho_q\langle\boldsymbol{\sigma}\rangle-\frac{1}{5}\rho_d\langle 2\boldsymbol{\sigma}\cdot \boldsymbol{n}\boldsymbol{n}+\boldsymbol{\sigma}\rangle\right]\cdot \nabla
\end{eqnarray}
Here we use \footnote{The following arguments are indebted to discussions with T. Sato} \cite{Bohr}
\begin{eqnarray}
\boldsymbol{\sigma}\cdot \boldsymbol{n}\boldsymbol{n}&=&\frac{1}{3}\boldsymbol{\sigma}-\frac{\sqrt{8\pi}}{3}[Y_2\otimes\boldsymbol{\sigma}]_{(1)}\nonumber\\
\boldsymbol{\sigma}&=&\sqrt{4\pi}[Y_0\otimes\boldsymbol{\sigma}]_{(1)}\\
2\boldsymbol{\sigma}\cdot\boldsymbol{n}\boldsymbol{n}+\boldsymbol{\sigma}&=&\sqrt{4\pi}\left(\frac{5}{3}\left[Y_0\otimes\boldsymbol{\sigma}\right]_{(1)}-\frac{2\sqrt{2}}{3}\left[Y_2\otimes \boldsymbol{\sigma}\right]_{(1)}\right)\nonumber
\end{eqnarray}
where
\begin{equation}
[Y_l\otimes\chi]_{j,m}=\sum_{m_l,m_s}\langle l,m_l;\frac{1}{2},m_s|j,m\rangle Y_{l,m_l}(\theta,\phi)\chi_{m_s}
\end{equation}
with the Clebsch-Gordan coefficient $\langle l,m_l;\frac{1}{2},m_s|j,m\rangle$ related with 3j-symbol 
\begin{equation}
\langle k_1,q_1,k_2,q_2|K,Q\rangle=(-1)^{k_1-k_2+Q}\sqrt{2K+1}
\left(
\begin{array}{ccc}
   
  k_1 & k_2 & K\\
  q_1& q_2 & -Q
     \end{array}
 \right).
\end{equation}
The following equation is the Wigner-Eckart theorem (the definition of reduced matrix element) $\langle~||~||~\rangle$,
\begin{equation}
\langle \kappa m|O_{JM}|\kappa m'\rangle =\frac{1}{\sqrt{2j+1}}\langle jmJM|jm'\rangle \langle\kappa||O_J||\kappa\rangle
\end{equation}
where $\kappa$ is defined by (\ref{kappa}) and
\begin{equation}
|\kappa\rangle\equiv|[Y_l(\boldsymbol{n})\otimes \chi]_{(j)}\rangle.
\end{equation}
It should be noted that the reduced matrix element has no dependence on $m,~m',~\mbox{nor} M$.

For $J=1$ case
\begin{equation}
\langle\kappa|\vec{O}|\kappa\rangle=\langle \kappa|\vec{J}|\kappa\rangle\frac{\langle\kappa||\vec{O}||\kappa\rangle}{\langle\kappa||\vec{J}||\kappa\rangle}.
\end{equation}
\begin{equation}
\langle \kappa||\vec{J}||\kappa\rangle=\sqrt{J(J+1)(2J+1)}
\end{equation}
\begin{eqnarray}
\langle\kappa||[Y_l\otimes \boldsymbol{\sigma}]_{(1)}||\kappa\rangle&=&2|\kappa|(-1)(-1)^{|\kappa|}(j,\frac{1}{2};j,-\frac{1}{2}|1,0)\nonumber\\
&&\times\left\{
  \begin{array}{l}
   \displaystyle
  \frac{1-2\kappa}{\sqrt{3}}~~~~~~\mbox{for}~l=0
\\
   \displaystyle
   -\frac{2(1+\kappa)}{\sqrt{6}}~~\mbox{for}~l=2
  \end{array}
 \right.\\
&&=\frac{1}{2}\sqrt{\frac{2j+1}{j(j+1)}}\times\left\{
  \begin{array}{l}
   \displaystyle
  1-2\kappa~~\mbox{for}~~l=0
\\
   \displaystyle
   -\sqrt{2}(1+\kappa)~~\mbox{for}~l=2.
  \end{array}
 \right.
\end{eqnarray}
Thus we obtain
\begin{eqnarray}
&&\left\{\rho_d\langle\sigma_ln_mn_n\rangle-\rho_q\frac{1}{3}\delta_{mn}\langle\sigma_l\rangle\right\}\frac{1}{5}\left[\delta_{lm}\nabla_m+\delta_{mn}\nabla_l+\delta_{nl}\nabla_m\right]\nonumber\\
&&=\left[\frac{1}{3}\rho_q(\kappa-\frac{1}{2})-\frac{1}{5}\rho_d(\kappa-\frac{3}{2})\right]\frac{\langle{\bf j}\rangle\cdot\nabla}{j(j+1)}.
\end{eqnarray}
So
\begin{equation}
V_s=\frac{ed_{p,n}}{2}\left[r_q^2\frac{1}{3}(\kappa -\frac{1}{2})-r_d^2\frac{1}{5}(\kappa -\frac{3}{2})\right]\frac{{\bf j}\cdot \nabla}{j(j+1)}4\pi\delta({\bf r}),
\end{equation}
where the mean squared radii are defined by
\begin{equation}
r_{q,d}^2\equiv \int d^3r'~r'^2\rho_{q,d}({\bf r}').
\label{r^2}
\end{equation}
In this derivation we assumed that the nuclear charge is uniformly distributed over a sphere of radius $r_0=1.2\times 10^{-13}A^{1/3}$ cm, and $r_q^2=\frac{3}{5}r_0^2$. Also we may assume $r_d^2=r_q^2$ \cite{Khriplovich}.
Then we get the final expression for the Schiff moment.
\begin{equation}
{\bf S}=d_{p,n}r_0^2\frac{4\pi}{25}\frac{(\kappa+1){\bf j}}{j(j+1)}.
\end{equation}
\section{Effective Hamiltonian in molecule \label{Das}}
We have said that there appears huge internal electric field ${\bf E}_{int}$ in polar molecule. Here we consider how to estimate $E_{int}$.
The Dirac-Coulomb Hamiltonian is
\be
H_0=\sum_i\{c\boldsymbol{\alpha}_i\cdot {\bf p}_i+\beta_imc^2+V_{nucl}({\bf r}_i)\}+\sum_{i<j}\frac{1}{r_{ij}}
\ee
and P,T-odd perturbation (the intrinsic part of $H_{PTV}$) is
\be
H'=-d_e\sum_i\beta_i\boldsymbol{\sigma}\cdot {\bf E}_i^{int}
\ee
with
\be
{\bf E}_{i, int}=-\nabla_i\left(V_{nucl}({\bf r}_i)+2\sum_{i>j}\frac{e^2}{r_{ij}}\right).
\ee
Here electric field is given by Eq.\bref{EM} with $\phi=\sum_i\{V_{nucl}({\bf r}_i)\}+\sum_{i<j}\frac{1}{{\bf r}_{ij}}\}$.
We are considering a static field, $\frac{1}{c}\frac{\pa {\bf A}}{\pa t}=0$, and $H'$ is represented as
\be
H'=d_e\sum_i\left[\beta_i\boldsymbol{\sigma}_i\cdot\nabla_i,~H-T\right].
\label{intrinsic}
\ee
Here $T$ is kinetic term of electron
\be
T=\sum_i\{c\boldsymbol{\alpha}_i\cdot {\bf p}_i+\beta_imc^2\}.
\ee
The expectation value w.r.t. the eigen function of $H_0$ gives
\be
\la \Psi |\sum_i\left[\beta_i\boldsymbol{\sigma}_i\cdot\nabla_i,~H\right] |\Psi\ra=0.
\ee
Whereas,
\be
\sum_i\left[\beta_i\boldsymbol{\sigma}_i\cdot\nabla_i,~T\right]\nonumber\\
=i\sum_i\sum_j\{\left[\beta_i\boldsymbol{\sigma}_i\cdot {\bf p}_i, \boldsymbol{\alpha}_j\cdot{\bf p}_j\right]
+\left[\beta_i\boldsymbol{\sigma}_i\cdot {\bf p}_i,~ (\beta_j-1)m_jc\right]\}.
\ee
Here the second term vanishes and the first term gives
\be
i\sum_i\sum_j\left[\beta_i\boldsymbol{\sigma}_i\cdot {\bf p}_i, \boldsymbol{\alpha}_j\cdot{\bf p}_j\right]\nonumber\\
=
 \left\{
  \begin{array}{l}
   \displaystyle
   \sum_i2i\beta_i\gamma_5{\bf p}_i^2~~\mbox{for}~ i=j.
\\
   \displaystyle
   0~~\mbox{for}~i\neq j.
  \end{array}
 \right.
\ee
%So 
%\be
%\psi=\left(
%   \begin{array}{c}
%    \varphi\\
%    \chi
%   \end{array}
%   \right)
%\ee
Thus $H'$ of \bref{intrinsic} is rewritten as \cite{Das}
\be
H'_{eff}=-2icd_e\sum_i\beta_i\gamma_5{\bf p}_i^2
\ee
and we obtain finally
\be
-2ic \la \psi_0|\beta\gamma_5p^2|\psi_0\ra=4cp^2\Im(\varphi^\dagger\chi).
\ee
The enhancement factor is given by
\be
K=\sum_n\frac{\la \psi |-2ic\beta_i\gamma_5{\bf p}_i^2 |\phi_n\ra \la\phi_n |\sum_iez_i |\psi\ra}{E-E_n}+h.c.
\ee
So the detailed calculations are reduced to the electron wave functions in atoms and molecules. For molecular case, unfortunately, only $H_2^+$ can be solved in the Born-Oppenheimer approximation \cite{L-L1}. However, its perturvation expansion around atomic level is also interesting since this method is applicable to the other diatomic molecule \cite{L-L2}.
For more detailed explanation for diatomic case, see \cite{Molecule}.
%For YbF, $Yb^+=[Xe]4f^{14}6s^1$, and we must calculate this wave function for obtaining $E%_{int}$.
\section{Spin motion in storage ring \label{storage}}
There are many ongoing and near future experiments measuring EDMs and anomalous MDMs of charged particles. There one of the most important equations is the following spin precession equation,
\be
\frac{d\bsigma}{dt}=\bomega_s\times \bsigma
\ee
where
\be
\bomega_s=-\frac{e}{m}\left[\left(G+\frac{1}{\gamma}\right){\bf H}-\left(G+\frac{1}{\gamma+1}\right){\bf v}\times{\bf E}+\frac{\eta}{2}\left({\bf v}\times {\bf H}+{\bf E}\right)\right].
\label{Nelson1}
\ee
(Notations will be explained shortly.) However, curiously enough, the explicit derivation of this equation has not been published \cite{Nelson}.
There are several confusions on the interpretations of this equation. In this appendix we give an explicit derivation of this equation. \footnote{For spinor case the above equation was derived by Silenko \cite{Silenko}. We are greatly indebted to Silenko for the discussions of this appendix.}

In relativistic theory, spin vector is not conserved. We must derive an equation of motion for the spin when the particle moves. For that purpose, it is convenient to introduce 4-pseudovector $a^\mu$ defined by \cite{Landau}
\be
a^\mu=(0, \boldsymbol{\zeta}), ~~p^\mu=(m,{\bf 0})
\ee
in the rest frame. So in any frame
\be
a^\mu p_\mu=0. ~~a_\mu a^\mu=-\boldsymbol{\zeta}^2.
\ee
In a moving frame with velocity ${\bf v}={\bf p}/\epsilon$, $a^\mu=(a^0,~{\bf a})$ is given by
\be
{\bf a}=\boldsymbol{\zeta}+\frac{{\bf p}(\boldsymbol{\zeta}\cdot {\bf p})}{m(\epsilon+m)},~~a^0=\frac{{\bf a}\cdot{\bf p}}{\epsilon}=\frac{{\bf p}\cdot\boldsymbol{\zeta}}{m},~~{\bf a}^2=\boldsymbol{\zeta}^2+\frac{({\bf p}\cdot \boldsymbol{\zeta})^2}{m^2}.
\label{spin}
\ee
Using this 4-pseudovector $a^\mu$, relativistic spin motion in electromagnetic field is given by
\be
\frac{d a^\mu}{d\tau}=\alpha F^{\mu\nu}a_\nu+\beta u^\mu F^{\nu\lambda}u_\nu a_\lambda+\gamma F^{*\mu\nu}a_\nu+\delta u^\mu F^{*\nu\lambda}u_\nu a_\lambda
\label{BMT}
\ee
with $F^{*\mu\nu}=\frac{1}{2}\epsilon^{\mu\nu\rho\sigma}F_{\rho\sigma}$.
Here $\alpha,~\beta,~\gamma,~\delta$ are coefficients whose meanings are determined as follows. 
In the rest frame, \bref{BMT} becomes
\be
\frac{d a^i}{dt}=\frac{d\zeta^i}{dt}=\alpha F^{ij}\zeta_j+\gamma F^{*ij}\zeta_j=\alpha (\boldsymbol{\zeta}\times {\bf H})^i+\gamma ({\bf E}\times \boldsymbol{\zeta})^i.
\label{NRspin}
\ee
In nonrelativistic case, Hamiltonian is
\be
H=H'-\mu\boldsymbol{\sigma}\cdot{\bf H}-d\boldsymbol{\sigma}\cdot{\bf E},
\ee
where $H'$ includes all terms independing of spin terms. The time variation of spin ${\bf s}=\boldsymbol{\sigma}/2$ is
\be
\dot {\bf s}=i(H{\bf s}-{\bf s}H)=2\mu {\bf s}\times {\bf H}+2d{\bf s}\times {\bf E}.
\ee
Comparing this equation with \bref{NRspin}, we obtain 
\be
\alpha=2\mu, ~~\gamma=-2d.
\ee
As for $\beta$ term, it goes from the equation of motion (up to P-odd term)
\be
m\frac{d u^{\mu}}{d\tau}=eF^{\mu\nu}u_\nu
\label{Lorentz}
\ee
and from $a_\mu u^\mu=0$ that
\be
u_\mu\frac{da^\mu}{d\tau}=-a_\mu\frac{du^\mu}{d\tau}=\frac{e}{m}F^{\mu\nu}u_\mu a_\nu.
\ee
On the other hand, multiplying $u_\mu$ on \bref{BMT} and taking $u_\mu u^\mu=1$ into account, we obtain 
\be
u_\mu\frac{da^\mu}{d\tau}=(2\mu+\beta) F^{\mu\nu}u_\mu a_\nu+(-2d+\delta)F^{*\mu\nu}u_\mu a_\nu
\ee
and then
\be
\beta=-2\left(\mu-\frac{e}{2m}\right)\equiv -2\mu ',~~\delta=2d.
\ee

Thus we obtain
\be
\frac{da^\mu}{d\tau}=2\mu F^{\mu\nu}a_\nu-2\mu'u^\mu F^{\nu\lambda}u_{\nu}a_\lambda-2d(F^{*\mu\nu}a_\nu-u^\mu F^{*\nu\lambda}u_\nu a_\lambda).
\ee
This equation is the generalized Bargmann-Michel-Telegdi (BMT) equation \cite{BMT}. 
The spatial part of this equation is decribed as
\bea
\frac{d{\bf a}}{dt}&=&\frac{2\mu m}{\epsilon}{\bf a}\times {\bf H}+\frac{2\mu m}{\epsilon}({\bf a}\cdot{\bf v}){\bf E}-\frac{2\mu'\epsilon}{m}{\bf v}({\bf a}\cdot {\bf E})+\frac{2\mu'\epsilon}{m}{\bf v}({\bf v}\cdot({\bf a}\times{\bf H}))+\frac{2\mu'\epsilon}{m}{\bf v}({\bf a}\cdot{\bf v})({\bf v}\cdot{\bf E})\nonumber\\
&-&\frac{2d m}{\epsilon}\left[({\bf a}\cdot {\bf v}){\bf H}-{\bf a}\times {\bf E}+\gamma^2{\bf v}\left\{-{\bf a}\cdot {\bf H}-{\bf v}\cdot ({\bf a}\times {\bf E})+({\bf a}\cdot{\bf v})({\bf v}\cdot{\bf H})\right\}\right].
\label{BMT2}
\eea
%\gamma ({\bf v}\cdot \boldsymbol{\zeta}){\bf H}+\boldsymbol{\zeta}\times {\bf E}+\frac{\gamma^2}{1+\gamma}
%(\boldsymbol{\zeta}\cdot {\bf v}){\bf v}\right)
%\eea
We consider the time development of $\bz$. Let us first consider in the absence of EDM.
Using the equation of motion \bref{Lorentz} or equivalently its decompositions into spatial and temporal components,
\be
\frac{d{\bf p}}{dt}=e{\bf E}+e{\bf v}\times {\bf H},~~\frac{d\epsilon}{dt}=e{\bf v}\cdot {\bf E},
\label{eqm}
\ee
Eq. \bref{BMT2} is described in terms of the rest frame spin $\boldsymbol{\zeta}$ as
\bea
\frac{d\boldsymbol{\zeta}}{dt}&=&\frac{2\mu m+2\mu'(\epsilon-m)}{\epsilon}\boldsymbol{\zeta}\times {\bf H}+\frac{2\mu'\epsilon}{\epsilon +m}({\bf v}\cdot{\bf H})({\bf v}\times \boldsymbol{\zeta})+\frac{2\mu m+2\mu'\epsilon}{\epsilon+m}\boldsymbol{\zeta}\times ({\bf E}\times {\bf v})\nonumber\\
%&-&\frac{2dm}{\epsilon}\left(\frac{\epsilon}{m}({\bf v}\cdot\boldsymbol{\zeta}){\bf H}-\bo%ldsymbol{\zeta}\times{\bf E}-\frac{\epsilon^2}{m(\epsilon+m)}(\boldsymbol{\zeta}\cdot {\bf% v}){\bf v}\times {\bf E}\right)\\
&=&\frac{e}{2m}\left(g-2+2\frac{m}{\epsilon}\right)\boldsymbol{\zeta}\times {\bf H}+\frac{e}{2m}(g-2)\frac{\epsilon}{\epsilon+m}({\bf v}\cdot{\bf H}){\bf v}\times \boldsymbol{\zeta}+\frac{e}{2m}\left(g-\frac{2\epsilon}{\epsilon+m}\right)\boldsymbol{\zeta}\times({\bf E}\times {\bf v})\nonumber\\
%&-&\frac{2dm}{\epsilon}\left(\gamma({\bf v}\cdot\boldsymbol{\zeta}){\bf H}-\boldsymbol{\ze%ta}\times{\bf E}-\frac{\gamma^2}{1+\gamma}(\boldsymbol{\zeta}\cdot {\bf v}){\bf v}\times {%\bf E}\right)\nonumber\\
%&-&\gamma^2{\bf v}\left({\bf a}\rightarrow \bz in \gamma^2{\bf v} term of (K14) \right)
%cdot {\bf H})+\frac{\gamma^2}{1+\gamma}(\bz\cdot{\bf v})(\bz\cdot {\bf H})+{\bf v}\cdot{\bf%\bz\times{\bf E})+\frac{\gamma^2}{1+\gamma}(\bz\cdor{\bf v})({\bf v}\times{\bf E})\cdot{\b%f v}-\gamma(\bz \cdot {\bf v})({\bfv}\cdot{\bf H}\right\}\right]
\label{bmt3}
\eea
with
\be
\mu=\frac{g}{2}\frac{e}{2m}.
\ee
In order to obtain the spin precession, we must subtract the rotation of particles moving around the storage ring. 
It goes from \bref{eqm} that
\be
\frac{d{\bf v}}{dt}=\frac{e}{m\gamma}\left({\bf E}+{\bf v}\times {\bf H}-{\bf v}({\bf v}\cdot{\bf E})\right).
\label{Lorentz}
\ee
Hereafter we consider the experimental situation where 
\be
{\bf H}\cdot{\bf v}=0,~~{\bf E}\cdot{\bf v}=0.
\label{transversal}
\ee
Then \bref{Lorentz} is rewritten as
\be
\frac{d{\bf v}}{dt}={\bf \Omega}_p\times {\bf v},
\ee
where
\be
{\bf \Omega}_p=\frac{e}{m\gamma}\left(\frac{{\bf v}\times {\bf E}}{v^2}-{\bf H}\right).
\ee

Eq.\bref{bmt3} is reduced to
\bea
\frac{d\boldsymbol{\zeta}}{dt}&=&\frac{e}{m}\left[\left(G+\frac{1}{\gamma}\right)\boldsymbol{\zeta}\times{\bf H}+\left(G+\frac{1}{\gamma +1}\right)\bz\times({\bf E}\times {\bf v})\right]\nonumber\\
&=&{\bf \Omega}_s\times \boldsymbol{\zeta}
\eea
with
\be
G=\frac{g-2}{2}
\ee 
and
\be
{\bf \Omega}_s\equiv -\frac{e}{m}\left[\left((G+\frac{1}{\gamma}\right){\bf H}+\left(G+\frac{1}{\gamma +1}\right){\bf E}\times {\bf v}\right].
\ee

Consequently we obtain
\be
{\bf \Omega}(d=0)={\bf \Omega}_s-{\bf \Omega}_p=-\frac{e}{m}\left[G{\bf H}+\left(G-\frac{1}{\gamma^2-1}\right){\bf E}\times {\bf v}\right].
\ee
In the presence of EDM $d\neq 0$, it should be remarked that all effects of \bref{Lorentz} (CP-even) are imposed on the MDM part. The change of $\frac{d{\bf a}}{dt}$ to $\frac{d\bz}{dt}$ in EDM case are obtaind by
\be
\frac{d\bz}{dt}=\frac{d{\bf a}}{dt}-\frac{\gamma{\bf v}}{\gamma+1}\left(\frac{d{\bf a}}{dt}\cdot
{\bf v}\right).
\label{deq}
\ee
Substituting \bref{BMT2} into \bref{deq}, EDM part is given by
\bea
&&\frac{d{\bf a}}{dt}=2d\left[\frac{1}{\gamma}\left(\bz\times
{\bf E}\right)+\frac{\gamma}{\gamma+1}({\bf v}\times {\bf E})(\bz\cdot{\bf v})-(\bz\cdot{\bf v}){\bf H}+
\gamma{\bf v}(\bz\cdot{\bf H})\right.\nonumber\\ 
&&\left. -\frac{\gamma^2}{\gamma+1}{\bf v}({\bf v}\cdot{\bf H})(\bz\cdot{\bf v})+\gamma{\bf v}({\bf v}\cdot
(\bz\times{\bf E}))\right]
\eea
and
\be
-\frac{\gamma{\bf v}}{\gamma+1}\left(\frac{d{\bf a}}{dt}\cdot
{\bf v}\right)=-2d\frac{\gamma{\bf v}}{\gamma+1}\left[\gamma({\bf v}\cdot(\bz\times{\bf E}))
+\frac{\gamma^2-1}{\gamma}(\bz\cdot{\bf H})-\gamma(\bz\cdot{\bf v})({\bf v}\cdot{\bf H})\right].
\label{HFW}
\ee
So $\bz$ spin precession due to EDM is given by
\be
\frac{d\bz}{dt}=2d\left[\bz\times
{\bf E}+\frac{\gamma}{\gamma+1}({\bf v}\cdot{\bf E})({\bf v}\times\bz)+\bz\times({\bf v}\times
{\bf H})\right].
\label{HFZ}
\ee
Finally we obtain $\bomega_s$ \bref{Nelson1} with both MDM and EDM and the subtracted rotation angular velocity, 
\be
\bomega(d\neq 0)=\bomega_s-\bomega_p=-\frac{e}{m}\left[G{\bf H}+\left(G-\frac{1}{\gamma^2-1}\right){\bf E}\times {\bf v}+\frac{1}{2}\eta\left({\bf E}+{\bf v}\times{\bf H}\right)\right]
\label{BMTg}
\ee
with
\be
d=\frac{\eta}{2}\frac{e}{2m}.
\ee

\end{document}